\documentclass{aa} 
\usepackage{graphicx}
\usepackage{url}
\usepackage[breaklinks=true]{hyperref}
\hypersetup{colorlinks=true,linkcolor=blue,citecolor=blue,filecolor=blue,urlcolor=blue}
\usepackage{twoopt}
\usepackage{natbib}
\bibpunct{(}{)}{;}{a}{}{,} 
\usepackage{amssymb}
\usepackage{float}
\usepackage{color}
\usepackage{epstopdf}
\usepackage{ctable}
\usepackage{multirow}
\usepackage[font=small]{caption}
\usepackage{multirow}
\usepackage{comment}
\usepackage{txfonts}
\usepackage{longtable}
\usepackage{moresize}

\usepackage{xcolor}

\newcommand{\jybeam}{${\rm Jy~beam}^{-1}$}
\newcommand{\rms}{{\sigma_{\rm rms}}}

\begin{document} 

\title{Limits and challenges of the detection of cluster-scale diffuse radio emission at high redshift}
\subtitle{The Massive and Distant Clusters of \textit{WISE} Survey (MaDCoWS) in LoTSS-DR2}

\author{
G.~Di Gennaro\inst{\ref{inst:ira},\ref{inst:hamb}}
\and
M,~Br\"uggen\inst{\ref{inst:hamb}}
\and
E. Moravec\inst{\ref{inst:gbo}}
\and 
L. Di Mascolo\inst{\ref{inst:kapteyn},\ref{inst:nice}}
\and 
R.J. van Weeren\inst{\ref{inst:leiden}} 
\and 
G. Brunetti\inst{\ref{inst:ira}} 
\and 
R. Cassano\inst{\ref{inst:ira}} 
\and 
A. Botteon\inst{\ref{inst:ira}} 
\and 
E. Churazov\inst{\ref{inst:mpa},\ref{inst:iki}} 
\and 
I. Khabibullin\inst{\ref{inst:usm},\ref{inst:mpa},\ref{inst:iki}} 
\and
N. Lyskova\inst{\ref{inst:iki},\ref{inst:kazan}}
\and 
F. de Gasperin\inst{\ref{inst:ira}} 
\and 
M.J. Hardcastle\inst{\ref{inst:car}} 
\and 
H.J.A. R\"ottgering\inst{\ref{inst:leiden}} 
\and 
T. Shimwell\inst{\ref{inst:astron},\ref{inst:leiden}} 
\and 
R. Sunyaev\inst{\ref{inst:iki},\ref{inst:mpa}} 
\and 
A. Stanford\inst{\ref{inst:cal}}
}

\institute{
{INAF - Istituto di Radioastronomia, via P. Gobetti 101, 40129 Bologna, Italy}\label{inst:ira}
\and
{Hamburger Sternwarte, Universit\"at Hamburg, Gojenbergsweg 112, 21029 Hamburg, Germany}\label{inst:hamb}
\and
{Green Bank Observatory, P.O. Box 2, Green Bank, WV 24944, USA}\label{inst:gbo} 
\and 
{Kapteyn Astronomical Institute, University of Groningen, Landleven 12, 9747 AD, Groningen, The Netherlands}\label{inst:kapteyn}
\and
{Universit\'e C\^ote d'Azur, Observatoire de la C\^ote d’Azur, CNRS, Laboratoire Lagrange, France}\label{inst:nice} 
\and 
{Leiden Observatory, Leiden University, PO Box 9513, 2300 RA Leiden, The Netherlands}\label{inst:leiden} 
\and 
{Max Planck Institute for Astrophysics, Karl-Schwarzschild-Str. 1, D-85741 Garching, Germany}\label{inst:mpa} 
\and 
{Space Research Institute (IKI), Profsoyuznaya 84/32, Moscow 117997, Russia}\label{inst:iki} 
\and 
{Universitäts-Sternwarte, Fakultät für Physik, Ludwig-Maximilians-Universität München, Scheinerstr.1, 81679 München, Germany}\label{inst:usm} 
\and 
{Centre for Astrophysics Research, University of Hertfordshire, College Lane, Hatfield AL10 9AB, UK}\label{inst:car} 
\and 
{ASTRON, Netherlands Institute for Radio Astronomy, Oude Hoogeveensedijk 4, 7991 PD, Dwingeloo, The Netherlands}\label{inst:astron} 
\and
{Kazan Federal University, Kremlevskaya Str. 18, 420008 Kazan, Russia}\label{inst:kazan} 
\and
{University of California, Davis, CA 95616, USA}\label{inst:cal}
}

\date{Received 28/11/2024; Accepted 26/02/2025}

\abstract
{Diffuse radio emission in galaxy clusters is a tracer of ultra-relativistic particles and $\mu$G-level magnetic fields, and is thought to be triggered by cluster merger events. 
In the distant Universe (i.e. $z>0.6$), such sources have been observed only in a handful of systems, and their study is important to understand the evolution of large-scale magnetic fields over the cosmic time. Previous studies of nine {\it Planck} clusters up to $z\sim0.9$ suggest a fast amplification of cluster-scale magnetic fields, at least up to half of the current Universe's age, and steep spectrum cluster scale emission, in line with particle re-acceleration due to turbulence. 
In this paper, we investigate the presence of diffuse radio emission in a larger sample of galaxy clusters reaching even higher redshifts (i.e. $z\gtrsim1$). 
We selected clusters from the Massive and Distant Clusters of {\it WISE} Survey (MaDCoWS) with richness $\lambda_{15}>40$ covering the area of the second data release of the LOFAR Two-Meter Sky Survey (LoTSS-DR2) at 144 MHz. These selected clusters are in the redshift range $0.78-1.53$ (with a median value of 1.05).
We detect the possible presence of diffuse radio emission, with the largest linear sizes of $350-500$ kpc, in 5 out of the 56 clusters in our sample. If this diffuse radio emission is due to a radio halo, these radio sources lie on or above the scatter of the $P_\nu-M_{500}$ radio halo correlations (at 150 MHz and 1.4 GHz) found at $z<0.6$, depending on the mass assumed. We also find that these radio sources are at the limit of the detection by LoTSS, and therefore deeper observations will be important for future studies.}

\keywords{
galaxies: clusters: general -- galaxies: clusters: intracluster medium -- cosmology: large-scale structure of Universe -- radiation mechanisms: non-thermal
}

\titlerunning{MaDCoWS in LOFAR-DR2}
\authorrunning{G. Di Gennaro et al.}
\maketitle

%

\section{Introduction}

In the $\rm\Lambda CDM$ cosmology, galaxy clusters grow via accretion of matter along the filaments of the cosmic web, and via mergers with other clusters and groups of galaxies \citep{press+schecter74,springel+06}. Mergers involving these large-scale structures are the most energetic events in the Universe, releasing up to $10^{64}$ erg into the intracluster medium (ICM) within a cluster crossing time \citep[i.e. $\sim1$ Gyr;][]{markevitch+sarazin+vikhlinin99}. These events affect the dynamics of the cluster galaxies \citep[e.g.][]{golovich+19a} and their properties \citep{stroe+17}, and trigger turbulence and shocks \citep{markevitch+vikhlinin07}, with most of the energy eventually transferred to the ICM. Turbulence and shocks are thought to play an important role in (re-)accelerating particles up to relativistic energies (Lorentz factor $\gamma_L\gg10^3$) and in amplifying magnetic fields up to a few $\rm\mu G$ \citep{brunetti+jones14,carilli+taylor02}. Extended, cluster-centric, non-thermal radiation in the form of radio halos is observed in a large number of merging clusters \citep{vanweeren+19,botteon+22}, especially at low radio frequencies ($\nu\lesssim100$ MHz) due to their steep-spectra\footnote{Here, we define the spectral index $\alpha\lesssim-1$, with $S_\nu\propto\nu^\alpha$.}.
The halos are proposed to be generated via stochastic Fermi-II particle re-acceleration mechanisms due to turbulence \citep[e.g.][]{brunetti+01, petrosian01,brunetti+lazarian07,brunetti+lazarian16}. An additional, although sub-dominant \citep{adam+21}, contribution to the radio halo emission could be provided by proton-proton collisions, which generate secondary electrons \citep{brunetti+lazarian11,pinzke+17,brunetti+17}. Since the turbulent energy budget is set by the cluster masses \citep{cassano+brunetti05}, more massive merging clusters are likely to host more powerful radio halos. Less powerful radio halos are also expected to have steeper spectral indices \citep[i.e. $\alpha\lesssim-1.5$, see also][]{pasini+24}. These properties have been observed by correlations in the halo power-mass diagram \citep{cuciti+21b,cuciti+23} and with ultra-steep spectrum sources \citep[e.g.][]{brunetti+08}.

Merger-induced turbulence associated with radio halos is thought to drive a small-scale dynamo, which amplifies magnetic fields after several eddy turnover times \citep[i.e. several Gyr;][]{beresnyak+miniati16}. Estimates of cluster magnetic fields in the local Universe come from Faraday Rotation Measures, source depolarisation, inverse Compton (IC) upper limits, and equipartition arguments \citep{govoni+04,bonafede+10,osinga+22}. These techniques agree in setting an average magnetic field level of a few $\mu$G, with a decreasing radial profile \citep{bonafede+10}. These values have been found to remain roughly constant up to $z\sim0.9$, at least in massive systems, implying fast magnetic amplification during the formation of the first large-scale structures in the Universe \citep{digennaro+21a,digennaro+21c}. The presence of diffuse radio emission on the Mpc scale was also recently reported in an extremely distant cluster, at $z=1.23$ \citep[i.e. ACT-CLJ0329.2-2330;][]{sikhosana+24}.
The origin of the ``seeds'' of cluster magnetic fields remains unclear, and it is still unknown whether they have a primordial (i.e. generated during the first phases of the Universe) or an astrophysical (i.e. injected by galactic winds, active galactic nuclei, and/or starbursts) origin  \citep{subramanian+06,tjemsland+23}. Although numerical simulations suggest that a small-scale dynamo erases this information \citep{dolag+99,cho14,donnert+18,dominguez-fernandez+19}, observing synchrotron emission in distant galaxy clusters still provides constraints on the mechanisms of magnetic amplification and particle acceleration. Particle re-acceleration mechanisms predict a low occurrence fraction of these high-$z$ radio sources \citep{cassano+23} and steep spectral index (i.e. $\alpha\lesssim-1.5$) because of the stronger losses due to the IC effect on the emitting particles.

In this paper, we investigate diffuse radio emission in a large sample of distant (i.e. $z>0.7$) galaxy clusters selected from the Massive and Distant Clusters of Wise Survey \citep[MaDCoWS;][]{gonzalez+19} using data from the second data release of the LOw Frequency Array \citep[LOFAR;][]{vanhaarlem+13} Two-Meter Sky Survey \citep[LoTSS-DR2;][]{shimwell+22}. The combination of these two surveys represents a unique opportunity to study the cosmic evolution of the cluster-scale synchrotron emission, as MaDCoWS collects more than 2,000 clusters at high redshift ($z\geq0.7$) and LoTSS-DR2 currently provides the most sensitive ($\rm 100~\mu$\jybeam\ at $6''$ resolution) low-frequency ($\sim150$ MHz) large survey. 
The manuscript is organised as follows: In Section \ref{sect:sample} we define the sample selection; in Section \ref{sect:obs} we describe the observations and the data calibration and imaging; results are presented in Section \ref{sect:results}, and discussed in Section \ref{sect:disc}; finally, a summary is presented in Section \ref{sect:concl}. Throughout the paper, we assume a standard $\Lambda$CDM cosmology, with $H_0 = 70$ km s$^{-1}$ Mpc$^{-1}$, $\Omega_m = 0.3$ and $\Omega_\Lambda = 0.7$.

\section{Sample and cluster selection criteria}\label{sect:sample}

\begin{figure}
\centering
\includegraphics[width=0.5\textwidth]{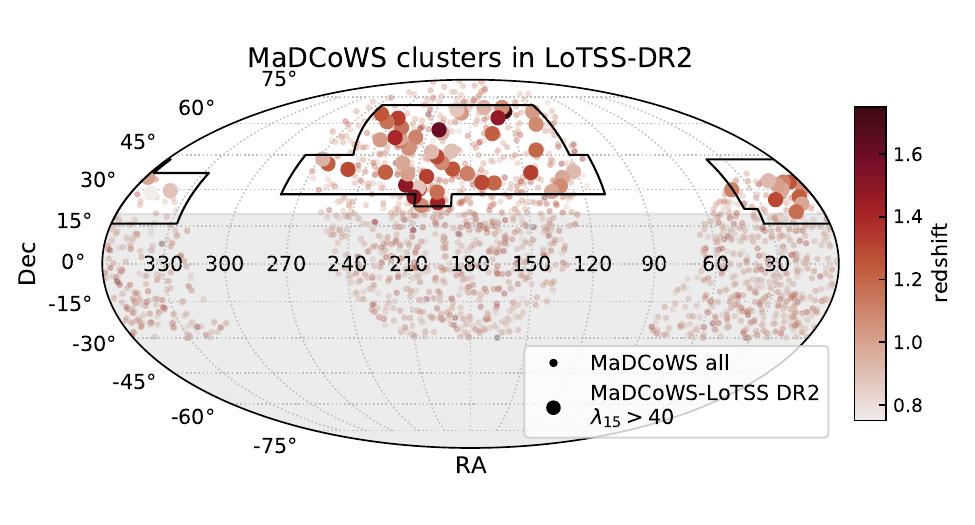}
\caption{Distribution of the MaDCoWS clusters in the PanSTARRS region (small dots), colour-coded based on their redshift. The grey area shows the sky region excluded because of the LoTSS sensitivity and sky coverage (i.e. $\rm Dec\leq20^\circ$). Large circles show the positions of the clusters in LoTSS-DR2 \citep[see black outlines;][]{shimwell+22} with a richness $\lambda_{15}>40$.}
\label{fig:madcows_lofar-dr2}
\end{figure}

\begin{figure}
\centering
\includegraphics[width=0.45\textwidth]{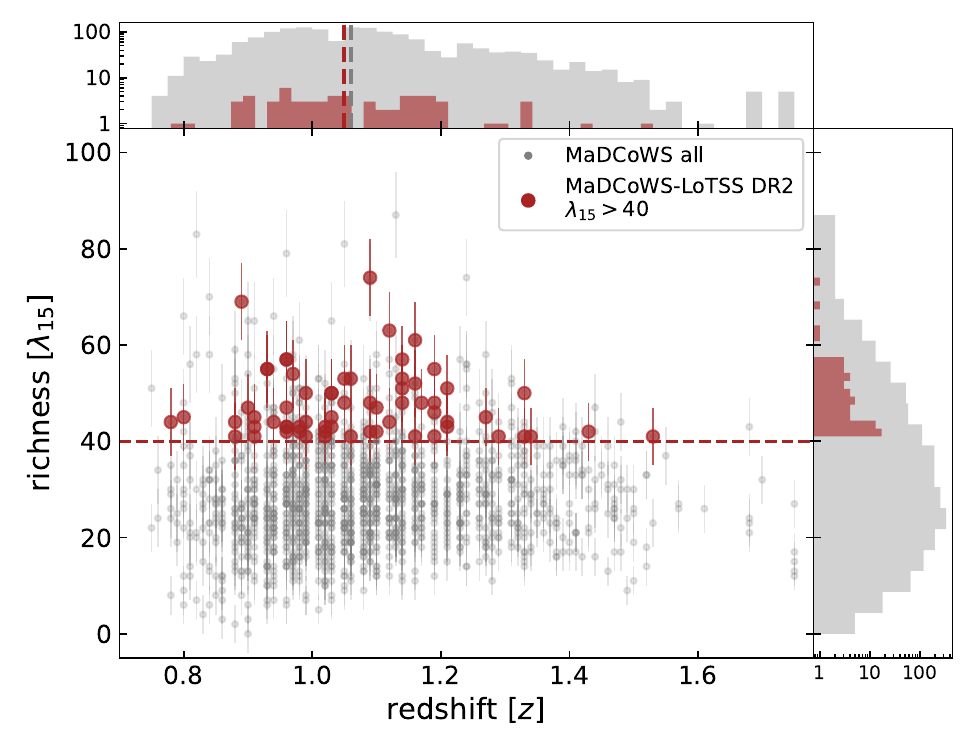}
\caption{Redshift-richness distribution of all the MaDCoWS clusters (small grey dots). The red, filled circles display the clusters in our sample (i.e. with $\lambda_{15}>40$, see dashed line). The histograms on the top and on the right show their distribution in comparison with the full MaDCoWS sample. Dashed grey and red lines in the top-panel histogram show the median redshift of the two distributions ($\langle z_{\rm all}\rangle\sim1.06$ and $\langle z_{\rm LoTSS}\rangle\sim1.05$, respectively).}
\label{fig:madcows_lofar-dr2_rich_z}
\end{figure}

The Massive and Distant Clusters of {\it WISE} Survey \citep[MaDCoWS;][]{gonzalez+19} is a catalogue of galaxy clusters in the redshift range $0.70\lesssim z \lesssim 1.75$, based upon {\it Wide-field Infrared Survey Explorer} \citep[{\it WISE};][]{wright+10} observations and complemented with data from the Panoramic Survey Telescope and Rapid Response System \citep[PanSTARRS;][]{panstarss16} at $\rm Dec>-30^\circ$ and from the SuperCOSMOS Sky Survey \citep{hambly+01} at $\rm Dec<-30^\circ$. From the {\it WISE}-PanSTARSS region, 1,676 of the 2,433 galaxy clusters detected by the survey have a photometric redshift ($z$) and a cluster richness \citep[$\lambda_{15}$; here, $\lambda_{15}$ corresponds to the excess number density of galaxies selected by {\it Spitzer} color cuts as possible cluster members with a brightness cut-off of 15 $\rm \mu Jy$; see][]{gonzalez+19}. 
Subsequent studies on the full MaDCoWS sample have attempted to calibrate the mass-richness relation. 
Particularly, comparisons with the Sunyaev-Zeldovich \citep[SZ;][]{sunyaev+zeldovic} measurements using the Combined Array for mm-wave Astronomy \citep[CARMA;][]{brodwin+15}, the Atacama Compact Array \citep[ACA;][]{dimascolo+20}, the MUSTANG2 camera on the Green Bank Telescope \citep{dicker+20}, and the Atacama Cosmology Telescope \citep[ACT;][]{orlowski-scherer+21} have revealed that these clusters are in the $M=0.1-6\times10^{14}~{\rm M}_\odot$ mass range, depending on the scaling relation used \citep{dicker+20}.

In the LoTSS sky area with the best sensitivity, i.e. $\rm Dec \geq 20^\circ$, the total number of clusters in MaDCoWs is 588. In this paper, we decided to focus on objects within the LoTSS-DR2 area \citep{shimwell+22} with a richness $\lambda_{15}>40$, resulting in a final number of 64 clusters (see Fig.~\ref{fig:madcows_lofar-dr2}). The richness threshold of 40 was chosen in order to include the most massive clusters in the MaDCoWS sample, while also still retaining a large sample of clusters (Fig. \ref{fig:madcows_lofar-dr2_rich_z}).
The final sample spans a wide photometric redshift range\footnote{Six of the 64 MaDCoWS clusters also have a spectroscopic redshift ($z_{\rm spec}$; Tab. \ref{tab:all} in Appendix \ref{apx:all}). If available, we use $z_{\rm spec}$ over the photometric redshift $z$.}, i.e. $0.78\leq z\leq1.53$ (median $\langle z_{\rm LoTSS}\rangle\sim1.05$), and richness, i.e. $40<\lambda_{15}<74$ (see Table \ref{tab:all} in Appendix \ref{apx:all} for the selected sample). This work thus extends the redshift and mass limits of our previously published work using the {\it Planck} PSZ2 catalogue \citep[i.e. $M_{500}=4-8\times10^{14}~{\rm M_\odot}$ and $0.6\leq z\leq0.9$,][]{digennaro+21a}.

\section{LOFAR data reduction and imaging}\label{sect:obs}
We have made use of the products of the LoTSS second data release (DR2). Therefore, we refer to \cite{shimwell+22} for a detailed description of the radio data reduction. We apply the standard calibration pipeline, which corrects for direction-independent \citep[\texttt{prefactor};][]{vanweeren+16,williams+16,degasperin+19} and direction-dependent \citep[\texttt{ddf-pipeline}, which includes \texttt{killMS} and \texttt{DDFacet};][]{tasse14,smirnov+tasse15,tasse+18,tasse+21} effects, and performs self-calibration of the entire field of view. To refine the solutions near the target, we also applied the ``extraction \& recalibration'' strategy described by \cite{vanweeren+21}. This procedure takes into account the local direction-dependent effects, by using the products of the pipeline, subtracting from the {\it uv}-plane all the sources outside a square region that includes the cluster (typically $\sim0.3-0.9~{\rm deg^2}$), and performing additional rounds of phase and amplitude self-calibration. At the end of the calibration, we assume conservative residual uncertainties on the relative amplitude calibration of $f=0.15$ \citep{shimwell+22}. We also employed flux-scale alignment due to the uncertainties in the LOFAR beam modeling during the calibration, as described by \cite{botteon+22} and \cite{hoang+22}. All the images and flux densities reported in the manuscript have these corrections applied.

Final, deep imaging was made using \texttt{WSClean v2.10} \citep{offringa+14,offringa+17}, with \texttt{Briggs} \citep{briggs95} weighting and \texttt{robust=-0.5}, and using \texttt{multiscale} deconvolution with scales of $[1,4,8,16]\times{\rm pixelscale}$ (${\rm pixelscale}=1.5''$) and {\tt channelsout=6}. An inner {\it uv}-cut at $80\lambda$ was applied to exclude the contribution of the Galactic emission. We produced images at different resolutions, tapering the {\it uv}-plane at 25 kpc, 50 kpc, and 100 kpc (see Appendix \ref{apx:all}). Additional higher-resolution images were created with \texttt{robust=-1.25}. To emphasise the possible presence of diffuse emission, we removed the contribution of compact sources: first we created a clean model only including compact sources (i.e. {\tt compact-only} image), by excluding data below the {\it uv}-range corresponding to linear sizes $\geq400$ kpc \citep[e.g. the typical size of radio halos; see][]{vanweeren+19} at the cluster's redshift. Then we subtracted this model and re-image the data, at different resolutions (i.e. without tapering, and with a taper of 25 kpc, 50 kpc, and 100 kpc; see Appendix \ref{apx:all}).
For all the images, the final reference frequency is 144 MHz.

\section{Results}\label{sect:results}

\begin{figure}
\centering
\includegraphics[width=0.45\textwidth]{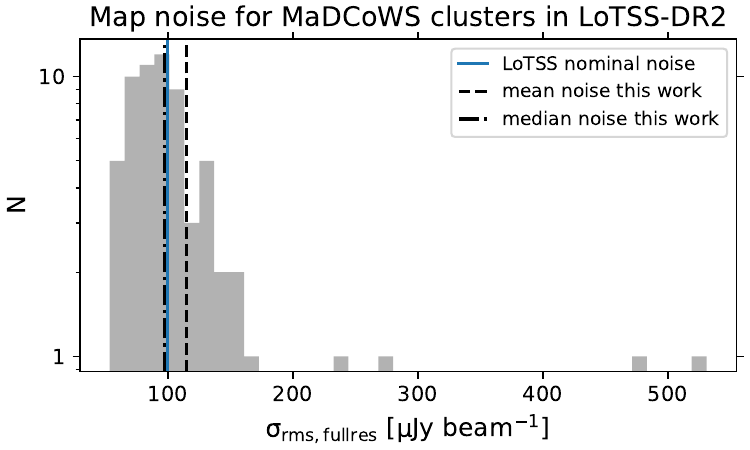}
\caption{Distribution of the map noise of the MaDCoWS clusters in LoTSS-DR2. The blue, solid line shows the nominal map noise from LoTSS, while the dot-dashed and dashed lines represent the median and mean values, respectively, for the 144 MHz cluster images in this work.}\label{fig:mapnoise}
\end{figure}

We inspected the {\tt full-resolution}, {\tt compact-only} and {\tt low-resolution source-subtracted} images by eye in order to investigate the presence of diffuse radio emission, similarly to what has been done by \cite{botteon+22}. A visual inspection was also made to exclude bad-quality data, i.e., those affected by artefacts or poor calibration. We excluded from our final sample those clusters with a map noise that is greater than twice the nominal LoTSS value (with $\sigma_{\rm rms,LoTSS}=100~\mu$\jybeam). These systems are: 
MOO\,J0907+2908 ($\rms=239$ $\mu$\jybeam), 
MOO\,J1110+6838 ($\rms=531$ $\mu$\jybeam), 
MOO\,J1135+3256 ($\rms=278$ $\mu$\jybeam),
MOO\,J1336+4622 ($\rms=10^8$ $\mu$\jybeam) and 
MOO\,J1616+6053 ($\rms=474$ $\mu$\jybeam). 
Despite a favourable noise level ($\rms=64~\mu$\jybeam), MOO\,J1319+5519 was also excluded because it is located nearby a bright compact radio source that creates strong artefacts and therefore prevents any detection of diffuse radio emission. 
For similar reasons, we excluded MOO\,J1248+6723 and MOO\,J1506+5137, which are close to extended lower-redshift radio galaxies. In particular, the radio galaxy on the line of sight of MOO\,J1506+5137 \citep[$z=0.611$ at ${\rm RA_{J2000}= 15^h06^m12.81^s}$ and ${\rm Dec_{J2000}=+51^\circ37^\prime73''}$,][]{aguado+19} was extensively studied at radio frequencies using LOFAR, the Karl Jansky Very Large Array (VLA), and the Robert C. Byrd Green Bank Telescope (GBT) observations by \cite{moravec+20a}.
After excluding these eight clusters, the map noise in our sample ranges between $54-162~\mu$\jybeam, with a median value of $98~\mu$\jybeam\ (see Fig. \ref{fig:mapnoise} and Appendix \ref{apx:all}).

We found that about 80\% (44/56) of the clusters in the sample host at least one radio source (i.e. radio galaxy or extended diffuse radio source) at 144 MHz. Among these, we detect diffuse radio emission in the source-subtracted images with {\tt taper=100kpc} covering the $0.5R_{500}$ region\footnote{We estimated $R_{500}$ from $M_{500}$ obtained by the mass-richness scaling relation from \cite{orlowski-scherer+21}, see Sect. \ref{sec:massesOS21}.} in five systems (see Fig. \ref{fig:clusters}), namely MOO\,J0123+2545 (hereafter MOOJ0123, $z_{\rm spec}=1.229$), MOO\,J1231+6533 (hereafter MOOJ1231, $z=0.99$), MOO\,J1246+4642 (hereafter MOOJ1246, $z=0.90$), MOO\,J1420+3150 (hereafter MOOJ1420, $z=1.34$), and MOO\,J2354+3507 (hereafter MOOJ2354, $z=0.97$). These radio sources have largest linear size (LLS) of roughly 350-500 kpc at the clusters' redshift (i.e. angular size of $\sim1^\prime$). For two systems, i.e. MOOJ0123 and MOOJ1231, SZ CARMA observation at 30 GHz \citep{decker+19} are also available, and show that the extended radio emission sits on the ICM (see Fig. \ref{fig:sz_radio}). 

\subsection{Flux density measurements}
We measure the flux density of the extended radio emission by integrating over the $2.5\rms$ radio contours from the source-subtracted image with {\tt taper=100kpc} ($S_{\rm 144MHz,sub}$). 
The total uncertainty for the flux density measurements is given by
\begin{equation}\label{eq:fluxerr}
\Delta S_{\rm 144MHz,sub} = \sqrt{(f\,S_{\rm 144MHz,sub})^2 + \rms^2N_{\rm beam} + \sigma_{\rm sub}^2} \, ,
\end{equation}
where $\rms$ is the map noise, and $N_{\rm beam}$ is the number of beams covering the diffuse radio emission. The term $\sigma^2_{\rm sub}$ describes the goodness of the subtraction from the visibilities, and is equal to $\sum_{i}N_{{\rm beams},i}\,\sigma^2_{\rm rms}$, namely the sum over all the $i$ sources that were subtracted within the cluster region.

The source subtraction in the $uv$-plane can be imperfect. This is due to the presence of foreground radio galaxies with angular sizes similar to the cluster (which for this reason are excluded from the model of the subtracted sources), or because the source components are not entirely included in the model. In such cases, we exclude the extended emission of the radio galaxy from the source-subtracted flux density and/or manually subtract the residual flux density from those sources ($S_{\rm 144MHz,RG}$). The residual flux of the radio galaxies was estimated by comparing the emission from the full-resolution and {\tt compact-only} images, following the $1\rms$ radio contours of the sources within the cluster region in the latter map.
The final flux density on the diffuse emission is therefore defined as:
\begin{equation}
\begin{aligned}
S_{\rm 144MHz,diff} =& \, S_{\rm 144MHz,sub} - \textstyle\sum_i S_{{\rm 144MHz,RG}_i} \\
& \pm \sqrt{ \Delta S_{\rm 144MHz,sub}^2 + \textstyle\sum_i \Delta S^2_{{\rm 144MHz,RG}_i} }  \, ,
\end{aligned}
\end{equation}
with the sum $\sum_i S_{{\rm 144MHz,RG}_i}$ over all the $i$th-subtracted sources and $\Delta S_{\rm 144MHz,RG}$ calculated similarly to Eq. \ref{eq:fluxerr}, but with $\sigma^2_{\rm sub}=0$.

Below, we describe the flux density measurements for each candidate cluster with extended diffuse radio emission, which are summarised in Tab. \ref{tab:flux}.

\subsubsection{MOOJ0123} This cluster shows extended radio emission in the full-resolution images, both in the original and source-subtracted images (see first row in Fig. \ref{fig:clusters}). This emission is enhanced in the low-resolution image (i.e. {\tt taper=100kpc}, corresponding to a resolution of $\sim18''\times15''$). From this map, we measure a flux density of $S_{\rm 144MHz,sub}=2.5\pm0.6$ mJy within the area covering the $2.5\rms$ level (with $\rms=150~\mu$\jybeam, see the yellow region in the last column in Fig. \ref{fig:clusters}). From this measured flux density, we additionally removed the residual contribution of the point sources visible in the {\tt compact-only} image, i.e. $\sum S_{\rm 144MHz,RG}=0.4\pm0.2$ mJy. The final flux density for the diffuse radio emission in MOOJ0123 is $S_{\rm144MHz,diff}=2.1\pm0.6$ mJy.

\subsubsection{MOOJ1231} Hints of diffuse radio emission for this cluster are only visible in the source-subtracted {\tt taper=100kpc} image (corresponding to a resolution of $19''\times16''$) at the $2.5\rms$ level, with $\rms=82~\mu$\jybeam\ (see the second row in Fig. \ref{fig:clusters}). However, the image is still contaminated by residual compact sources, which were excluded from the area of the diffuse radio emission
(see yellow region in the last column). We measure a flux density of $S_{\rm 144MHz,sub}=1.3\pm0.3$ mJy, from which we additionally subtract $\sum S_{\rm 144MHz,RG}=0.3\pm0.1$ mJy of residual flux from a radio galaxy, leading to $S_{\rm 144MHz,diff}=1.0\pm0.3$ mJy.

\subsubsection{MOOJ1246} Hints of extended diffuse radio emission for this cluster are visible in the full-resolution image, south of two compact sources (see the third row in Fig. \ref{fig:clusters}). These two compact sources are not fully removed in the source subtraction process, we therefore exclude them from the area of the extended diffuse emission at low resolution (i.e. $26''\times17''$, with $\rms=340~\mu$\jybeam; see yellow region in the last column in Fig. \ref{fig:clusters}). Here, we measure $S_{\rm 144MHz,sub}=1.7\pm0.6$ mJy and $\sum S_{\rm 144MHz,RG}=0.4\pm0.2$ mJy, and thus $S_{\rm 144MHz,diff}=1.3\pm0.6$ mJy.

\subsubsection{MOOJ1420} The diffuse radio emission in the cluster is clearly visible in the source-subtracted {\tt taper=100kpc} image (corresponding to a resolution of $\sim18''\times16''$; see fourth row in Fig. \ref{fig:clusters}). We measure a flux density within the $2.5\rms$ area (with $\rms=160~\mu$\jybeam, see yellow region in the last column in Fig. \ref{fig:clusters}) of $S_{\rm 144MHz,sub}=3.1\pm0.8$ mJy. We estimate a residual flux density from the compact sources of $\sum S_{\rm 144MHz,RG}=1.3\pm0.5$ mJy. This corresponds to a flux density of the diffuse component of $S_{\rm 144MHz,diff}=1.8\pm0.8$ mJy. 

\subsubsection{MOOJ2354} 
\begin{figure*}
\centering
\includegraphics[width=0.85\textwidth]{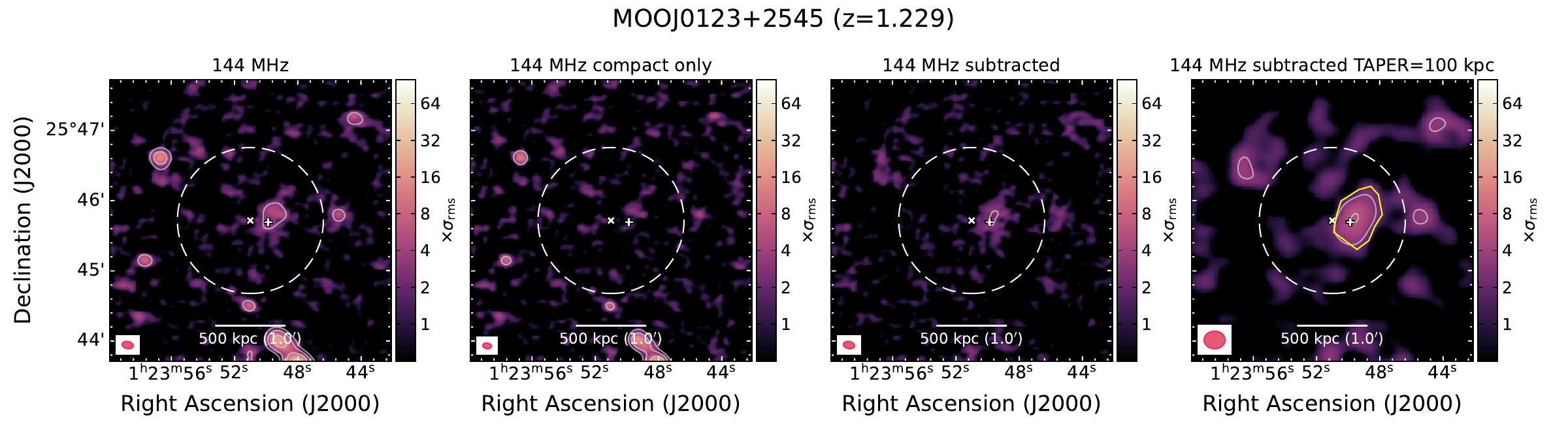}
\includegraphics[width=0.85\textwidth]{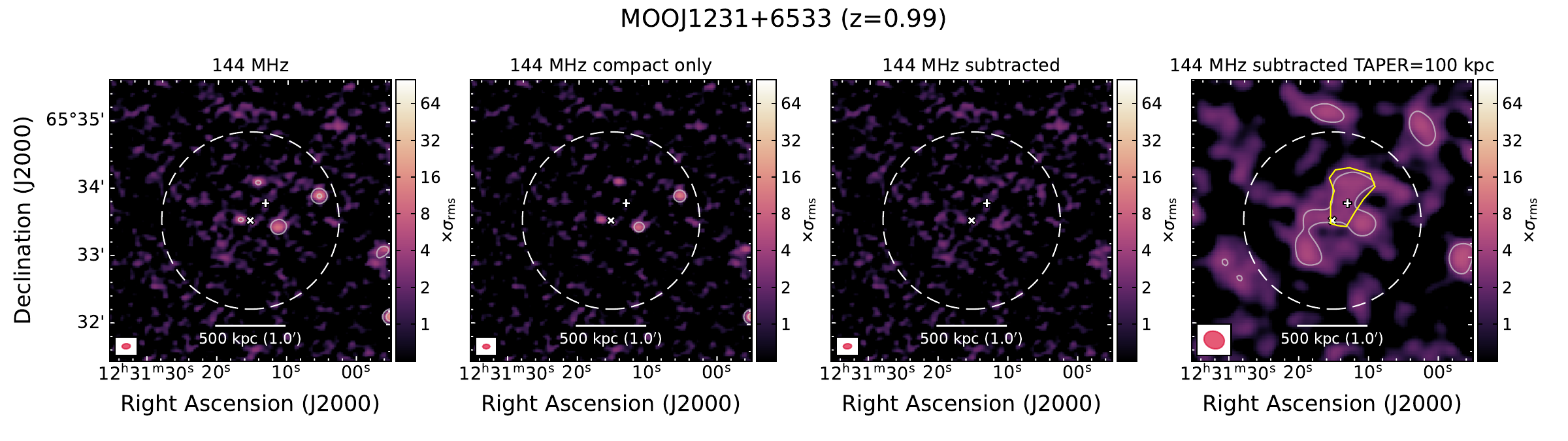}
\includegraphics[width=0.85\textwidth]{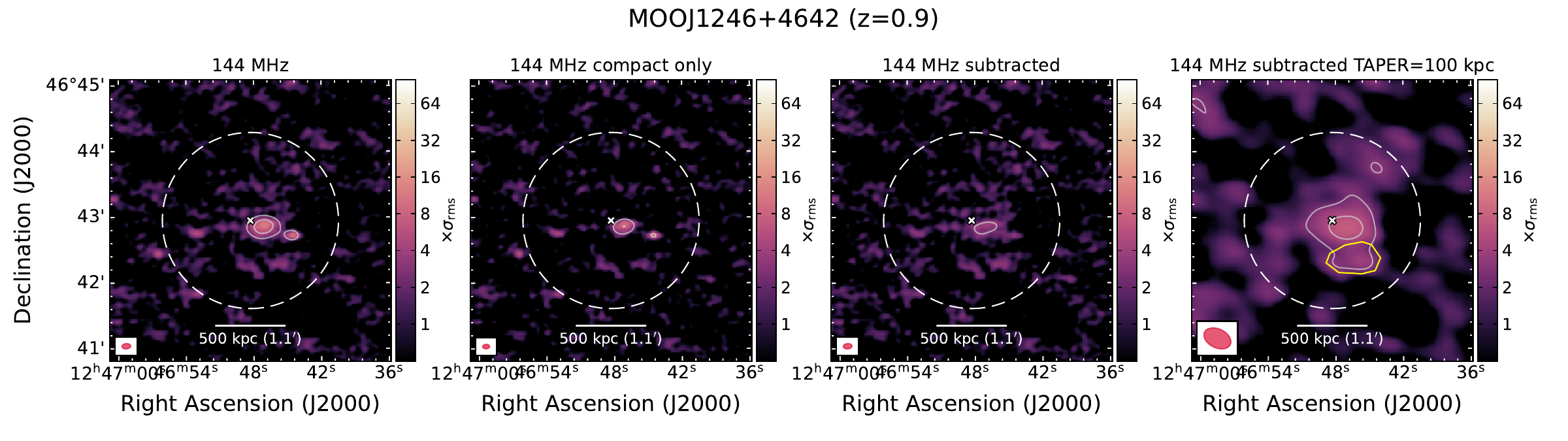}
\includegraphics[width=0.85\textwidth]{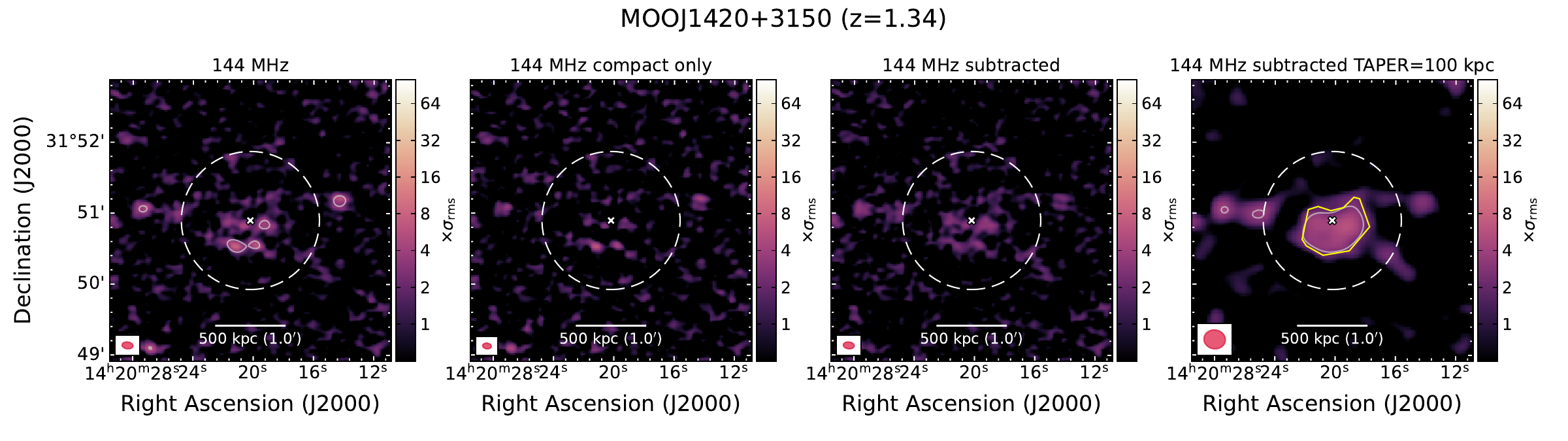}
\includegraphics[width=0.85\textwidth]{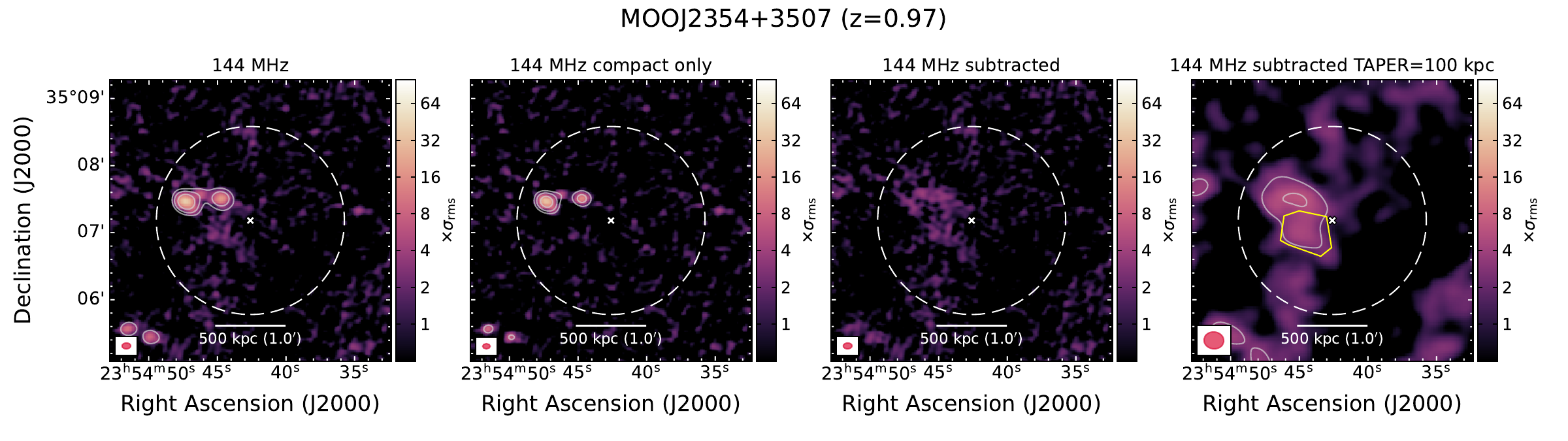}
\caption{LOFAR 144 MHz images of the MaDCoWS clusters with cluster-scale diffuse emission. The cluster name and redshift are stated at the top of each row. From left to right: full-resolution image; full resolution, compact only (i.e. after applying an inner $uv$-cut of 400 kpc; full-resolution source-subtracted image; same as previous panel, but with {\tt taper=100kpc}. Yellow regions in the right panel of each row show the area where we measure the radio flux densities. Radio contours are displayed in white, solid lines, starting from $2.5\rms\times[2,4,8,16,32]$ and negative contours at $-2.5\rms$ are shown in white, dashed lines. The beam shape is shown at the bottom left corner of each panel. The dashed white circle shows the $R_{500}$ kpc area, with the cross marking the MaDCoWS coordinates reported by \cite{gonzalez+19} and, when available, the plus marking the peak of the CARMA SZ observation \citep{decker+19}.}\label{fig:clusters}
\end{figure*}

\begin{figure*}
\centering
\includegraphics[width=0.45\textwidth]{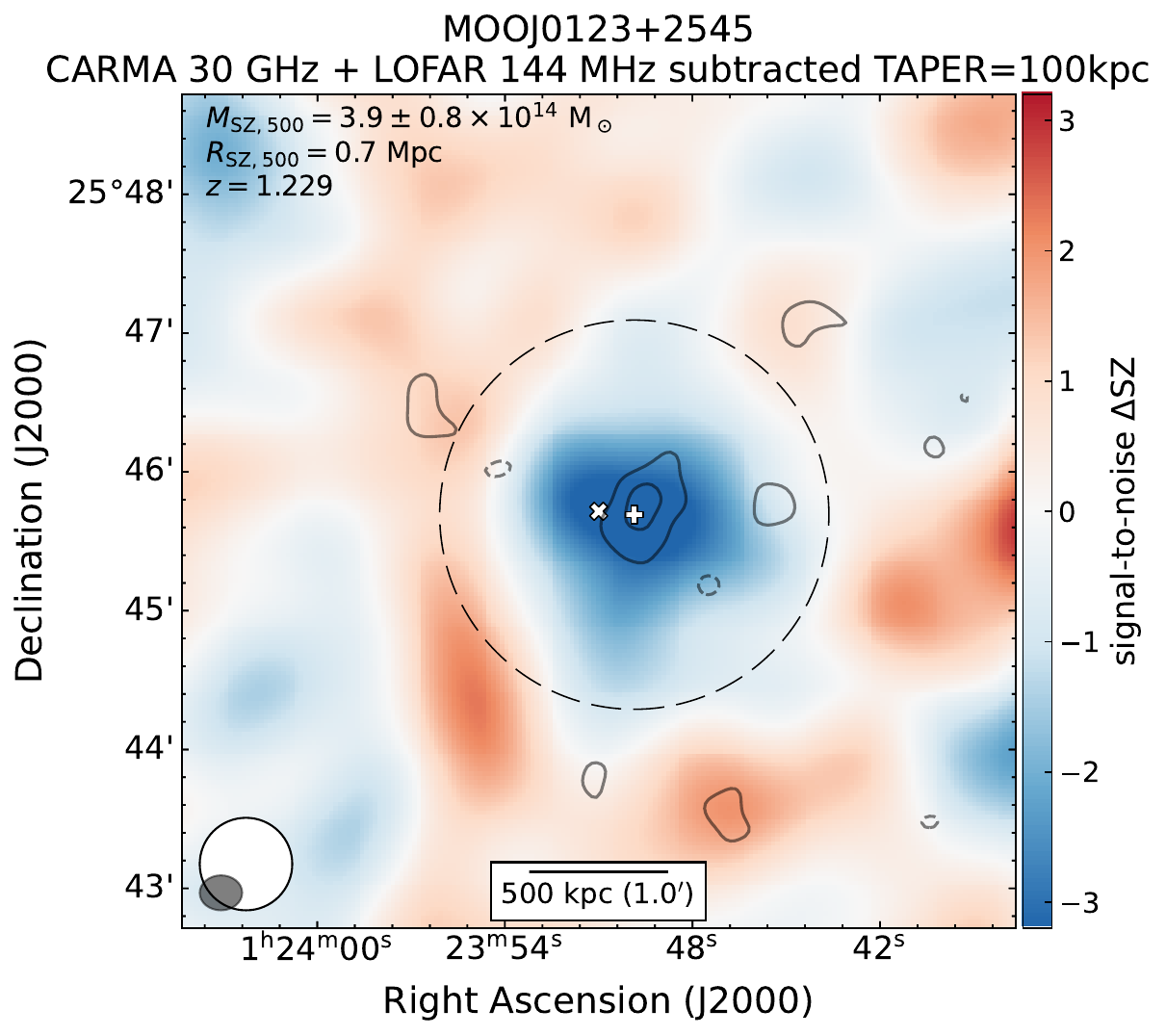}
\includegraphics[width=0.45\textwidth]{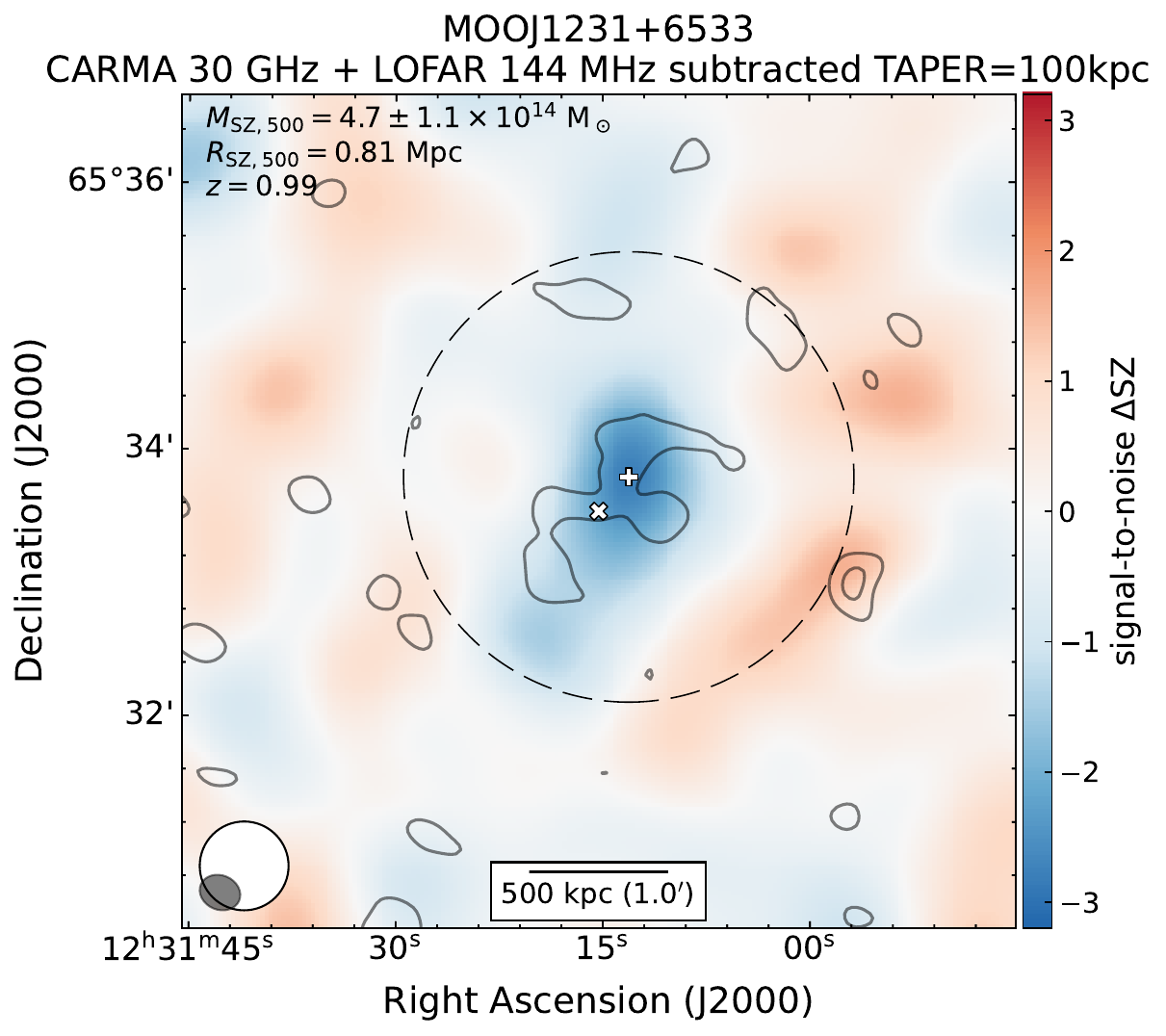}
\caption{CARMA 30 GHz observations \citep{decker+19} with radio source-subtracted LOFAR low-resolution contours of the available MaDCoWS clusters in LoTSS-DR2. The different beam sizes are shown the on the bottom left corner of each panel, with the solid grey corresponding to LoTSS-DR2 and open white to the CARMA 30 GHz data. The colour map represents the SZ variation in units of signal-to-noise, therefore negative values reveal the presence of the cluster \citep[with the centre marked by the white `plus';][]{decker+19}. The black circle places the $R_{500}$ area given the SZ coordinates, and the white cross provides the MaDCoWS centre \citep{gonzalez+19}. The cluster mass, $R_{500}$ and redshift from the CARMA observations are indicated in the upper left corner of each panel.}\label{fig:sz_radio}
\end{figure*}

The radio emission in the cluster in the full-resolution image is dominated by an extended radio galaxy, although we observe hints of faint diffuse radio emission south of it (see fifth row in Fig. \ref{fig:clusters}). Hence, we define the area of the extended diffuse emission in the source-subtracted {\tt taper=100kpc} image (corresponding to a resolution of $18''\times15''$) excluding the region covered by the radio galaxy (see the yellow region the last column in Fig. \ref{fig:clusters}). We measure  $S_{\rm 144MHz,sub}=1.6\pm0.8$ mJy. 

\begin{table*}[]
\caption{Flux densities of the cluster-scale diffuse radio emission at 144 MHz. Radio powers are calculated assuming a spectral index $\alpha=-1.5\pm0.3$ \citep{digennaro+21a,digennaro+21c}.}
\vspace{-5mm}
\begin{center}
\begin{tabular}{lccccc}
\hline
\hline
Cluster name & Redshift & LLS & 144 MHz Flux Density &  150 MHz Radio Power &  1.4 GHz Radio Power \\
& $z$ & [kpc] &  $S_{\rm 144MHz,diff}$ [mJy] & $P_{\rm 150MHz}~[\rm \times10^{25}~W\,Hz^{-1}]$  &  $P_{\rm 1.4GHz}~[\rm \times10^{23}~W\,Hz^{-1}]$ \\ 
\hline
MOOJ0123+2545 &  1.229$^\dagger$ & 420 &  $2.1\pm0.6$ & $2.6\pm1.1$  &  $9.8\pm5.6$ \\
MOOJ1231+6533 &  0.99  & 430 &   $1.0\pm0.3$  & $0.7\pm0.2$  &  $2.8\pm1.3$ \\
MOOJ1246+4642 &  0.90  & 390 &   $1.3\pm0.6$  & $0.7\pm0.4$  &  $2.8\pm2.1$ \\
MOOJ1420+3150 &  1.34  & 475 &   $1.8\pm0.8$  & $2.9\pm1.6$  &  $10.8\pm6.7$ \\
MOOJ2354+3507 &  0.97  & 360 &   $1.6\pm0.8$  & $1.1\pm0.5$  &  $4.3\pm3.2$ \\
\hline
\end{tabular}
\end{center}
\vspace{-5mm}
\tablefoot{$^\dagger$Spectroscopic redshift. 
}
\label{tab:flux}
\end{table*}

\subsection{Radio power estimation}\label{sec:radiopow}

For all the aforementioned clusters, we calculated the $k$-corrected radio luminosities at frequencies $\nu=150$ MHz and $\nu=1.4$ GHz, in order to compare to literature values \citep{cassano+13,cuciti+21b,cuciti+23}, as follows:

\begin{equation}\label{eq:radiopow}
P_\nu = \frac{4\pi\,D_L(z)^2}{(1+z)^{\alpha+1}} \, \left (\frac{\nu}{\rm 144MHz} \right)^{\alpha}S_{\rm 144MHz} \quad [{\rm W\,Hz^{-1}}]\,,
\end{equation}
where $S_{\rm 144MHz}$ is the flux density of the diffuse emission measured at 144 MHz, $\alpha$ is the spectral index of the diffuse emission, $D_L$ is the luminosity distance at the redshift $z$, and the factor $(1+z)^{-(\alpha+1)}$ is the $k-$correction. Since we do not have information on the spectral index for the clusters in the presented work, following \cite{digennaro+21a,digennaro+21c} we assume $\alpha=-1.5\pm0.3$. 
Uncertainties in the radio luminosities are obtained with 150 Monte-Carlo simulations, which include both the uncertainties associated with the flux densities and the spectral indices.
The flux densities at 144 MHz and the radio luminosities at 150 MHz and 1.4 GHz are listed in Tab.~\ref{tab:flux}.

\subsection{Upper limits}\label{sec:ul}
For 51 galaxy clusters in our sample, we did not detect extended diffuse radio emission in the cluster volume, and hence only upper limits can be provided. 
Following \cite{bruno+23}, we calculate our upper limits as:
\begin{equation}
\log \left ( \frac{S_{\rm UL}}{\sigma_{\rm rms}} \right) = m \log(N_{\rm beam})+q \, .
\end{equation}
Here, $\rms$ is the source-subtracted {\tt taper=100kpc} map noise, and $N_{\rm beam}$ is the number of beams covering the radio halo region. We define the area of the halo region equal to $3r_e$, being $r_e=75$ kpc the cluster $e$-folding radius, which corresponds to a physical size comparable to those we detect in our sample (i.e 450 kpc, $\sim1^\prime$). 
Given the low number of beams covering the radio halo region ($N_{\rm beam}\lesssim10$), we adopt the best-fit parameters of $m=0.5$ and $q=0.155$ \citep{digennaro+21c,bruno+23}.
As for the detected diffuse extended emission, we then derive radio luminosities of the upper limits at 150 MHz and 1.4 GHz assuming a spectral index of $\alpha=-1.5\pm0.3$ in Eq. \ref{eq:radiopow}.

\subsection{Cluster mass}\label{sec:masses}
In order to investigate the properties of the extended diffuse radio emission in the MaDCoWS clusters, and to have a comparison with diffuse sources at lower redshifts, it is crucial to have an estimate of the cluster mass. Literature mass measurements are available only for MOOJ0123 and MOOJ1231, through CARMA 30 GHz observations \citep{decker+19}, being $M_{500}=(3.9\pm0.8)\times10^{14}~{\rm M_\odot}$ and $M_{500}=(4.7\pm1.1)\times10^{14}~{\rm M_\odot}$, respectively (Tab. \ref{tab:masses}). For the other clusters, we can make use of scaling relations via optical-IR, SZ and X-ray observations.

\subsubsection{Mass-richness relation}\label{sec:massesOS21}
Masses for the MaDCoWS clusters can be retrieved from their galaxy richness \citep{brodwin+15,gonzalez+19,dimascolo+20,dicker+20,orlowski-scherer+21}, according to the relation:
\begin{equation}
\log_{10}\frac{M_{500}}{10^{14}~{\rm M_\odot}}=A\log_{10}\lambda_{15}+B  \, . 
\end{equation}
In this work we assume the relation found by \cite{orlowski-scherer+21}, where an extensive calibration of the relation was performed by analysing the MaDCoWS-selected clusters on forced-photometry estimates from ACT observations and resulting in $A=-6.08^{+0.51}_{-0.48}$ and $B=1.81^{+0.14}_{-0.13}$, with an intrinsic scatter $\sigma_{\ln M|\lambda}=0.21^{+0.08}_{-0.11}$. We report the resulting cluster mass in Tab. \ref{tab:masses}, including the uncertainties due to the scatter of the relation (numbers in brackets in the third column).

Comparing the only two mass estimates from CARMA 30 GHz observations with the mass we would obtain using the $M_{500}-\lambda_{15}$ scaling relation, we find the corresponding ones from the scaling relation are a factor of $\sim2$ lower, although consistent within $1\sigma$ errorbar including the scatter of the relation.
This difference is probably associated with the different assumptions regarding the integrated SZ signal to mass scaling relation adopted by \cite{decker+19} and \cite{orlowski-scherer+21}.

\begin{table}[]
\caption{Mass estimation of the clusters in our sample with diffuse radio emission.}
\vspace{-5mm}
\begin{center}
\resizebox{0.5\textwidth}{!}{
\begin{tabular}{lcccccc}
\hline
\hline
Cluster name & \multicolumn{3}{c}{Cluster mass ($M_{500}~[\rm \times10^{14}~M_\odot]$)} \\
& CARMA 30 GHz & $M_{500}-\lambda_{15}$ & $M_{500}-F_X$ \\ 
\hline
MOOJ0123+2545 & $3.9\pm0.8$ & $1.9^{+0.6(+0.4)}_{-0.5(-0.4)}$ & $1.9^{+0.7}_{-1.0}$  \\
MOOJ1231+6533 & $4.7\pm1.1$ & $2.8^{+0.8(+0.3)}_{-0.7(-0.3)}$  & $4.3^{+0.4}_{-0.5}$ \\
MOOJ1246+4642 & N/A &  $2.5^{+0.7(+0.4)}_{-0.6(-0.4)}$ & $3.6^{+0.5}_{-0.5}$ \\
MOOJ1420+3150 & N/A & $1.9^{+0.5(+0.4)}_{-0.5(-0.4)}$ & $1.5^{+0.9}_{-1.5}$ \\
MOOJ2354+3507 & N/A & $3.2^{+0.8(+0.3)}_{-0.7(-0.3)}$ & $4.1^{+0.4}_{-0.5}$ \\
\hline
\end{tabular}}
\end{center}
\vspace{-5mm}
\tablefoot{First column: cluster name. Second to fourth columns: masses obtained from the literature \citep[CARMA observations at 30 GHz, see][second column]{decker+19}, from the $M_{500}-\lambda_{15}$ relation calibrated with ACT clusters \cite[third column]{orlowski-scherer+21}, and from the $M_{500}-F_X$ scaling relation from eROSITA 0.4--2.3 keV observations \citep[fourth column]{sunyaev+21,predehl+21}.}
\label{tab:masses}
\end{table}

\begin{figure*}
\centering
\includegraphics[width=\textwidth]{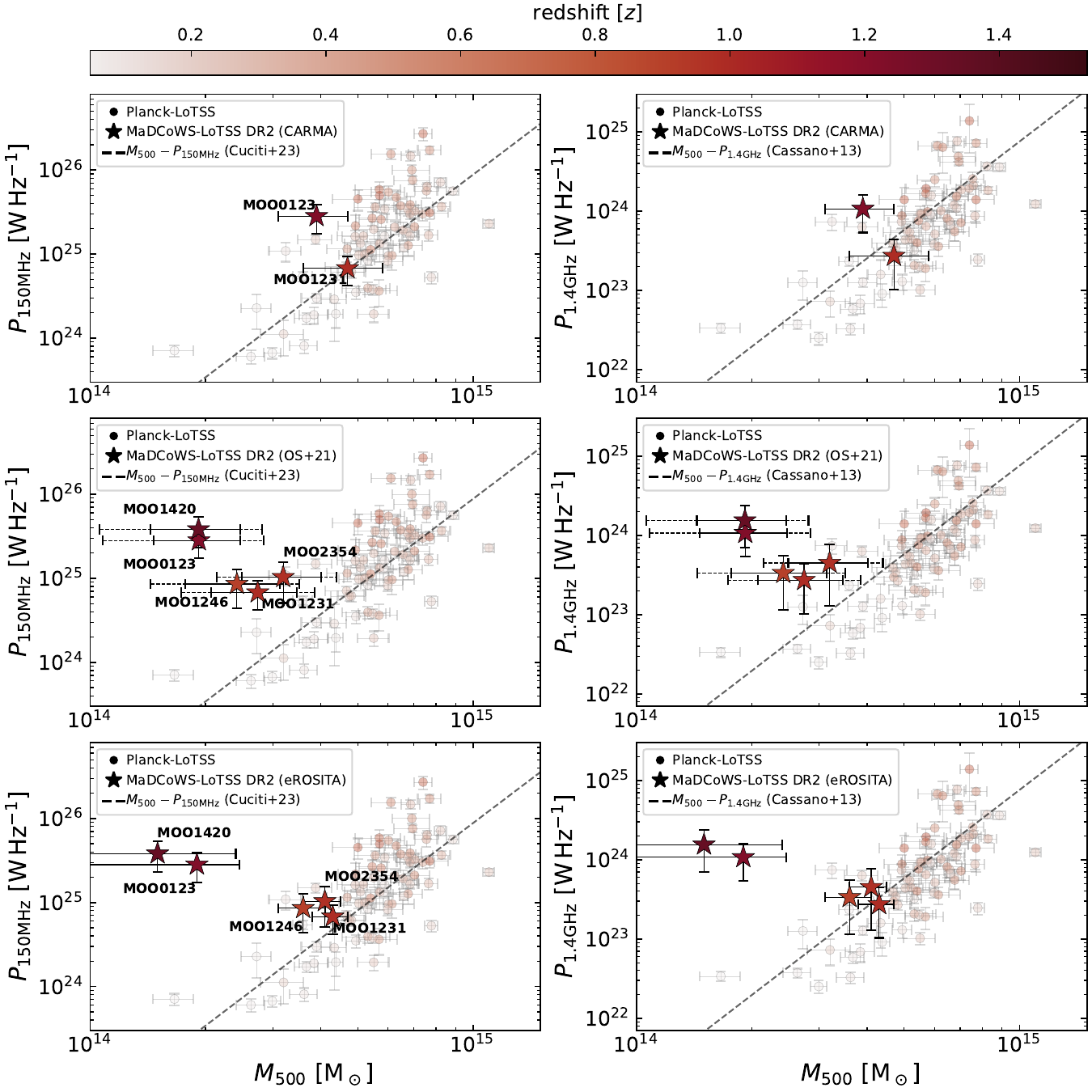}
\caption{Radio power versus mass diagrams (150 MHz, left column; 1.4 GHz, right column). Small shaded circles are from the literature at lower-redshifts \citep{digennaro+21a,botteon+22}, while stars display the detection from the MaDCoWS clusters in LoTSS-DR2 presented in this work. All markers are colour-coded according to their redshifts. We also display the $P_\nu-M_{500}$ correlations found by \cite{cuciti+23} and \cite{cassano+13}, at 150 MHz (left panels) and 1.4 GHz (right panles) respectively. Top row: masses from CARMA 30 GHz observations \citep{decker+19}. Middle row: masses from the richness-mass scale relation calibrated with ACT clusters \citep[OS+21]{orlowski-scherer+21}; solid errorbars reflect the uncertainties on the slope of the scale relation, while the dashed errorbars define the uncertainties associated with the scatter of the scale relation. Bottom row: masses from eROSITA observation.}
\label{fig:powermass}
\end{figure*}

\subsubsection{Masses from eROSITA observations}
The MaDCoWS clusters in the LoTSS-DR2 samples are covered in the SRG/eROSITA all-sky survey \citep{sunyaev+21,predehl+21}. We, therefore,  can use X-ray data to estimate their masses. For $z\sim 1$ clusters the X-ray flux turns out to be a useful mass proxy \citep[e.g.][]{churazov+15}. The X-ray flux was estimated from the 0.4--2.3 keV count rate within a circle with radius $R=2^\prime$ centred at the cluster position (see Tab. \ref{tab:all}), and using a wider ring from $6^\prime$ to $20^\prime$ to estimate the local X- ray background signal. For the latter, bright point and extended sources, with the 0.5--2 keV flux above $\rm 10^{-13}~erg\,s^{-1}\,cm^{-2}$ are detected and masked, following the strategy described by \cite{churazov+21} and \cite{khabibullin+23}. 
The variance in the background flux within the source aperture was estimated in a model-independent way, selecting 24 regions of the same size within a $30^\prime$ circle centred on the cluster candidate.  
The extracted count rates within the source apertures are corrected for the expected background contributions and finally converted to the 0.5--2 keV flux, $F_X$, using a constant factor, which weakly depends on the cluster temperature \citep[see, e.g. Fig. B1 in][for the temperature dependence of emissivity in the 0.3-2.3 keV band]{lyskova+23}. Uncertainties on the flux are set by the photon count statistics (Poisson noise, $\sigma_{\rm stat}$) and the average background variance contribution ($\sigma_{\rm var} = \rm 1.5\times10^{-14}~erg\,s^{-1}\,cm^{-2}$), i.e. $\sqrt{\sigma^2_{\rm stat} + \sigma^2_{\rm var}}$.

The cluster mass is then estimated via the following relation \citep{churazov+15}:
\begin{equation}
M_{500}=1.2\times10^{14}M_{\odot}\,\eta\left ( \frac{F_X}{10^{-14}} \right )^{0.57}\, z^{0.5} \, ,
\end{equation}
where the factor $z^{0.5}$ is expected to work well in the redshift range $\sim0.7-1.5$ when scaling relations from \cite{vikhlinin+09} are used, while $\eta$ encapsulates factors related to the method of flux estimation, the sample selection function, and the definition of the mass. We use $\eta=0.86$ found for a subset of ACT clusters with $z>0.7$ from \cite{orlowski-scherer+21} sample using {\tt ACT\_MASS} value as the mass proxy (for details, we refer to Lyskova et al. in prep.).
The retrieved masses are reported in Tab. \ref{tab:masses}. From these eROSITA observations, we found a good agreement (within $1\sigma$) with the masses obtained from the \cite{orlowski-scherer+21} mass-richness relation except for MOOJ1231, for which the X-ray mass is similar to the CARMA 30 GHz one \citep{decker+19}. This could suggest a more complicated morphology of this cluster than the others in the sample.

\section{Discussion}\label{sect:disc}
Given the radio power of the diffuse radio emission in our sample (Sect. \ref{sec:radiopow}) and the estimated masses of the host clusters (Sect. \ref{sec:masses}), we can place these systems in the canonical radio power-mass diagram for a comparison with the diffuse radio emission at lower redshift \citep[see Fig. \ref{fig:powermass};][]{cassano+13,cuciti+23}. In this context, we interpreted our detections as candidate radio halo emission, due to their central location with respect to the distribution of the cluster galaxies and taking into account the uncertainties on the radio galaxies subtraction and the lack of a clear overlay with the thermal emission of the ICM. 
Following the same criteria as in \cite{botteon+22}, we do not separate candidate mini-halos from giant halos based on the size of the radio emission in our targets. Specifically, in  Fig. \ref{fig:powermass} we show all the {\it Planck} clusters with a detected diffuse radio emission in LoTSS \citep[left column;][]{digennaro+21a,botteon+22} and their corresponding radio power at 1.4 GHz (right column) using a spectral index of $\alpha=-1.3$ for the clusters at $z<0.6$ and $\alpha=-1.5$ for the clusters at $z>0.6$ \citep{digennaro+21a}.
For these lower-redshift clusters, the existence of such a correlation has been extensively proved, both at 150 MHz and 1.4 GHz. In particular, the recent analysis of the {\it Planck} clusters in the LoTSS-DR2 area has confirmed the presence of a correlation between cluster mass and radio power over a wide mass and redshift range (i.e. $M_{500}\sim 3-10\times10^{14}~{\rm M_\odot}$ and $z\sim0.02-0.6$), although the analysis was focused only on clusters above the 50\% completeness level of the {\it Planck} clusters \citep[see dashed line in the left panels in  Fig. \ref{fig:powermass}]{cuciti+23}.
The parameters (i.e. slope and normalisation) of the correlation at 150 MHz and 1.4 GHz were found to be in line with previous literature studies based on smaller samples at the same frequency \citep[see dashed line in the right panels in  Fig. \ref{fig:powermass}]{cassano+13}.  
The existence of such a correlation is interpreted as the amount of energy injected by turbulence into the intracluster medium which then powers particle re-acceleration and the small-scale dynamo for the magnetic amplification. Therefore, the most massive clusters are expected to host the most powerful radio halos \citep{cassano+brunetti05}. 
The implication is that the properties of the extended radio emission lying on this correlation, such as the average magnetic fields ($\langle B \rangle$) and the turbulent energy ($\eta_t$, i.e. the fraction of the $PdV$ work done by the sub-clusters falling into the main cluster) can be assumed to be similar in clusters regardless their mass and redshift.
Recent work has shown a good agreement between theoretical models and observations, at least up to $z<0.4$ \citep{cassano+23}. 
Clusters with lower average magnetic fields would be placed below the correlation, including its scatter \citep[see Fig. 3a of][]{digennaro+21a}. High-redshift clusters hosting extended diffuse radio emission are hence expected to populate this region of the $P_\nu-M_{500}$ diagram, as magnetic fields are thought to evolve from weak (primordial or astrophysical) seeds through compression and turbulence \citep{vazza+18}.

Three of the five clusters in the the final sample -- namely MOOJ1231, MOOJ1246 and MOOJ2354 -- fall within the scatter of the $P_\nu-M_{500}$ distributions of the lower redshift systems. This is regardless of whether we use the mass obtained from the $M_{500}-\lambda_{15}$ or the $M_{500}-F_X$ scaling relations, although the former tends towards lower values. On the other hand, the only cluster among these three systems with also CARMA observations, MOOJ1231, has a mass that is more consistent with the $M_{500}-F_X$ scale relation (i.e. $\rm 4.7\pm1.1\times10^{14}~M_\odot$ and $\rm 4.3^{+0.4}_{-0.5}\times10^{14}~M_\odot$, respectively; see Tab. \ref{tab:masses}).
The two highest redshift clusters, instead, -- namely MOOJ0123 and MOOJ1420 -- are well above the radio power-mass correlation, using the masses obtained from the mass-richness scaling relation or the ones from the X-ray flux from eROSITA observations. 
However, similarly to MOOJ1231, the CARMA observation for MOOJ0123 points to a higher mass, i.e. $\sim2$ times higher than those from those obtained from the scale relations. Assuming this latter mass, this cluster would also lie within the scatter of the correlation found by \cite{cuciti+23}. It is worth noting that the CARMA high frequency observations (i.e. 30 GHz) are characterised by poor resolution (i.e. $40''$) and low sensitivity, combined with interferometric filtering, and single-frequency data. 

If we assume that the masses estimated from the scaling relations are reliable, considering the high redshifts of these clusters -- and therefore the stronger Inverse Compton energy losses (i.e. $dE/dt\propto (1+z)^4$) --, the location of these systems above the $P_{\nu}-M_{500}$ correlation is quite surprising. Following the reasoning by \cite{digennaro+21a}, at high redshift it is expected that the flux of turbulent energy ($\rho v_t^3/L_{\rm inj}$, where $\rho$ is the gas density, and $v_t$ and $L_{\rm inj}$ are the turbulent velocity and injection scale, respectively) is about 3 times higher than that dissipated at lower redshift ($z\sim0.2$) because of the larger impact velocities and virial densities of merging clusters. This translates to an average magnetic field level in clusters at $z\sim0.7$ that would be similar to that in the low-redshift ones, i.e. $\rm\sim few~\mu$G, if we measured similar radio power of the diffuse radio emission. Assuming the amount of flux of turbulent energy remains constant between $z\sim0.7$ and $z\sim1$ with respect to the low-redshift regime (and that $\eta_t$ is redshift-independent), the over-luminosity of 2 orders of magnitude we see in our sample suggests an average magnetic field level that is up to one order of magnitude higher than at low redshift.

This result would pose challenges in the understanding the evolution of magnetic fields in galaxy clusters over the cosmic time. The only other cluster known so far to clearly host a radio halo at $z>1$ is ACT-CL J0329.2-2330 \citep[$z=1.23$;][]{sikhosana+24}. Moreover, a putative claim for extended radio emission was also made for SPT-CL J2106-584 \citep[$z=1.13$;][]{dimascolo+21}, although with the data in hand it was not possible to unambiguously separate the contribution of the of individual cluster galaxies. These systems are both placed on the $P_{\rm 1.4GHz}-M_{500}$ correlation, therefore suggesting $\sim\mu$G-level magnetic fields, but, in contrast to those from our sample, they are extremely massive ($M_{\rm SZ,500}=9.7^{+1.7}_{-1.6}\times10^{14}~{\rm M_\odot}$ and $M_{\rm SZ,500}=8.3^{+0.8}_{-1.0}\times10^{14}~{\rm M_\odot}$, respectively). For these cases, it is plausible that the formation of the systems started earlier in the cosmic evolution of the Universe and, therefore, they would amplify their magnetic fields up to the $\sim\mu$G level earlier.

We finally note that from the full sample of the MaDCoWS clusters in LoTSS-DR2, we retrieve a detection rate for diffuse radio emission of $\sim9\%$ (i.e. 5 candidate extended radio emission over 56 clusters), in the redshift range 0.78--1.53. This is much smaller than the $\sim50\%$ value previously found by \cite{digennaro+21a}, for the $z=0.6-0.9$ redshift range and much larger cluster masses ($M_{\rm SZ,500}\sim4-8\times10^{14}~{\rm M_\odot}$).  
Although this is a simplistic comparison, which does not take into account the different sample selection (i.e. SZ versus optical), the decreasing detection rate with the redshift is not unexpected because of the largest energy losses due to IC effects on the relativistic particles ($dE/dt \propto (1+z)^4$) and because of the low masses of our MaDCoWS clusters, as also highlighted by theoretical models \citep{cassano+23}.

\subsection{Caveats}
The comparison with the low-redshift samples in the $P_{\nu}-M_{500}$ diagram is strongly affected by 
the uncertainties in the mass estimation, and on the discrepancies among the targeted observations \citep[with CARMA at 30 GHz, see][]{decker+19} and the values obtained through the scaling relations, both from the IR-selected richness and from the X-ray flux. This uncertainty reflects on the interpretation on the origin of the extended radio emission in these high-$z$ clusters, where a difference of a factor of 2 in mass strongly shifts the position of the cluster with the respect to the correlation. This is clearly seen for MOOJ0123, which is located within the scatter of the correlation if the mass estimated by the CARMA observations is taken, while it is more than one order of magnitude more radio luminous assuming the mass obtained from the two scale relations. A reliable estimation of the cluster mass, for instance with the MUSTANG-2, at the Green Bank Telescope (GBT) at 90 GHz \citep{dicker+14}, is therefore a crucial point for such studies. This is currently under investigatoin and will be part of a forthcoming work.

Moreover, the radio luminosities are estimated by assuming a given spectral index ($\alpha=-1.5\pm0.3$) which is taken from limited literature studies at high redshift \citep{digennaro+21a,digennaro+21c}. This, however, would only affect $P_{\rm 1.4GHz}$ and would not justify the position of two orders of magnitude above the scatter of the clusters at low redshift for MOOJ0123 and MOOJ1420. Following \cite{digennaro+21c}, higher-frequency observations with the uGMRT could help to determine a more precise spectral index of this extended diffuse radio emission, while lower-frequency observations with LOFAR LBA would be limited by poorer resolution (i.e. $15''$) and sensitivity \citep[$\rm\sim1~mJy\,beam^{-1}$; see][]{degasperin+23}.

Finally, we cannot completely exclude that part of the radio emission seen as extended in the cluster volume is actually due to blending of unresolved active galactic nuclei (AGN). 
To quantify this effect, we artificially masked all the observed radio galaxies in the full resolution source-subtracted images, and then successively smooth the data to lower resolutions (i.e. {\tt taper=100kpc}). Using this method, the flux densities of the diffuse emission decrease of 25--40\%. 
To better exploit the effect of the contamination of faint AGN in the full sample, observations with the International LOFAR Telescope (ILT) -- whose antennas are located throughout Europe -- are necessary to provide the necessary resolution (up to sub-arcsecond) to disentangle the two kinds of different radio emission.

\begin{figure*}
\centering
\includegraphics[width=0.8\textwidth]{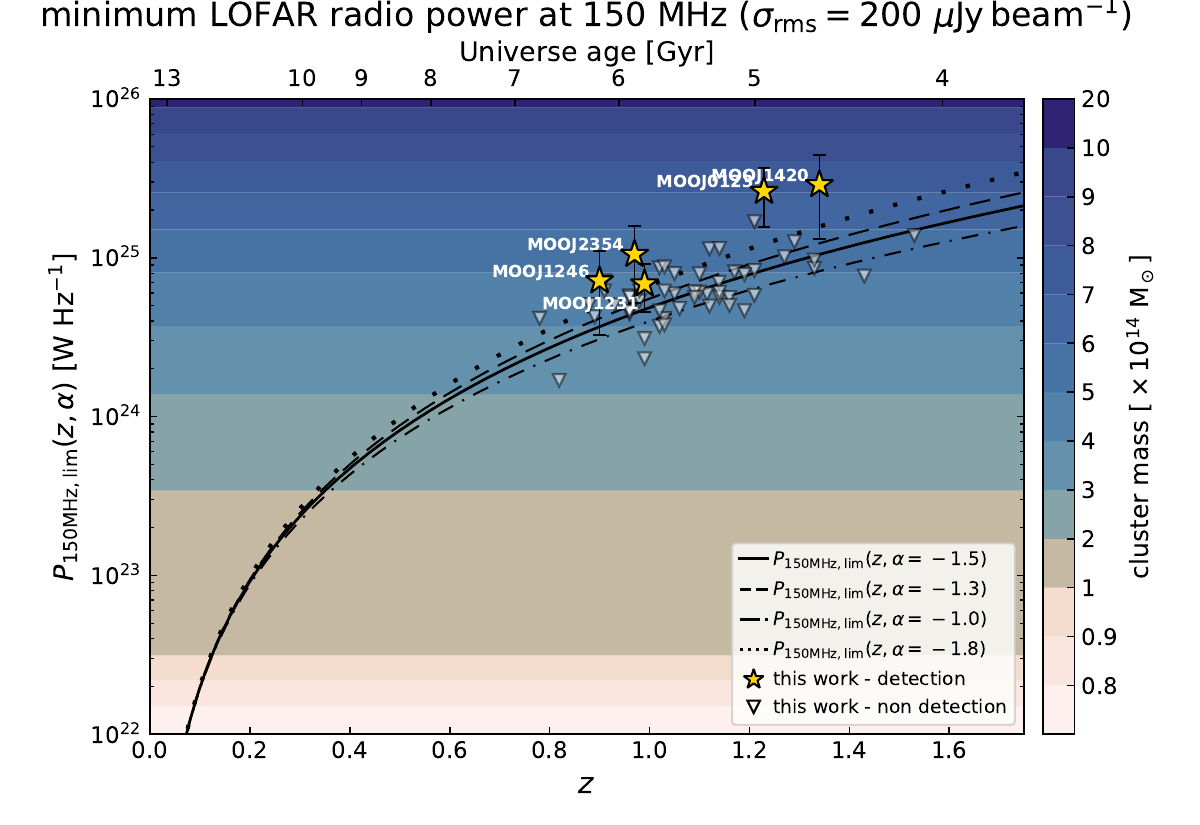}
\vspace{-5mm}
\caption{Detection limit as a function of the redshift ($z$), as detectable by a standard LoTSS observation (Eq. \ref{eq:power_limt}). Different lines show the dependence of the radio power on different spectral indices (solid, $\alpha=-1.5$; dot-dashed, $\alpha=-1.0$; dashed, $\alpha=-1.3$; dotted, $\alpha=-1.8$). 
The colour bar and the coloured bands refer to the mass that a galaxy cluster should have to lie {\it exactly} on the $P_{\rm 150MHz}-M_{500}$ correlation found by \cite{cuciti+23}.
Clusters from the MaDCoWS-LoTSS DR2 sample are also displayed (detections with golden stars, and non-detections with low-vertices triangles). }\label{fig:powerlim}
\end{figure*}

\subsection{Limits from LoTSS observations}
As mentioned in Sect. \ref{sec:ul}, most of the clusters in the MaDCoWS sample in the LoTSS-DR2 area do not show radio emission on the Mpc scale. To investigate whether this is a limit due to the observations, we derived the minimum flux detectable by LoTSS observation as presented by \cite{cassano+23}:
\begin{equation}\label{eq:power_limt}
S_{\rm 150MHz,lim}(<3\Theta_e,z)=4.44\times10^{-3} \xi \, \rms \left ( \frac{\Theta_e(z)}{\Theta_{\rm beam}} \right) \quad [{\rm mJy}] \, .
\end{equation}
Here, $\rms=200~\mu$\jybeam\ is the nominal map noise at low resolution (i.e. {\tt taper=100kpc}) of a standard LoTSS observation of 8 hours\footnote{This is also the median of our {\tt taper=100kpc} observations.} \citep{shimwell+22}, $\Theta_{\rm beam}$ is the observing resolution in arcsecond, $\Theta_e$ is the angular size of the $e$-folding radius $r_e$, being equal to 75 kpc (see Sect. \ref{sec:ul}). 
All the parameters described above are set to roughly describe the behaviour of the upper limits from our sample \citep{cassano+23}.
The minimum radio power detectable at 150 MHz was then calculated using Eq. \ref{eq:radiopow}, assuming different values for different spectral indices, i.e. $\alpha=[-1.0, -1.3, -1.5, -1.8]$.

In Fig. \ref{fig:powerlim}, we show the comparison of this theoretical limit and all the clusters in our sample. As expected, all the clusters with diffuse radio emission are above the $P_{\rm 150MHz,lim}(z,\alpha)$ curve, while the upper limits are all located on the theoretical limits. 
This means that at the high redshift ($z>0.8$) and relatively low mass ($M_{500}\lesssim4\times10^{14}~{\rm M_\odot}$) of the MaDCoWS clusters, we are limited by the LoTSS sensitivity \citep{shimwell+22}.
In Appendix \ref{apx:powerlimit}, we show the comparison of the evolution of the detectable radio power with deeper observations (i.e. observing time 100 hours), reaching a noise level $\rms=55~\mu$\jybeam\ at the same low resolution (e.g. $\Theta_{\rm beam}=100$ kpc, see Fig. \ref{fig:deepobspower}). 

We also show in colour shades the mass that a galaxy cluster should have to lie exactly on the $P_{\rm 150MHz}-M_{500}$ correlation presented in \cite{cuciti+23}, i.e.:
\begin{equation}
\log \left ( \frac{P_{\rm 150MHz}}{\rm 10^{24.5} ~ W\,Hz^{-1}} \right ) = B \log \left ( \frac{M_{500}}{\rm 10^{14.9}~M_\odot} \right ) + A \, ,
\end{equation}
with $A=1.1\pm0.09$ and $B=3.45\pm0.44$. Although we do not take into account the scatter of the correlation, this implies that at $z=0.8$ and at $z=1.4$ clusters with masses $M_{500}\gtrsim 4\times10^{14}~{\rm M_\odot}$ and $M_{500}\gtrsim 6\times10^{14}~{\rm M_\odot}$, respectively, could in principle be detectable to host extended radio emission and following the correlation. Clusters with such high masses are supposed to be rare, in the context of the $\Lambda$CDM cosmology, at such high redshift \citep{menanteau+12,katz+13,jee+14}. At the same time, the comparison with expected cluster masses and the $P_{\rm 150MHz,lim}$ curve challenges the chances to populate the region below the correlation, where the high-$z$ radio halos should lie, because of the expected lower magnetic field levels \citep{digennaro+21a}. Deeper LOFAR observations \citep[$>100$ hours on target; see][]{tasse+21} could in principle help to detect lower-mass clusters (see Appendix \ref{apx:powerlimit}, Fig. \ref{fig:deepobspower}), but they will be demanding, and therefore would be feasible only for a selected number of clusters and not for large surveys.

\section{Summary and future analysis}\label{sect:concl}
In this paper, we have attempted for the first time to investigate the presence of extended and diffuse radio emission in a large sample of galaxy clusters selected at high redshift (i.e. $z>0.75$). We have made use of the Massive and Distant Clusters of {\it WISE} Survey \citep[MaDCoWS;][]{gonzalez+19}, where we select clusters with richness $\lambda_{15}>40$ which are in the second data release of the LOFAR Two-Meter Sky Survey (LoTSS-DR2). 

The final sample collects 56 galaxy clusters with a median redshift $\langle z_{\rm LoTSS}\rangle=1.05$. Among these, only 5 systems show hints of diffuse radio emission on the cluster scale (i.e. a fraction of about 9\%). All these candidate radio halos have integrated flux densities that correspond to radio powers that are above the $P_{\rm 150MHz}-M_{500}$ and $P_{\rm 1.4GHz}-M_{500}$ correlations at lower redshifts \citep[respectively]{cuciti+23,cassano+13}. However, we stress that the mass values we report for the clusters in our sample are still very uncertain. Future targeted SZ observations with MUSTANG-2, at the Green Bank Telescope (GBT) at 90 GHz \citep{dicker+14}, or near-IR observations, with the James Webb Space Telescope \citep[JWST;][]{jakobsen+22,boker+23} and the ESA-{\it Euclid} mission \citep{laureijs+11,euclidcoll+22}, would provide a more reliable estimation of the mass values. 

We also investigated the limitations of our radio observations. Assuming a standard LoTSS setting (i.e. 8 hr on pointing) where a sensitivity of $200~\mu$\jybeam\ at low resolution ($\Theta_{\rm beam}=100$ kpc) is reached, we will only be able to detect the most powerful cluster-scale diffuse radio emission with radio powers at $z>0.8$ (i.e. $P_{\rm 150MHZ}>10^{25}~{\rm W\,Hz^{-1}}$). If we assume an exact relation between the luminosity of the diffuse radio emission and the cluster mass according to \cite{cuciti+23}, this would imply that clusters with masses above $6\times10^{14}~{\rm M_\odot}$ could be observed to host such extended diffuse radio sources.
Additionally, we should keep in mind that the fraction of \textit{Planck} clusters found to host a radio halo is only $\sim30\%$ \citep{botteon+22}, averaged for a large range of redshift ($z=0.016-0.9$, with a median of $0.280$) and mass ($M_{\rm SZ,500}=1.1-11.7\times10^{14}~{\rm M_\odot}$, with a median of $4.9\times10^{14}~{\rm M_\odot}$). This fraction is expected to decrease at higher redshift and, especially, for lower masses \citep{cassano+23}.

All these findings pose a limitation on the detection of diffuse radio emission from samples of high-redshift clusters. However, the forthcoming large high-redshift surveys with a reliable estimation of the cluster mass -- such as {\it Euclid}, where $>10^5$ clusters are expected to be found up to $z\sim2$ -- will provide interesting systems to target with deep LOFAR HBA observations. 


\begin{acknowledgements}
We thank the referee for the suggestions which improved the quality of the manuscript.
GDG and FdG acknowledge support from the ERC Consolidator Grant ULU 101086378. GDG and MB acknowledge funding by the DFG under Germany's Excellence Strategy -- EXC 2121 ``Quantum Universe'' --  390833306. MJH thanks the UK STFC for support [ST/V000624/1, ST/Y001249/1]. RJvW acknowledges support from the ERC Starting Grant ClusterWeb 804208. IK acknowledges support by the COMPLEX project from the European Research Council (ERC) under the European Union’s Horizon 2020 research and innovation program grant agreement ERC-2019-AdG 882679.
LOFAR data products were provided by the LOFAR Surveys Key Science project (LSKSP; \url{https://lofar-surveys.org/}) and were derived from observations with the International LOFAR Telescope (ILT). LOFAR \citep{vanhaarlem+13} is the Low Frequency Array designed and constructed by ASTRON. It has observing, data processing, and data storage facilities in several countries, which are owned by various parties (each with their own funding sources), and which are collectively operated by the ILT foundation under a joint scientific policy. The ILT resources have benefited from the following recent major funding sources: CNRS-INSU, Observatoire de Paris and Université d'Orléans, France; BMBF, MIWF-NRW, MPG, Germany; Science Foundation Ireland (SFI), Department of Business, Enterprise and Innovation (DBEI), Ireland; NWO, The Netherlands; The Science and Technology Facilities Council, UK; Ministry of Science and Higher Education, Poland; The Istituto Nazionale di Astrofisica (INAF), Italy.
This research made use of the Dutch national e-infrastructure with support of the SURF Cooperative (e-infra 180169) and the LOFAR e-infra group. The J\"ulich LOFAR Long Term Archive and the German LOFAR network are both coordinated and operated by the J\"ulich Supercomputing Centre (JSC), and computing resources on the supercomputer JUWELS at JSC were provided by the Gauss Centre for Supercomputing e.V. (grant CHTB00) through the John von Neumann Institute for Computing (NIC).
This research made use of the University of Hertfordshire high-performance computing facility and the LOFAR-UK computing facility located at the University of Hertfordshire and supported by STFC [ST/P000096/1], and of the Italian LOFAR IT computing infrastructure supported and operated by INAF, and by the Physics Department of Turin university (under an agreement with Consorzio Interuniversitario per la Fisica Spaziale) at the C3S Supercomputing Centre, Italy.
This work is based on observations with eROSITA telescope on board SRG space observatory. The SRG observatory was built by Roskosmos in the interests of the Russian Academy of Sciences represented by its Space Research Institute (IKI) in the framework of the Russian Federal Space Program, with the participation of the Deutsches Zentrum f\"ur Luftund Raumfahrt (DLR). The eROSITA X-ray telescope was built by a consortium of German Institutes led by MPE, and supported by DLR. The SRG spacecraft was designed, built, launched and is operated by the Lavochkin Association and its subcontractors. The science data are downlinked via the Deep Space Network Antennae in Bear Lakes, Ussurijsk, and Baikonur, funded by Roskosmos. The eROSITA data used in this work were converted to calibrated event lists using the eSASS software system developed by the German eROSITA Consortium and analysed using proprietary data reduction software developed by the Russian eROSITA Consortium.
This research made use of {\tt APLpy}, an open-source plotting package for Python \citep{aplpy}, {\tt astropy}, a community-developed core Python package for Astronomy \citep{astropy+13,astropy+18}, {\tt matplotlib} (Hunter 2007), and {\tt numpy} \citep{harris+20}.
\end{acknowledgements}

\bibliographystyle{aa}
\bibliography{biblio.bib}

\clearpage
\onecolumn
\begin{appendix}
\clearpage

\section{LOFAR images of MaDCoWS clusters}\label{apx:all}
List of all the clusters in  in our MaDCoWS-LoTSS DR2 sample (Tab. \ref{tab:all}). 
Images at all the resolutions (with and without compact sources) are displayed only for those with diffuse radio emission (Fig. \ref{fig:madcows_lofar_images}). 

\begin{tiny}
\begin{longtable}{ccccccccc}
\caption{List of MaDCoWS clusters \citep{gonzalez+19} in LOFAR-DR2 \citep{shimwell+22} with richness $\lambda_{15}>40$.}\label{tab:all}\\
\hline\hline
Cluster name & Redshift & Right Ascension &	Declination	& Richness & Mass & \multicolumn{2}{c}{Map noise (full/low res)} & Flux density \\
(MOO)       & $z~(z_{\rm spec})$       & RA [deg]    & Dec [deg]   & $\lambda_{15}$ & $M_{500}~[\times10^{14}~{\rm M_\odot}]$ & \multicolumn{2}{c}{$\sigma_{\rm rms}~[\mu$\jybeam]} & $S_{\rm 144MHz,diff}$ [mJy] \\
\hline
\endfirsthead
\caption{Continued.}\\
\hline \hline
Cluster name & Redshift & Right Ascension &	Declination	& Richness & Mass & \multicolumn{2}{c}{Map noise (full/low res)} & Flux density \\
(MOO)       & $z~(z_{\rm spec})$       & RA [deg]    & Dec [deg]   & $\lambda_{15}$ & $M_{500}~[\times10^{14}~{\rm M_\odot}]$ & \multicolumn{2}{c}{$\sigma_{\rm rms}~[\mu$\jybeam]} & $S_{\rm 144MHz,diff}$ [mJy] \\
\hline
\endhead
\hline
\endfoot
J0006+3050 &	1.02 &	1.62208	  &  30.84861 & $42\pm7$ & $2.0^{+0.7(+0.4)}_{-0.6(-0.4)}$ & 132 & 319 & $<1.4$\\
J0024+3303 &	1.16 (1.115) & 	6.19375	  &  33.05917 & $61\pm8$  & $3.9^{+1.1(+0.3)}_{-0.9(-0.3)}$ & 77 & 163 & $<0.6$\\
J0029+2657 &	1.14 & 	7.45792	  &  26.96139 & $48\pm7$  & $2.5^{+0.8(+0.4)}_{-0.7(-0.3)}$ & 98 & 157 & $<0.6$\\ 	    
J0035+2358 &	0.96 & 	8.75958	  &  23.97139 & $42\pm7$  & $2.0^{+0.6(+0.4)}_{-0.6(-0.4)}$ & 108 & 176 & $<0.8$\\ 
J0037+3306 &	1.06 (1.139) & 	9.4425	  &  33.11611 & $53\pm7$ & $2.3\pm0.6^\dagger$ & 98 & 189 & $<0.6$\\ 
J0054+2959 &	0.96 & 	13.56125  &	29.99583 & $57\pm8$  & $3.5^{+1.0(+0.3)}_{-0.6(-0.3)}$ & 87 & 203 & $<0.7$\\  	    
J0056+2048 &	1.10 & 	14.0325	  &  20.81556 & $42\pm6$  & $2.0^{+0.6(+0.4)}_{-0.5(-0.4)}$ & 144 & 228 & $<1.2$\\	    
J0056+3202 &	0.98 & 	14.10917  &	32.0350 & $42\pm7$  & $2.0^{+0.7(+0.4)}_{-0.5(-0.4)}$ & 101 & 258 & $<0.7$\\
J0107+3341 &	0.91 & 	16.82375  &	33.68556 & $43\pm7$	 & $2.1^{+0.7(+0.4)}_{-0.6(-0.3)}$ & 129 & 383 & $<1.4$ \\
J0123+2545 &	1.19 (1.229) & 	20.9625	  & 25.76194 & $41\pm6$ & $3.9^{+0.9\dagger}_{-0.8}$ & 99 & 179 & $2.1\pm0.6$\\
J0242+2951 &	1.05 & 	40.61167  &	29.86083 & $48\pm7$	 & $2.5^{+0.7(+0.4)}_{-0.6(-0.4)}$ & 98 & 216 & $<0.7$\\	    
J0807+3732 &	0.91 & 	121.89667 &	37.53806 &  $41\pm7$  & $1.9^{+0.6(+0.4)}_{-0.5(-0.4)}$ & 152 & 312 & $<1.0$\\	     	 
J0824+6611 &	0.99 & 	126.20333 &	66.18417 & $41\pm7$ & $1.9^{+0.6(+0.4)}_{-0.6(-0.4)}$ & 54 & 119 & $<0.4$\\
J0836+3504 &	0.96 & 	129.20125 &	35.07472 & $43\pm7$ & $2.1^{+0.6(+0.4)}_{-0.6(-0.4)}$ & 129 & 246 & $<1.0$\\  	    
J0843+5933 &	1.03 & 	130.93875 &	59.56194 & $50\pm7$ & $2.8^{+0.8(+0.3)}_{-0.7(-0.3)}$ & 86 & 144 & $<0.6$\\
J0846+3128 &	1.03 & 	131.58625 &	31.47139 & $43\pm7$ & $2.1^{+0.6(+0.4)}_{-0.5(-0.4)}$ & 155 & 318 & $<1.1$\\  	    
J0907+2908 &	0.96 & 	136.83333 &	29.13583 & $47\pm7$ & $2.5^{+0.7(+0.4)}_{-0.6(-0.4)}$ & 239 & 475 & N/A \\	    
J0917+4710 &	1.12 & 	139.31542 &	47.17417 & $44\pm7$ & $2.2^{+0.7(+0.4)}_{-0.6(-0.4)}$ & 109 & 298 & $<1.2$\\  	    
J0944+3710 &	1.21 & 	146.09208 &	37.1725  & $43\pm6$ & $2.1^{+0.6(+0.4)}_{-0.5(-0.4)}$ & 85 & 199 & $<0.7$\\	    
J1001+6619 &	1.53 & 	150.34333 &	66.32306 & $41\pm6$ & $1.9^{+0.5(+0.4)}_{-0.5(-0.4)}$& 92 & 170 & $<0.6$\\	    
J1003+6836 &	1.02 & 	150.99708 &	68.60778 & $41\pm6$ & $1.9^{+0.6(+0.4)}_{-0.5(-0.4)}$ & 73 & 137 & $<0.5$\\	    
J1031+6255 &	1.33 & 	157.95208 &	62.92194 & $50\pm7$ & $2.7^{+0.8(+0.3)}_{-0.7(-0.3)}$ & 80 & 155 & $<0.5$\\	    
J1059+5454 &	1.14 & 	164.95583 &	54.91028 & $57\pm7$ & $3.4^{+0.9(+0.3)}_{-0.7(-0.3)}$ & 110 & 325 & $<1.0$\\	    
J1108+3242 &	1.12 & 	167.19333 &	32.71389 & $63\pm8$ & $4.2^{+1.1(+0.3)}_{-0.9(-0.3)}$ & 74 & 142 & $<0.7$\\ 
J1110+6838 &	0.93 & 	167.71625 &	68.63917 & $55\pm7$ & $3.3^{+0.8(+0.3)}_{-0.7(-0.3)}$ & 531 & 1125 & N/A\\	    
J1135+3256 &	1.19 & 	173.83667 &	32.93444 & $46\pm7$ & $2.4^{+0.7(+0.4)}_{-0.6(-0.4)}$ & 278 & 576 & N/A\\  
J1207+3643 &	1.06 & 	181.81	  & 36.72056 & $41\pm6$ & $1.9^{+0.6(+0.4)}_{-0.5(-0.4)}$ & 83  & 163 & $<0.7$\\	    
J1229+6521 &	0.80 (0.819) & 	187.4350  &	65.36361 & $45\pm7$  & $2.3^{+0.6(+0.4)}_{-0.6(-0.4)}$ & 64 & 111 & $<0.4$\\
J1231+6533 &	0.99 & 	187.81375 &	65.55889 & $50\pm7$ & $4.7^{+1.3\dagger}_{-0.9}$ & 64 & 107 & $1.0\pm0.3$\\ 
J1241+3842 &	1.16 & 	190.25292 &	38.70278 & $41\pm6$ & $1.9^{+0.6(+0.4)}_{-0.5(-0.4)}$ & 71 & 143 & $<0.5$\\ 
J1246+4642 &	0.90 & 	191.70125 &	46.71583 & $47\pm7$ & $2.5^{+0.7(+0.4)}_{-0.6(-0.4)}$ & 117 & 342 & $1.3\pm0.6$\\  	    
J1248+6723 &	0.88 & 	192.19083 &	67.39889 & $44\pm7$ & $2.2^{+0.6(+0.4)}_{-0.6(-0.4)}$ & 123 & 258 & N/A \\ 
J1307+4131 &	1.03 & 	196.82792 &	41.53222 & $50\pm7$ & $2.8^{+0.8(+0.3)}_{-0.7(-0.3)}$ & 74 & 141 & $<0.5$\\ 
J1308+2429 &	1.29 & 	197.06375 &	24.49778 & $41\pm6$ & $1.9^{+0.5(+0.4)}_{-0.5(-0.4)}$ & 104 & 247 & $<1.1$\\	    
J1310+2852 &	1.10 & 	197.7075  &	28.87111 & $47\pm7$ & $2.4^{+0.7(+0.4)}_{-0.7(-0.4)}$ & 95 & 204 & $<0.7$\\	    
J1319+5519 &	0.94 (0.936) & 	199.91625 &	55.31861 & $44\pm7$ & $2.2^{+0.7(+0.4)}_{-0.6(-0.3)}$ & 64 & 173 & N/A\\ 
J1321+4411 &	1.19 & 	200.25208 &	44.19306 & $55\pm7$ & $3.3^{+0.9(+0.3)}_{-0.7(-0.3)}$ & 74 & 149 & $<0.4$\\ 
J1329+5647 &	1.43 & 	202.45	  & 56.79417 & $42\pm6$ & $2.0^{+0.6(+0.4)}_{-0.5(-0.4)}$ & 69 & 106 & $<0.4$\\ 
J1336+4622 &	0.91 & 	204.0650  &	46.37944 & $45\pm7$ & $2.3^{+0.7(+0.4)}_{-0.6(-0.4)}$ & $10^8$ & $5\times10^8$ & N/A\\
J1337+3529 &	1.21 & 	204.38125 &	35.4850 & $51\pm7$  & $2.8^{+0.8(+0.3)}_{-0.7(-0.3)}$ & 66 & 139 & $<0.5$ \\
J1341+2320 &	1.03 & 	205.38208 &	23.33556 & $45\pm7$ & $2.3^{+0.7(+0.4)}_{-0.6(-0.4)}$ & 111 & 228 & $<0.8$\\
J1349+3008 &	1.17 & 	207.40208 &	30.14639 & $48\pm7$ & $2.5^{+0.8(+0.4)}_{-0.7(-0.4)}$ & 114 & 205 & $<0.7$\\ 
J1351+3044 &	0.88 & 	207.98333 &	30.73833 & $41\pm7$ & $1.9^{+0.6(+0.4)}_{-0.6(-0.4)}$ & 132 & 297 & $<1.0$\\	    
J1358+2654 &	1.33 & 	209.65333 &	26.90639 & $41\pm6$ & $1.9^{+0.5(+0.4)}_{-0.5(-0.4)}$ & 87 & 165 & $<0.7$\\ 
J1420+3150 &	1.34 & 	215.08417 &	31.84833 & $41\pm6$ & $1.9^{+0.5(+0.4)}_{-0.5(-0.4)}$ & 105 & 176 & $1.8\pm0.8$\\	    
J1420+3633 &	0.98 & 	215.14375 &	36.56167 & $43\pm7$ & $2.0^{+0.7(+0.4)}_{-0.6(-0.4)}$ & 98 & 258 & $<0.9$\\	    
J1427+5309 &	1.05 & 	216.93125 &	53.15333 & $53\pm7$ & $3.0^{+0.8(+0.3)}_{-0.7(-0.3)}$ & 98 & 277 & $<1.0$\\ 
J1435+4759 &	1.02 & 	218.84917 &	47.99694 & $43\pm7$ & $2.1^{+0.7(+0.4)}_{-0.6(-0.4)}$ & 70 & 179 & $<0.8$\\ 
J1438+4120 &	0.96 & 	219.7375  &	41.33917 & $43\pm7$ & $2.1^{+0.7(+0.4)}_{-0.6(-0.4)}$ & 90 & 180 & $<0.7$\\ 
J1506+5137 &	1.09 & 	226.58625 &	51.61694 & $74\pm8$ & $5.6^{+1.4(+0.3)}_{-1.1(-0.3)}$ & 77 & 207 & N/A \\ 
J1507+5234 &	0.96 & 	226.94333 &	52.57944 & $57\pm7$ & $3.5^{+1.0(+0.3)}_{-0.8(-0.3)}$ & 101 & 270 & $<0.9$\\  	    
J1511+3719 &	1.14 (1.090) & 	227.96292 &	37.32028 & $53\pm7$ & $3.1^{+0.8(+0.3)}_{-0.7(-0.3)}$ & 103 & 179 & $<0.7$\\ 
J1520+5751 &	1.09 & 	230.05667 &	57.85056 & $48\pm7$ & $2.5^{+0.8(+0.4)}_{-0.6(-0.4)}$ & 91 & 161 & $<0.7$\\ 
J1522+5259 &	1.27 & 	230.6175  &	52.98639 & $45\pm6$ & $2.3^{+0.6(+0.4)}_{-0.5(-0.4)}$ & 84 & 200 & $<0.7$\\ 
J1551+6245 &	1.21 & 	237.92208 &	62.75056 & $44\pm7$ & $2.2^{+0.6(+0.4)}_{-0.5(-0.4)}$ & 162 & 364 & $<1.2$\\  	    
J1558+5154 &	0.99 & 	239.67125 &	51.90861 & $44\pm7$ & $2.2^{+0.7(+0.4)}_{-0.6(-0.4)}$ & 59 & 124 & $<0.5$\\ 
J1616+6053 &	1.09 & 	244.24125 &	60.89944 & $42\pm7$ & $2.0^{+0.7(+0.4)}_{-0.6(-0.4)}$ & 474 & 1426 & N/A \\ 
J1639+3831 &	1.19 & 	249.77042 &	38.53111 & $48\pm6$ & $2.6^{+0.6(+0.3)}_{-0.6(-0.4)}$ & 88 & 219 & $<0.8$\\
J1659+6501 &	1.16 & 	254.96375 &	65.0225 & $52\pm7$  & $3.0^{+0.8(+0.3)}_{-0.7(-0.3)}$ & 81 & 123 & $<0.5$\\
J1732+4102 &	1.14 & 	263.0150  &	41.03722 & $51\pm7$ & $2.8^{+0.8(+0.3)}_{-0.7(-0.3)}$ & 84 & 161 & $<0.8$\\	    
J1751+4307 &	0.93 & 	267.78292 &	43.12917 & $55\pm8$ & $3.3^{+0.9(+0.3)}_{-0.8(-0.3)}$ & 88 & 196 & $<1.0$\\ 
J2239+2929 &	0.89 & 	339.87542 &	29.49306 & $69\pm8$ & $4.9^{+0.9(+0.3)}_{-1.0(-0.3)}$ & 135 & 238 & $<0.8$\\  	    
J2317+2829 &	0.78 & 	349.37375 &	28.49139 & $44\pm7$ & $2.2^{+0.7(+0.4)}_{-0.6(-0.4)}$ & 141 & 390 & $<1.2$\\
J2354+3507 &	0.97 & 	358.6775  &	35.11972 & $54\pm7$ & $3.2^{+0.9(+0.3)}_{-0.7(-0.3)}$ & 109 & 228 & $1.6\pm0.8$\\  
\end{longtable}
\end{tiny}
\tablefoot{First to fifth columns: cluster name, photometric redshift, coordinates and richness as reported in \cite{gonzalez+19}; if available, the spectroscopic redshift ($z_{\rm spec}$) is indicated within brackets in the second column, and used during the analysis.
Sixth column: cluster masses retrieved using the mass-richness scaling relation \citep[$M_{500}-\lambda_{15}$; see][]{orlowski-scherer+21}; $^\dagger$masses obtained from CARMA observations \citep{decker+19}. Seventh column: map noise of the full-resolution (left) and low-resolution (i.e. {\tt taper=100kpc}; right) LoTSS images, for the full $uv$-plane. Eighth column: 144 MHz flux density or upper limits of the diffuse radio emission.}

\begin{figure}[h!]
\centering
\includegraphics[width=\textwidth]{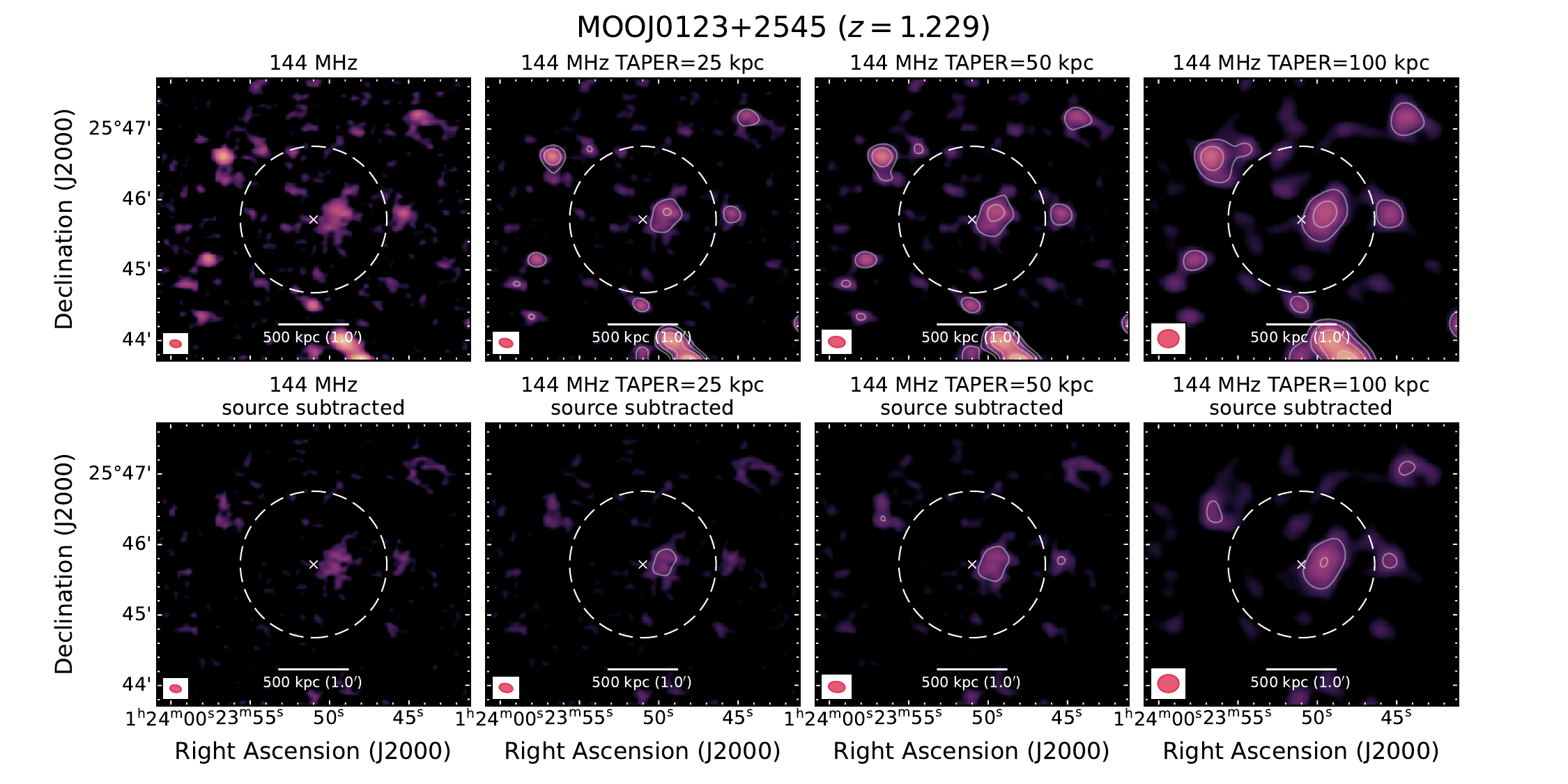}
\includegraphics[width=\textwidth]{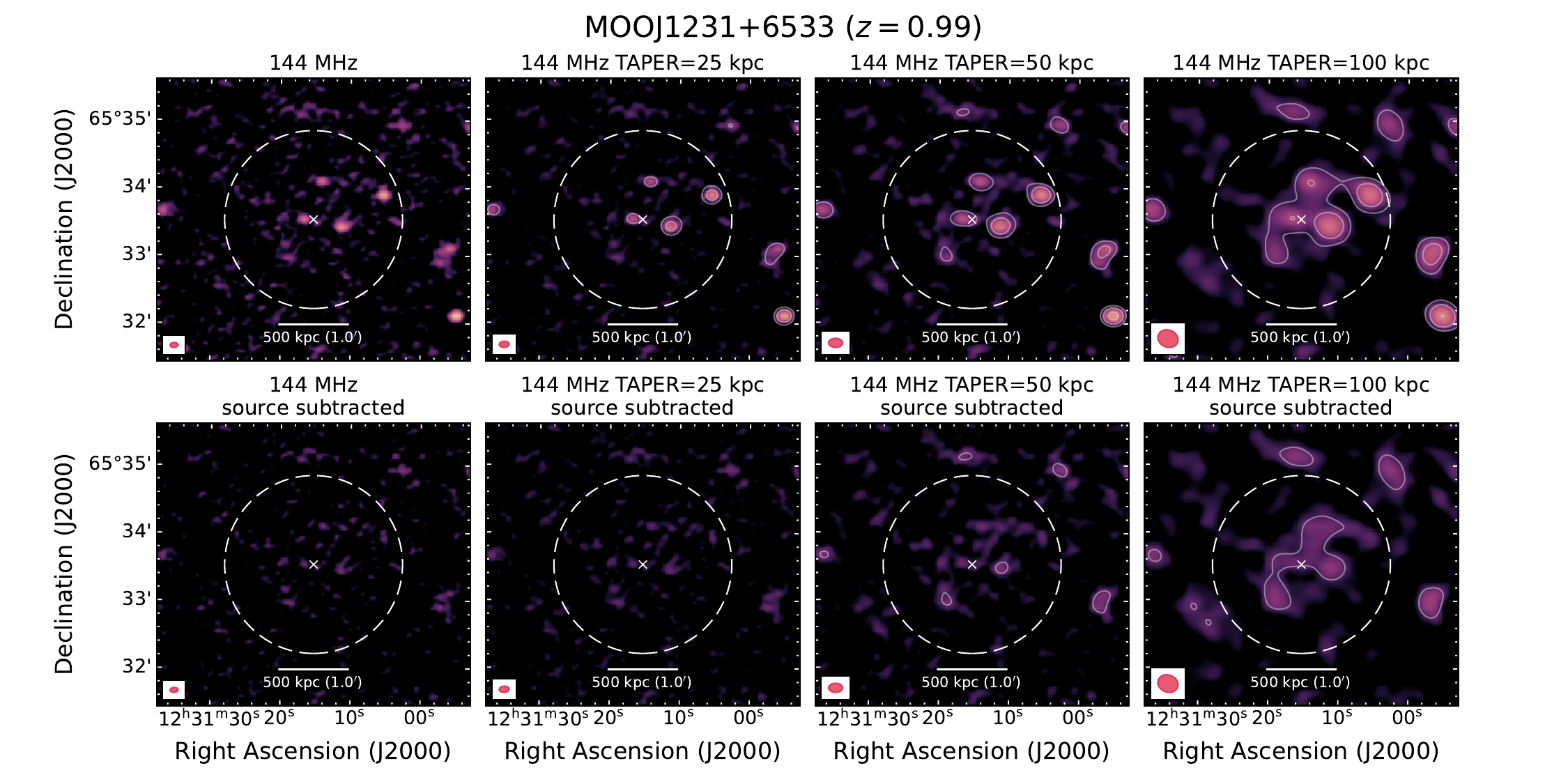}
\caption{LOFAR 144 MHz images of the MaDCoWS clusters before (top row) and after (bottom row) source subtraction. From left to right, full-resolution, {\tt taper=25kpc}, {\tt taper=50kpc} and {\tt taper=100kpc} images; each beam is displayed in the bottom-left corner. Radio contours are displayed in white solid lines, starting from $2.5\sigma_{\rm rms}\times[2,4,8,16,32]$; negative contours at $-2.5\sigma_{\rm rms}$ are shown in white dashed lines. The dashes white circle shows the $R=500$ kpc area, while the white cross marks the MaDCoWS coordinates reported in \cite{gonzalez+19}.}
\label{fig:madcows_lofar_images}
\end{figure}

\begin{figure}
\centering
\ContinuedFloat
\includegraphics[width=\textwidth]{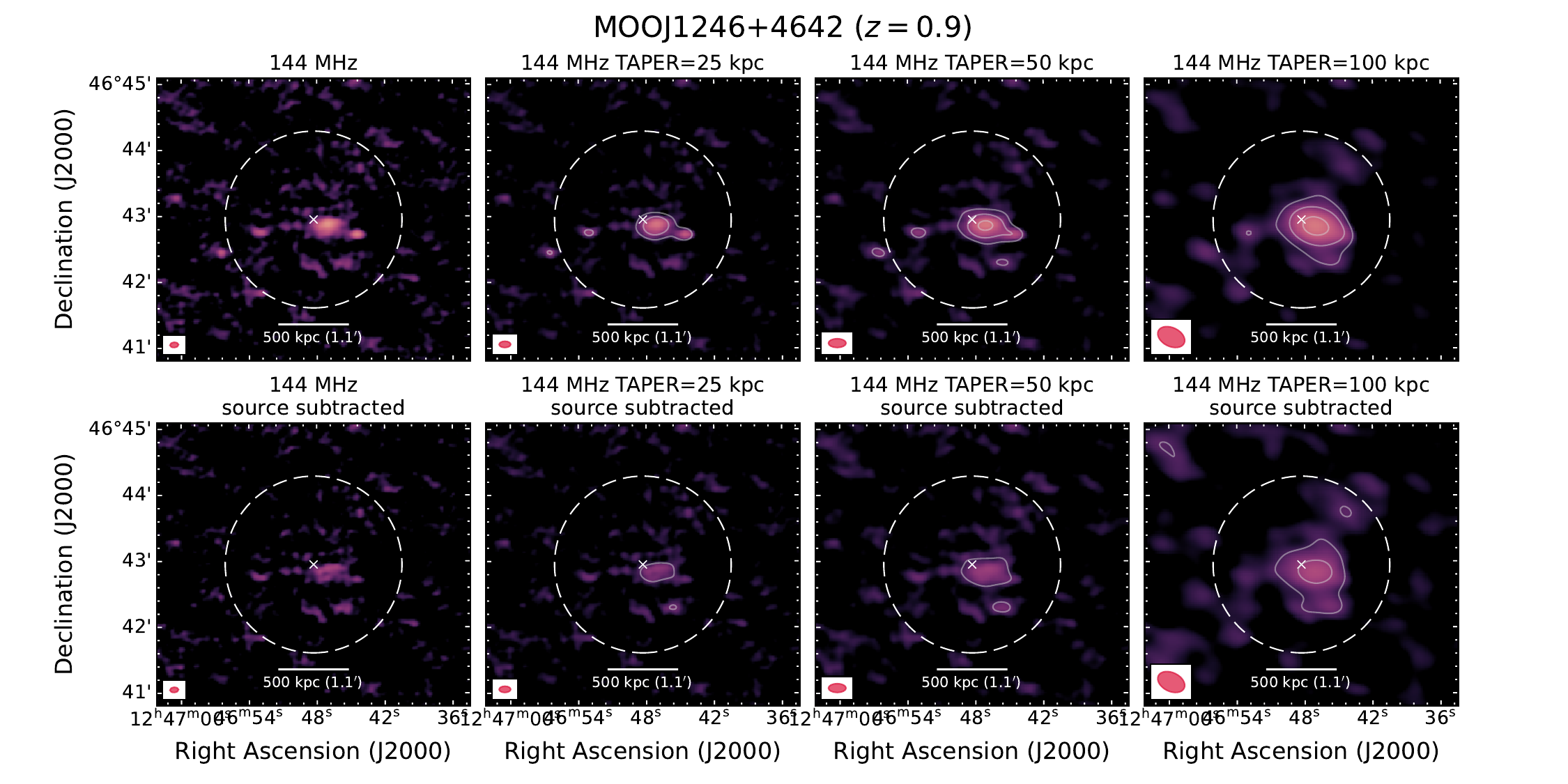}
\includegraphics[width=\textwidth]{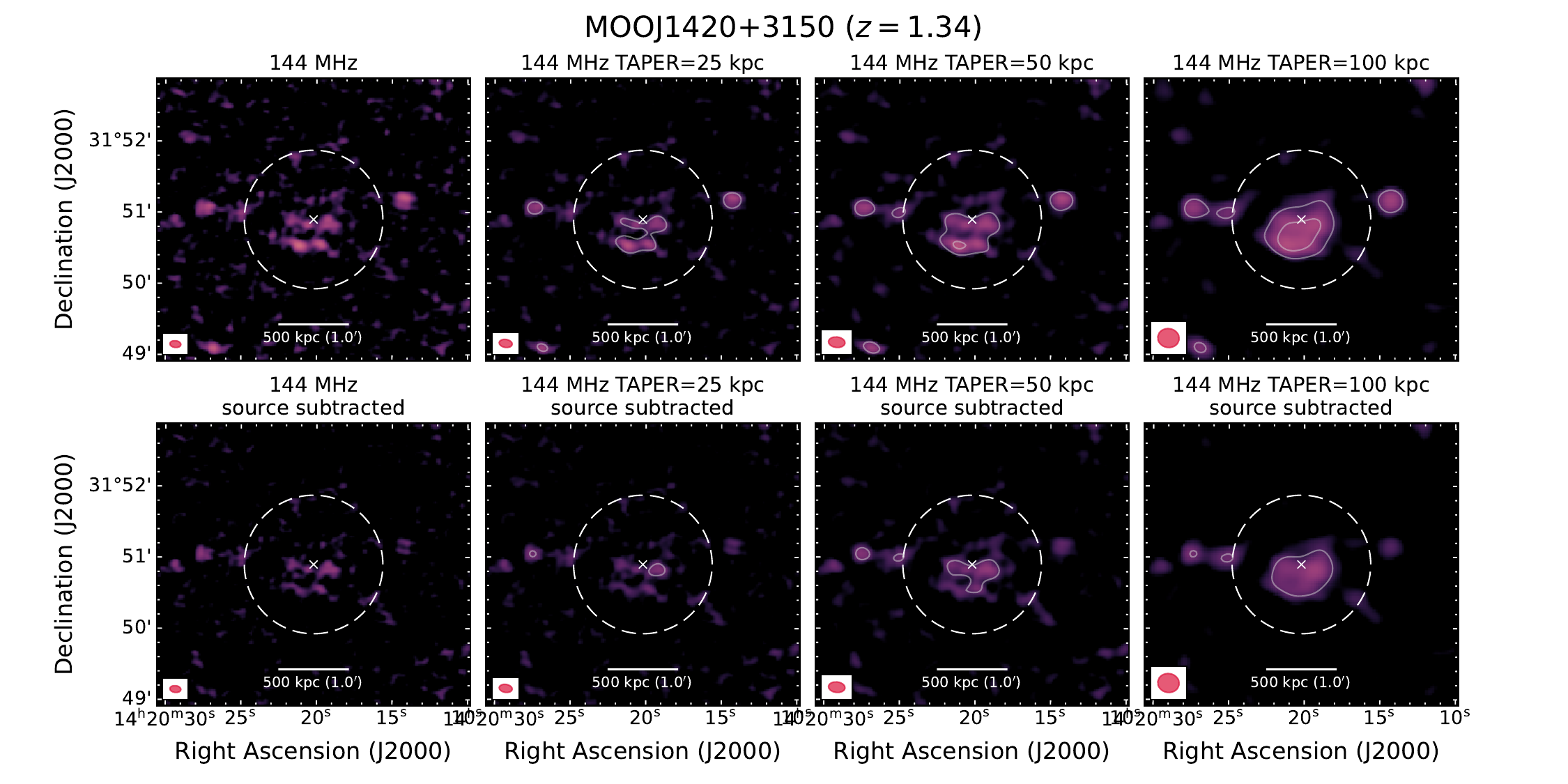}
\caption{Continued.}
\end{figure}

\begin{figure}
\centering
\ContinuedFloat
\includegraphics[width=\textwidth]{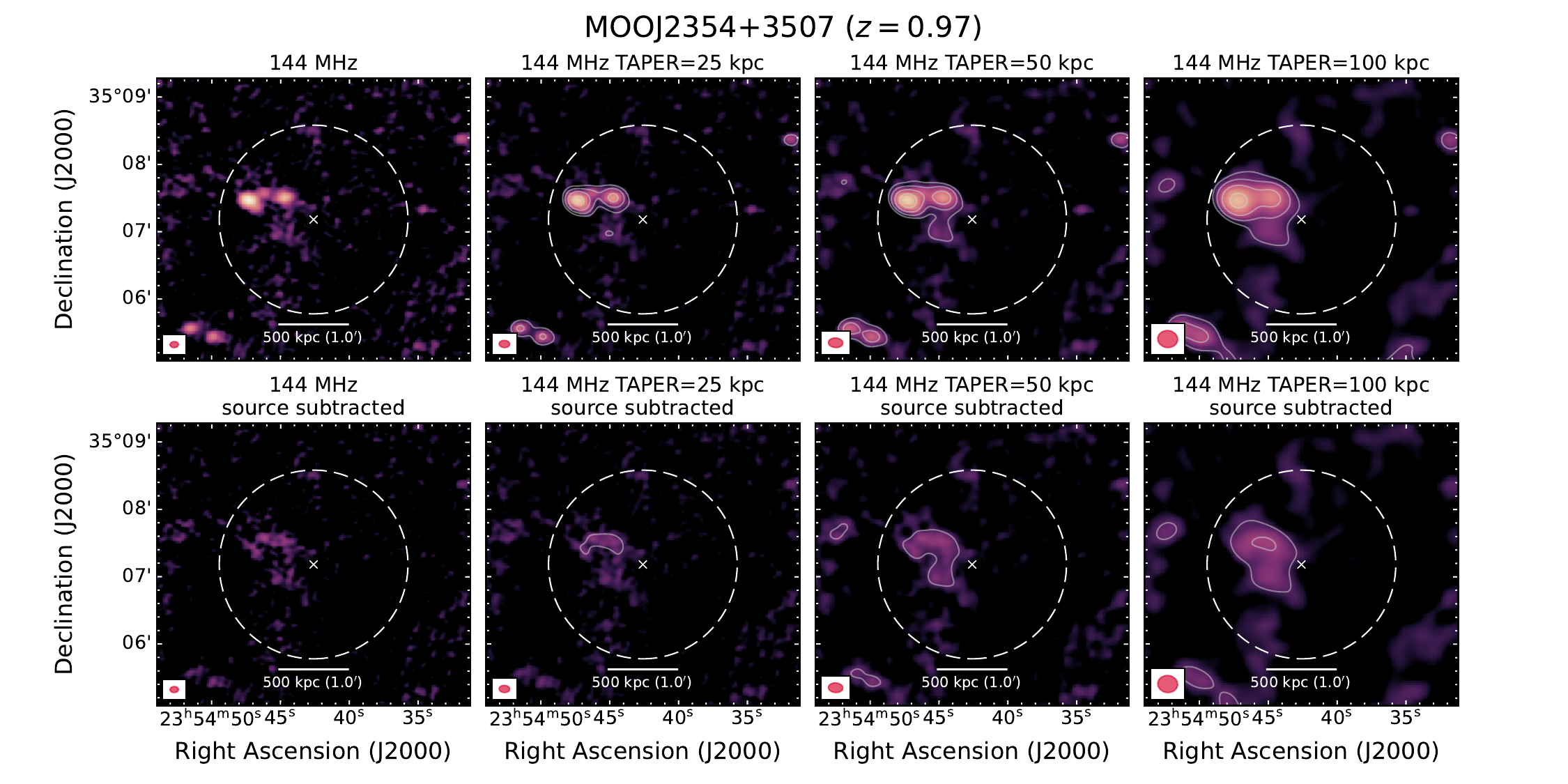}
\caption{Continued.}
\end{figure}

\clearpage

\section{Optical images}
In this section, we present the optical images of the MaDCoWS clusters in our sample (Fig. \ref{fig:optical}).

\begin{figure*}[h!]
\centering
\includegraphics[width=0.16\textwidth]{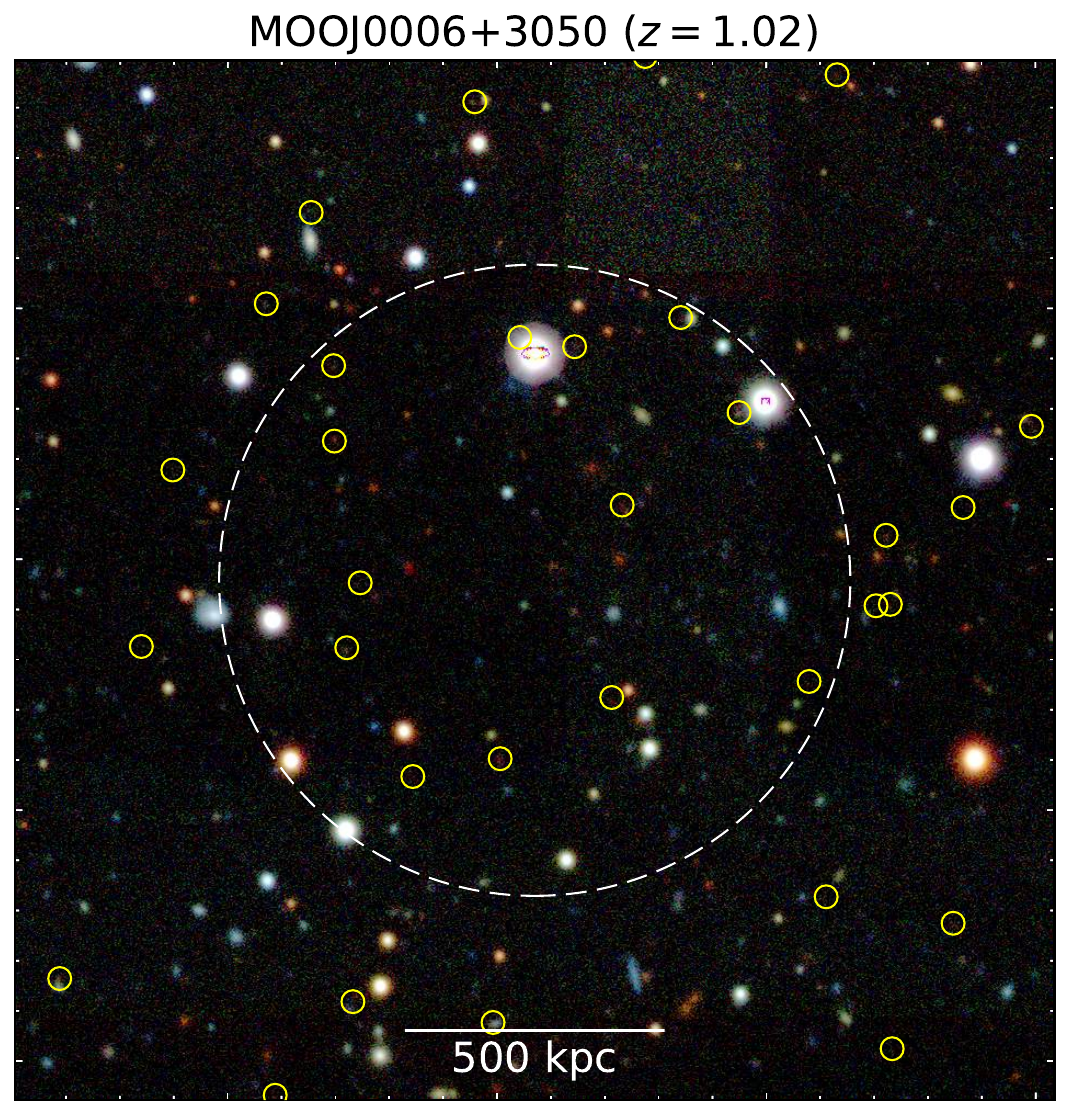}
\includegraphics[width=0.16\textwidth]{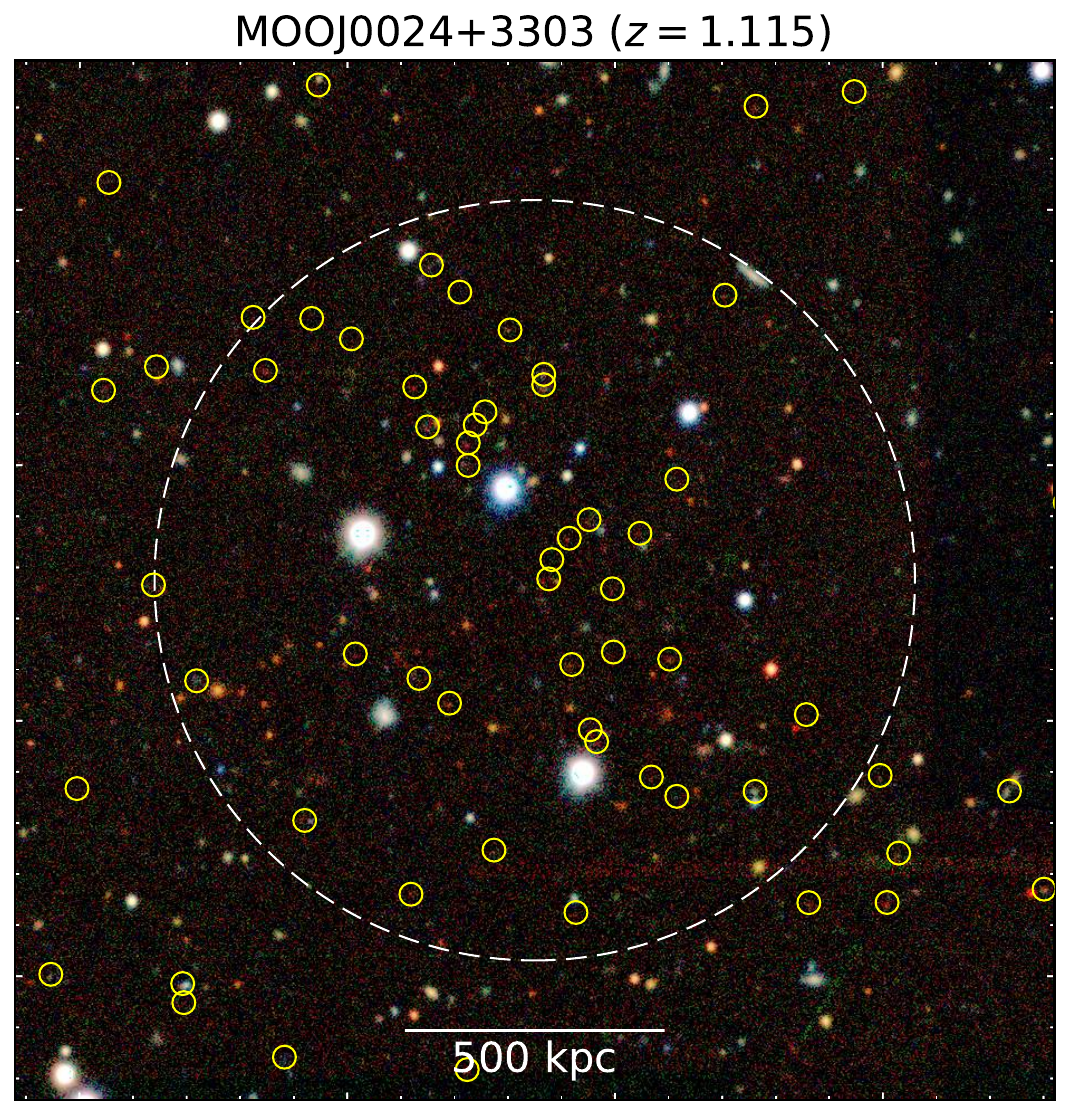}
\includegraphics[width=0.16\textwidth]{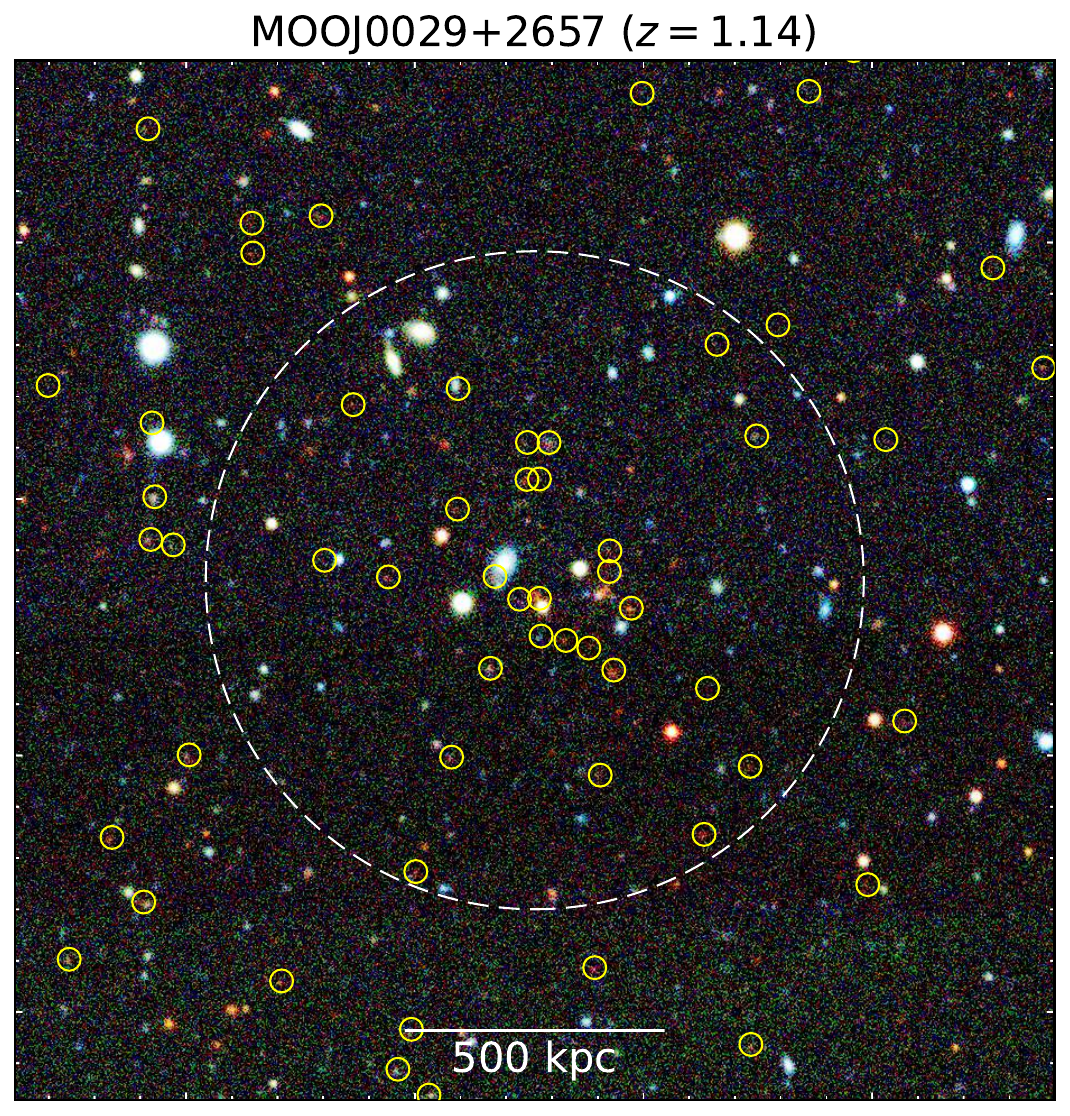}
\includegraphics[width=0.16\textwidth]{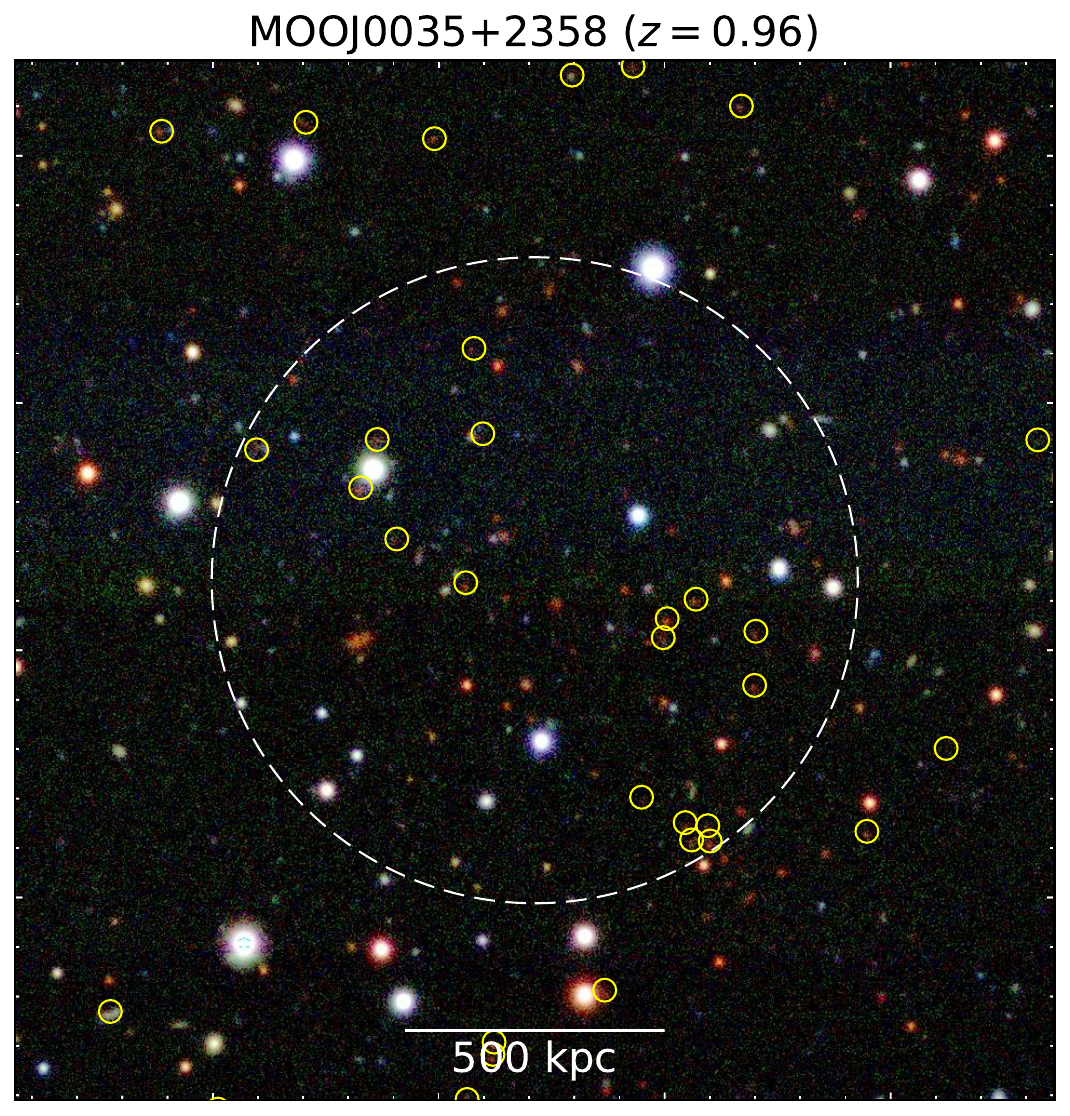}
\includegraphics[width=0.16\textwidth]{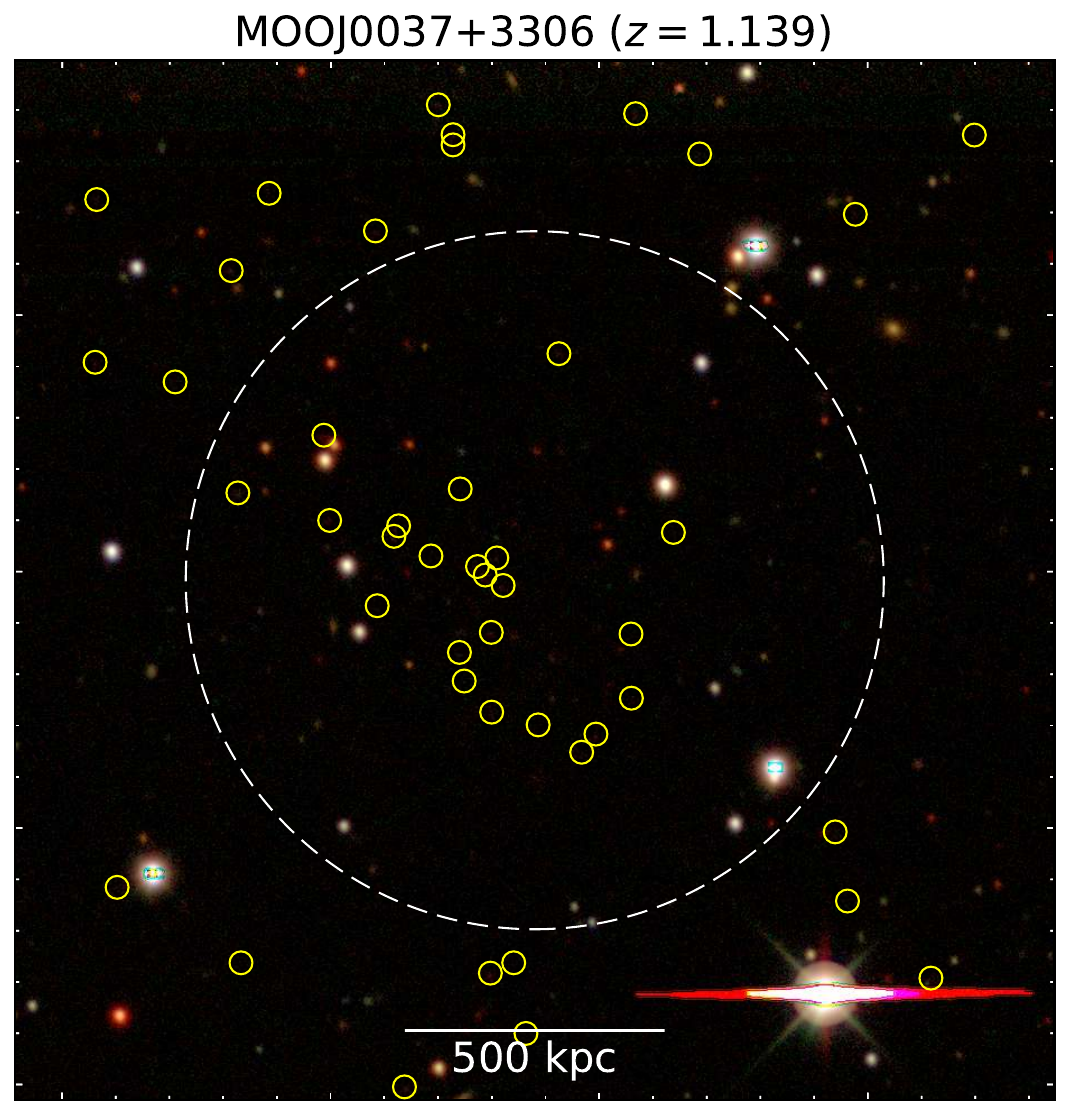}
\includegraphics[width=0.16\textwidth]{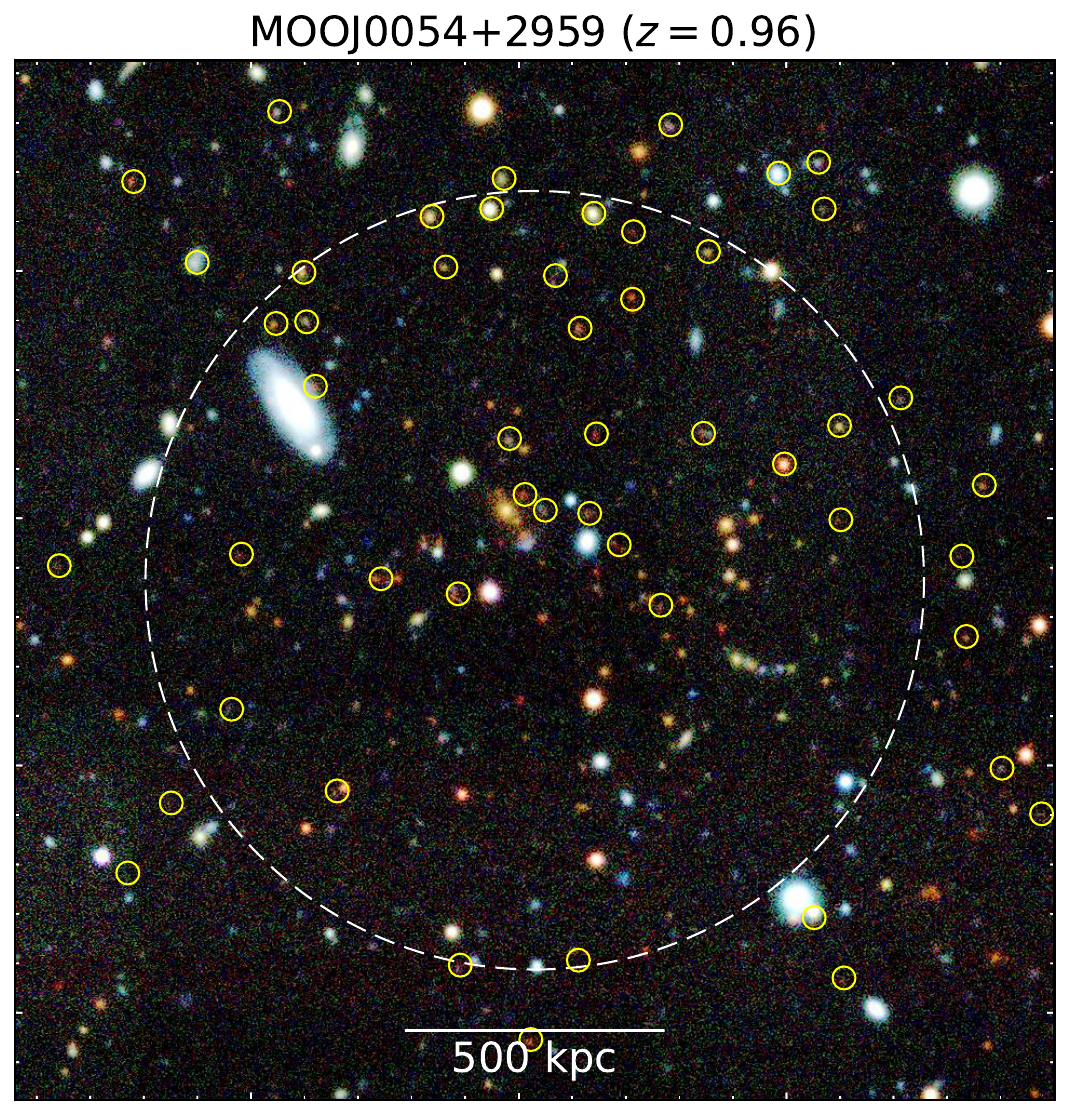}
\includegraphics[width=0.16\textwidth]{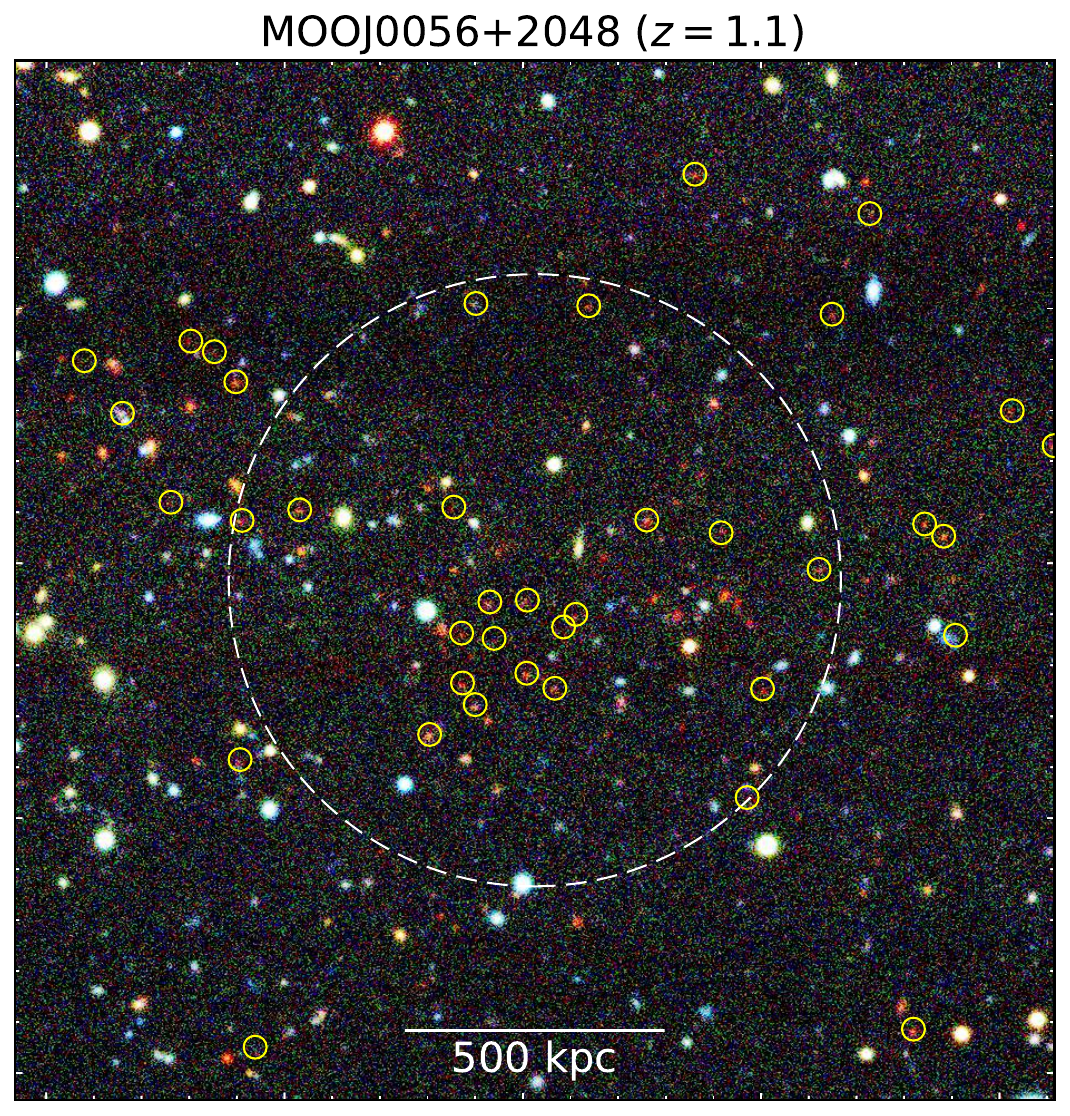}
\includegraphics[width=0.16\textwidth]{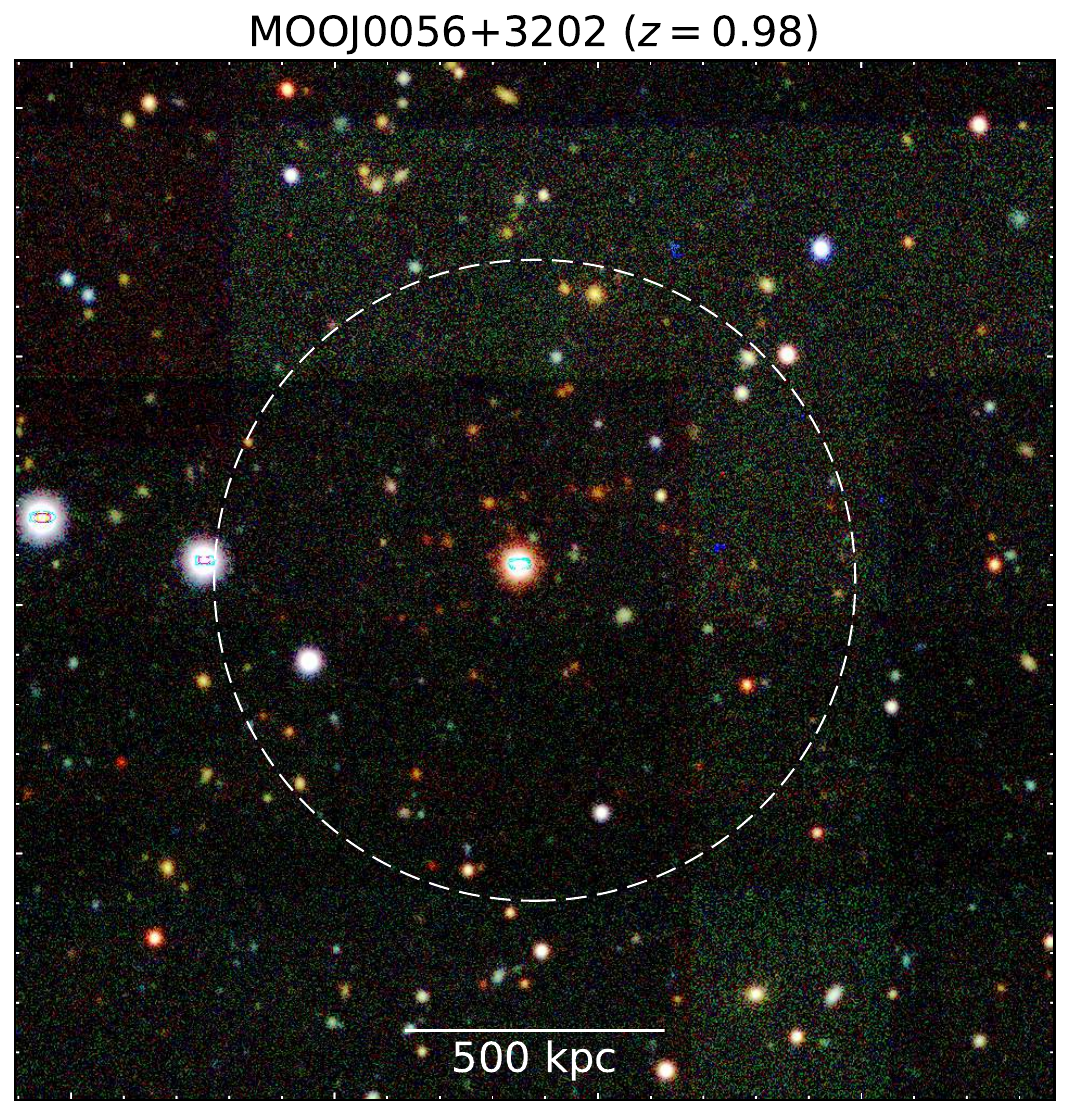}
\includegraphics[width=0.16\textwidth]{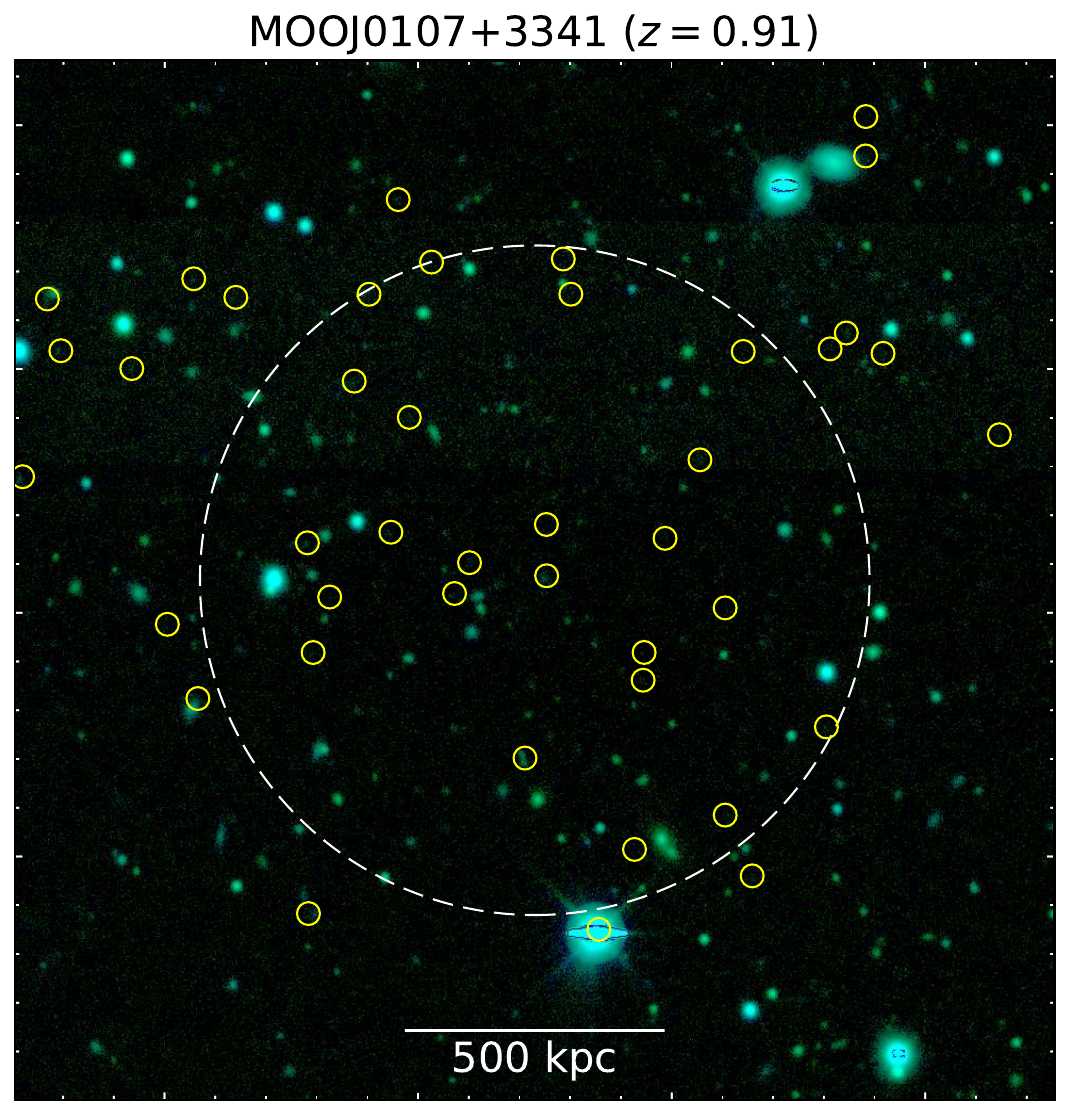}
\includegraphics[width=0.16\textwidth]{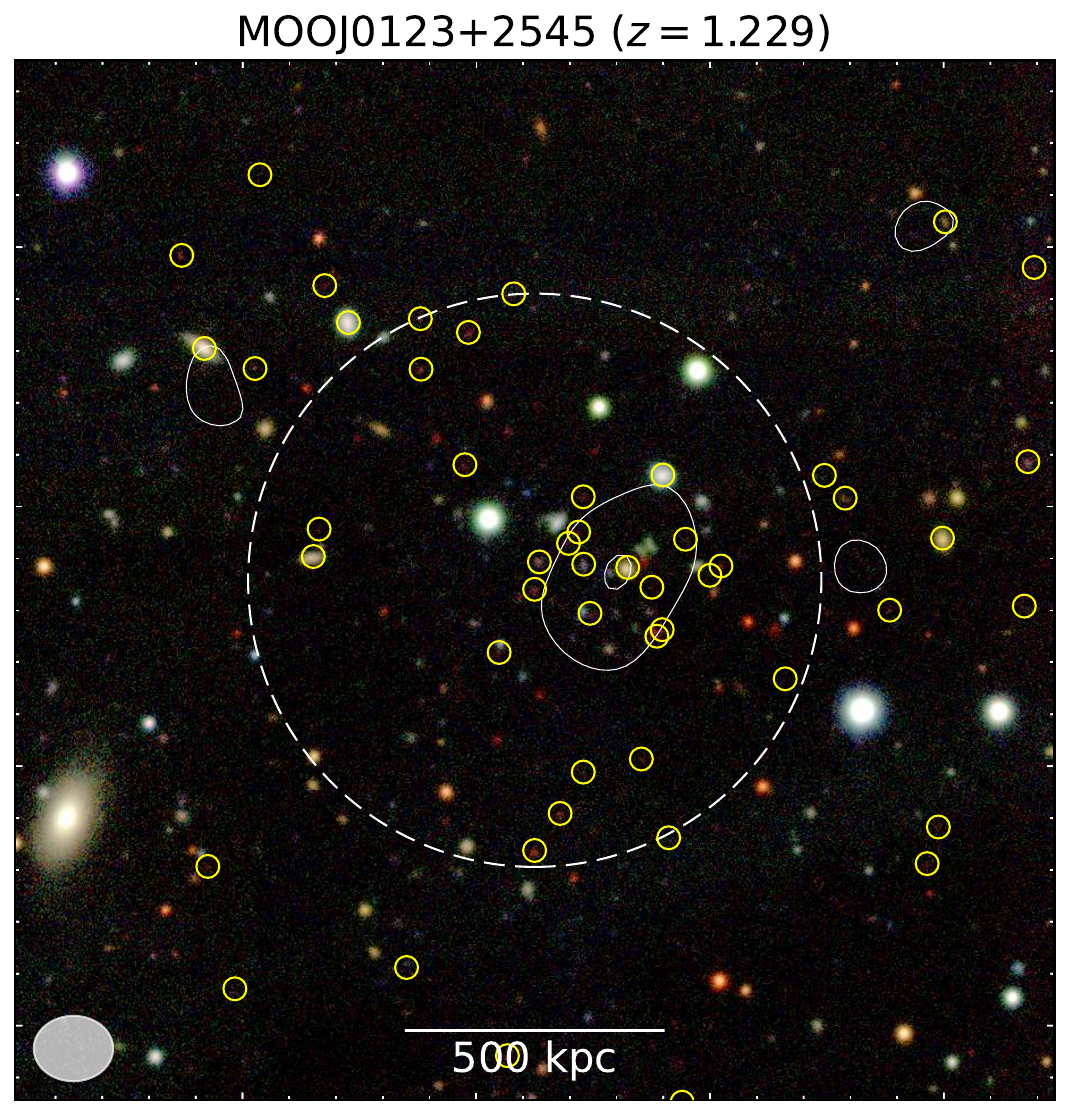}
\includegraphics[width=0.16\textwidth]{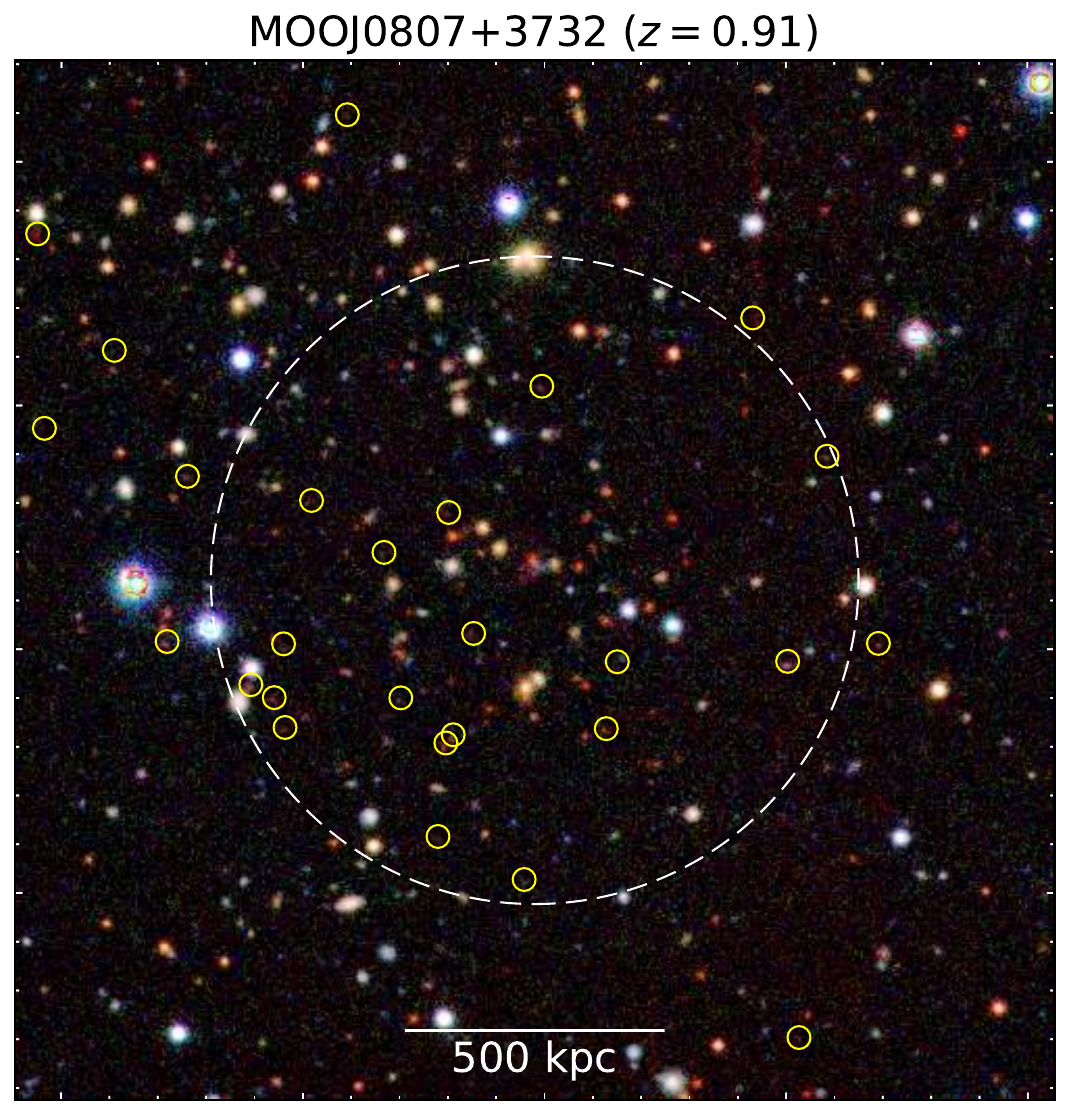}
\includegraphics[width=0.16\textwidth]{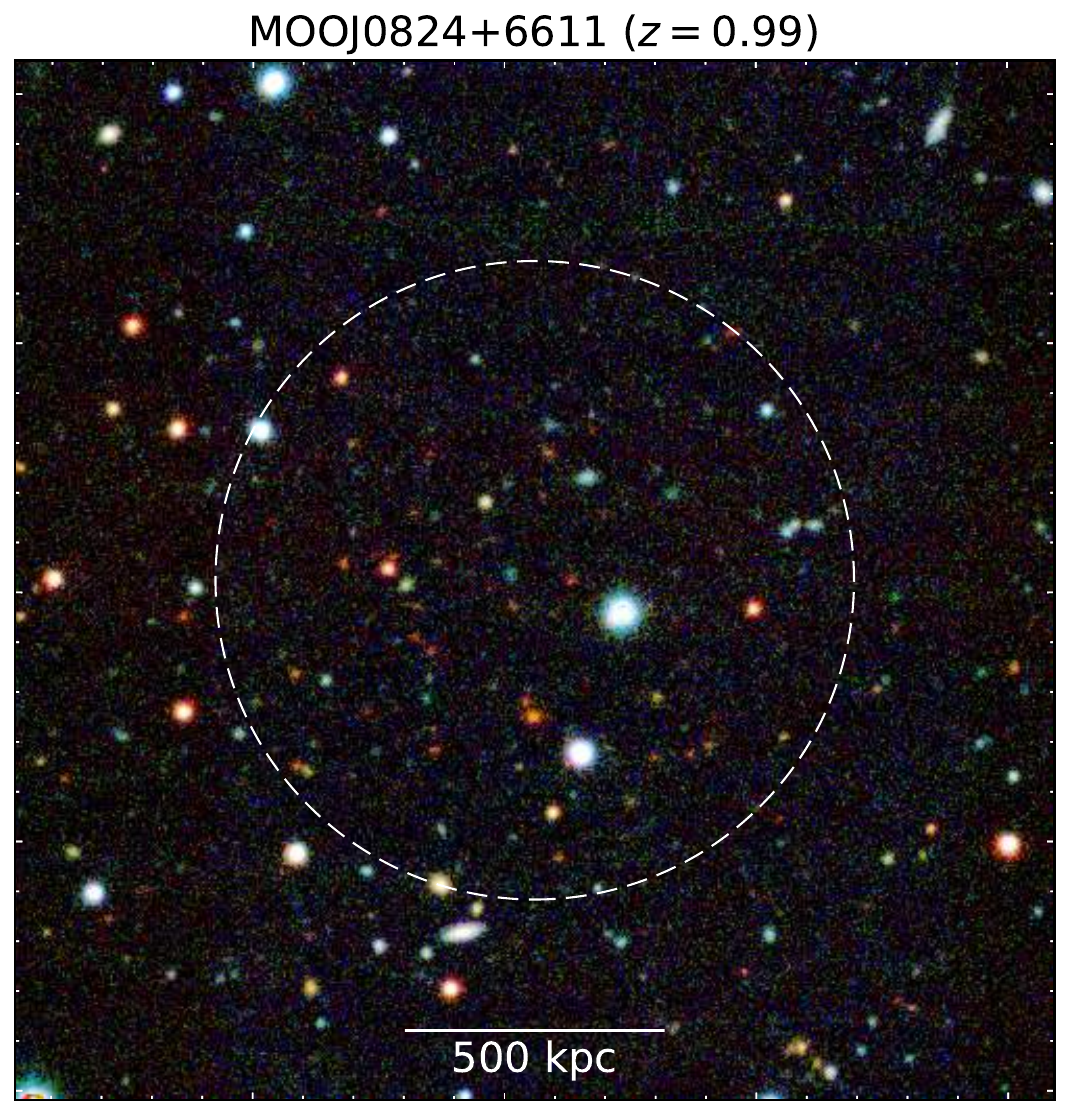}
\includegraphics[width=0.16\textwidth]{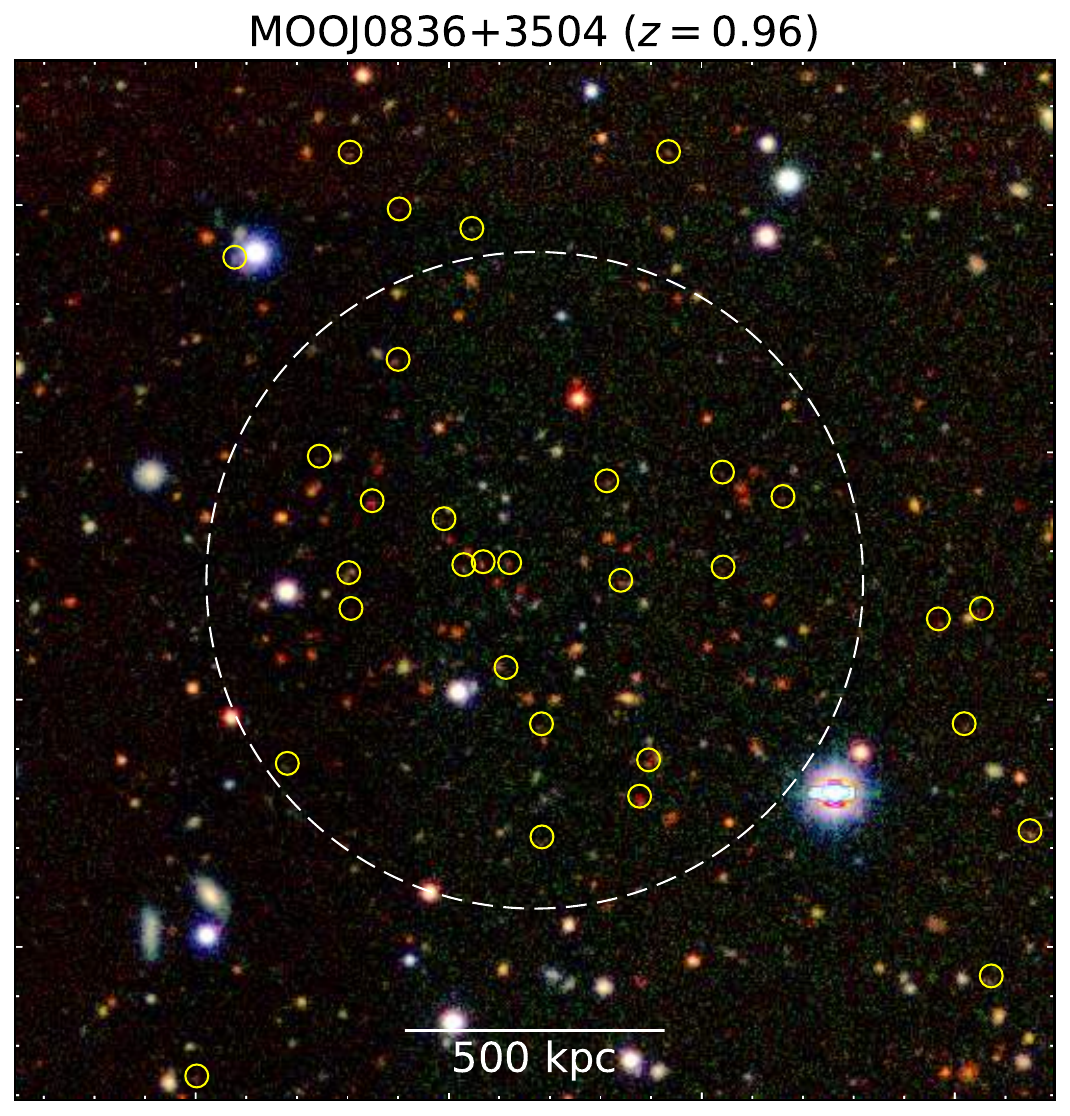}
\includegraphics[width=0.16\textwidth]{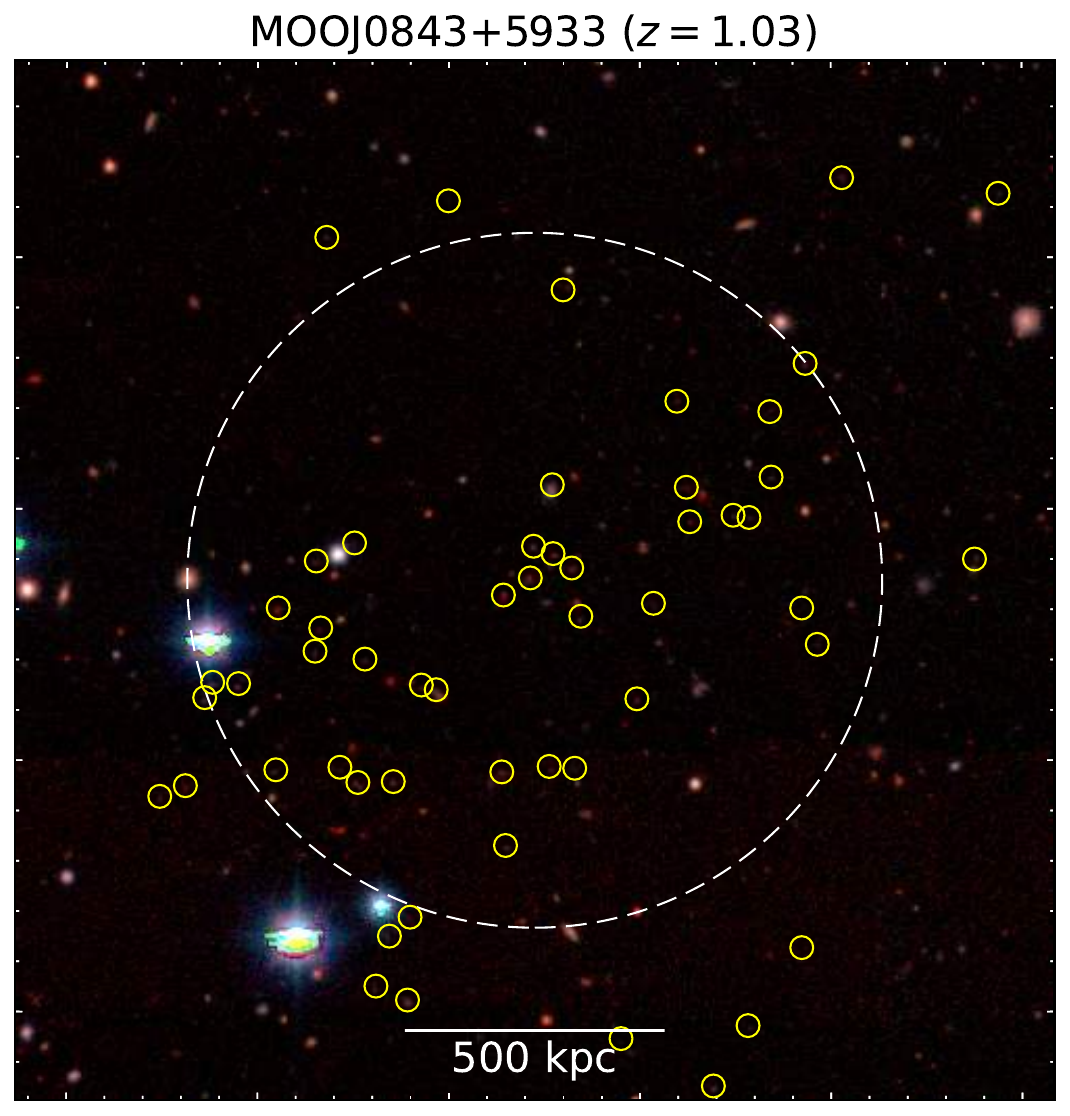}
\includegraphics[width=0.16\textwidth]{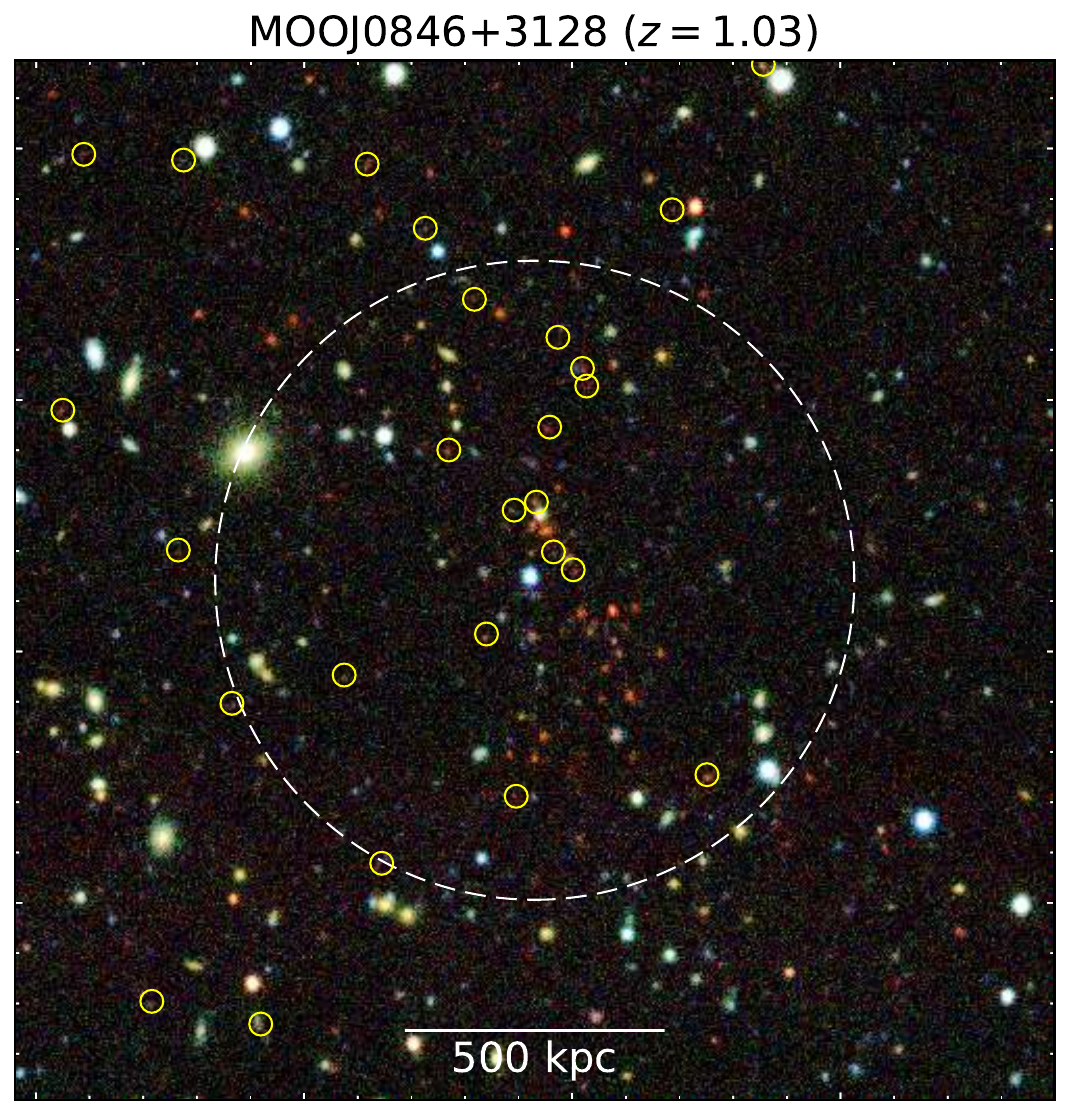}
\includegraphics[width=0.16\textwidth]{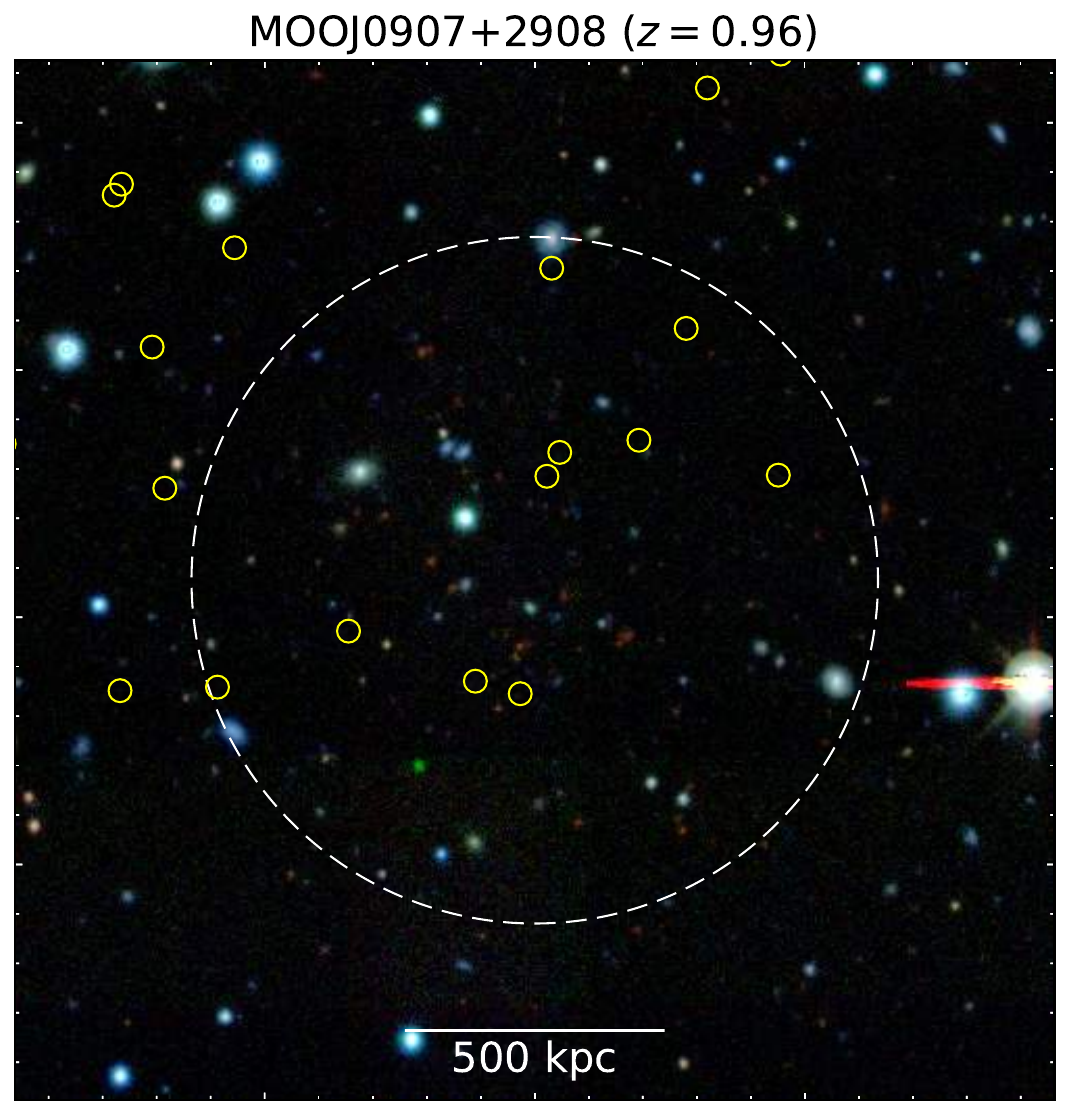}
\includegraphics[width=0.16\textwidth]{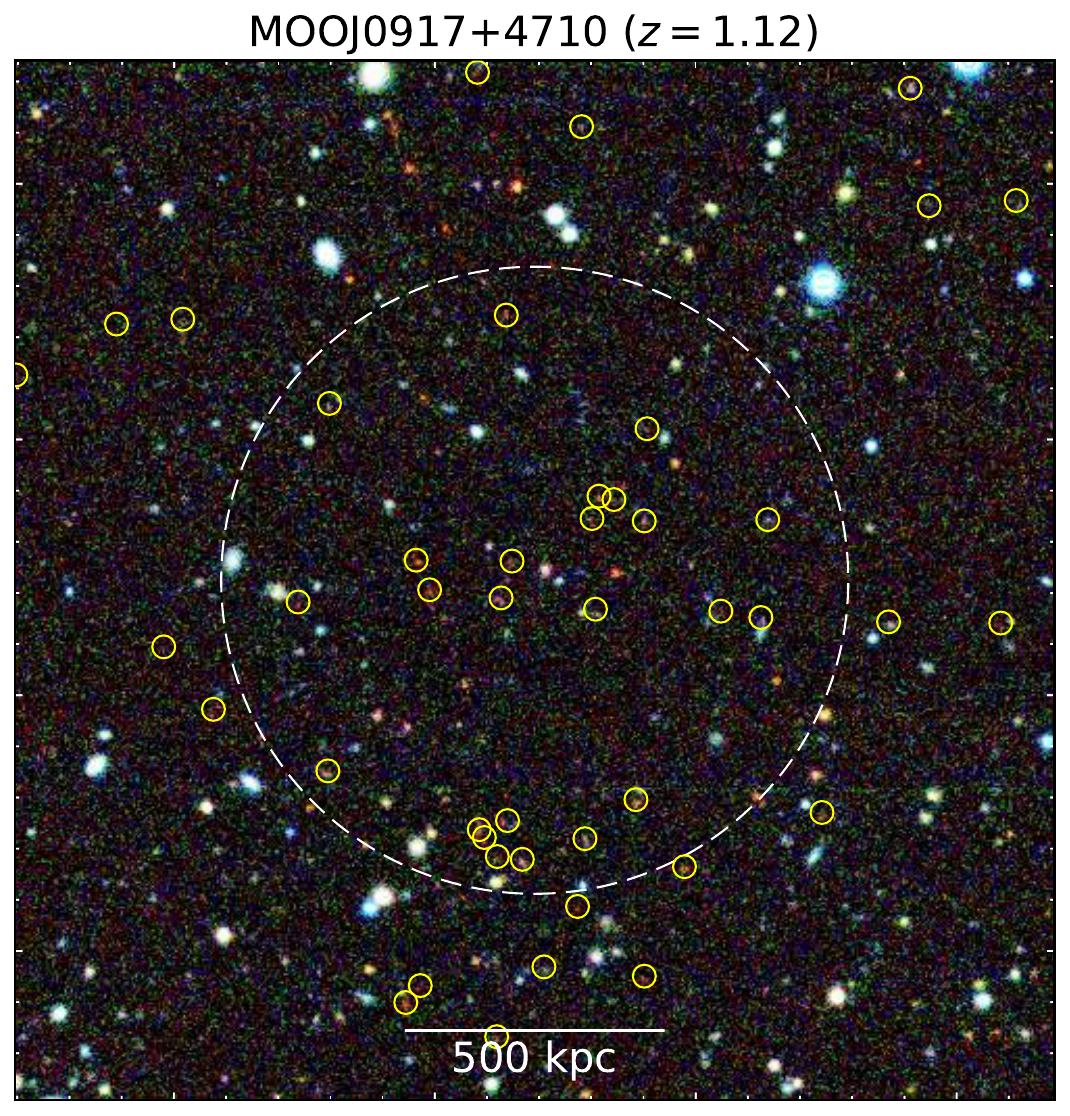}
\includegraphics[width=0.16\textwidth]{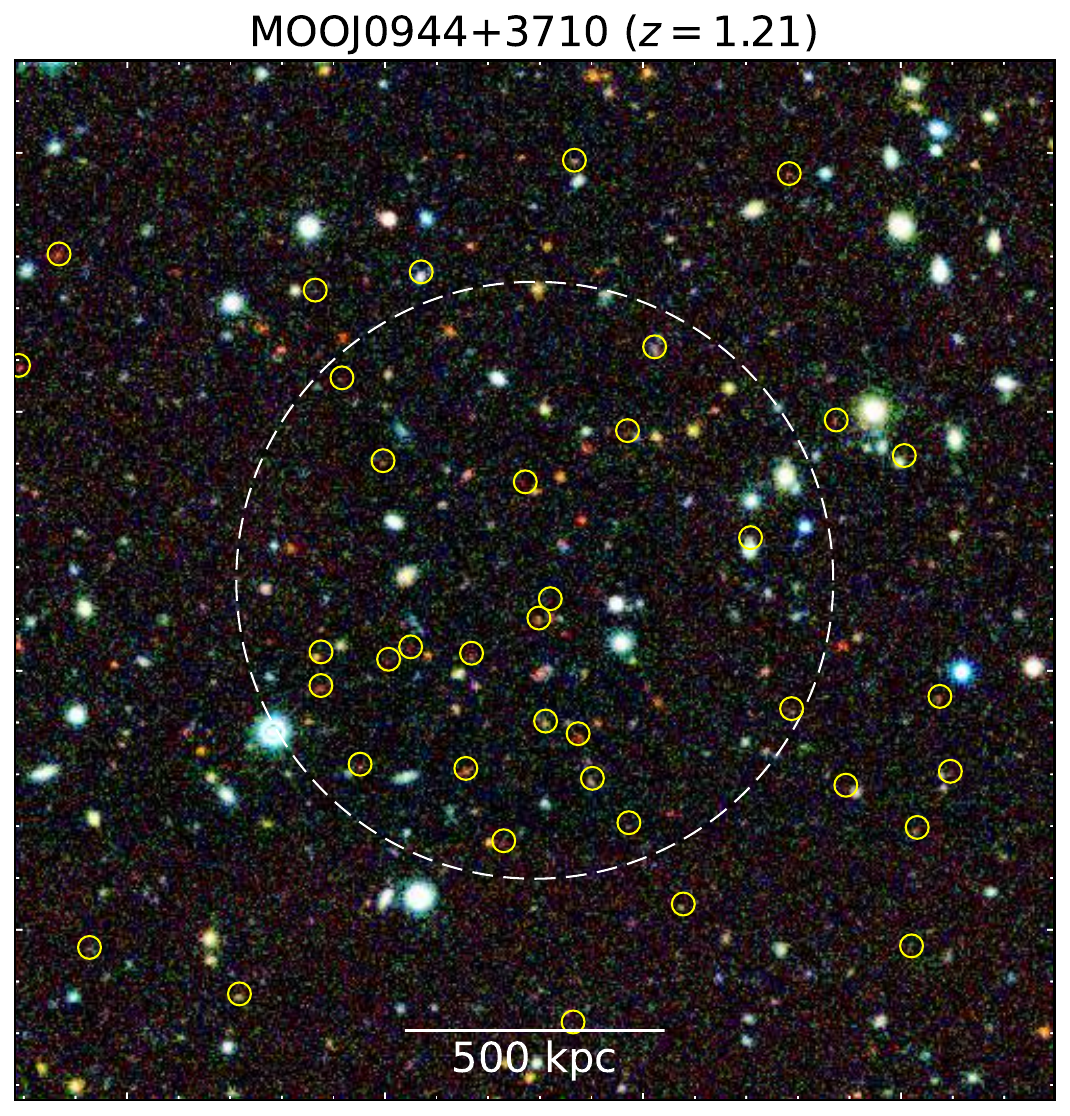}
\includegraphics[width=0.16\textwidth]{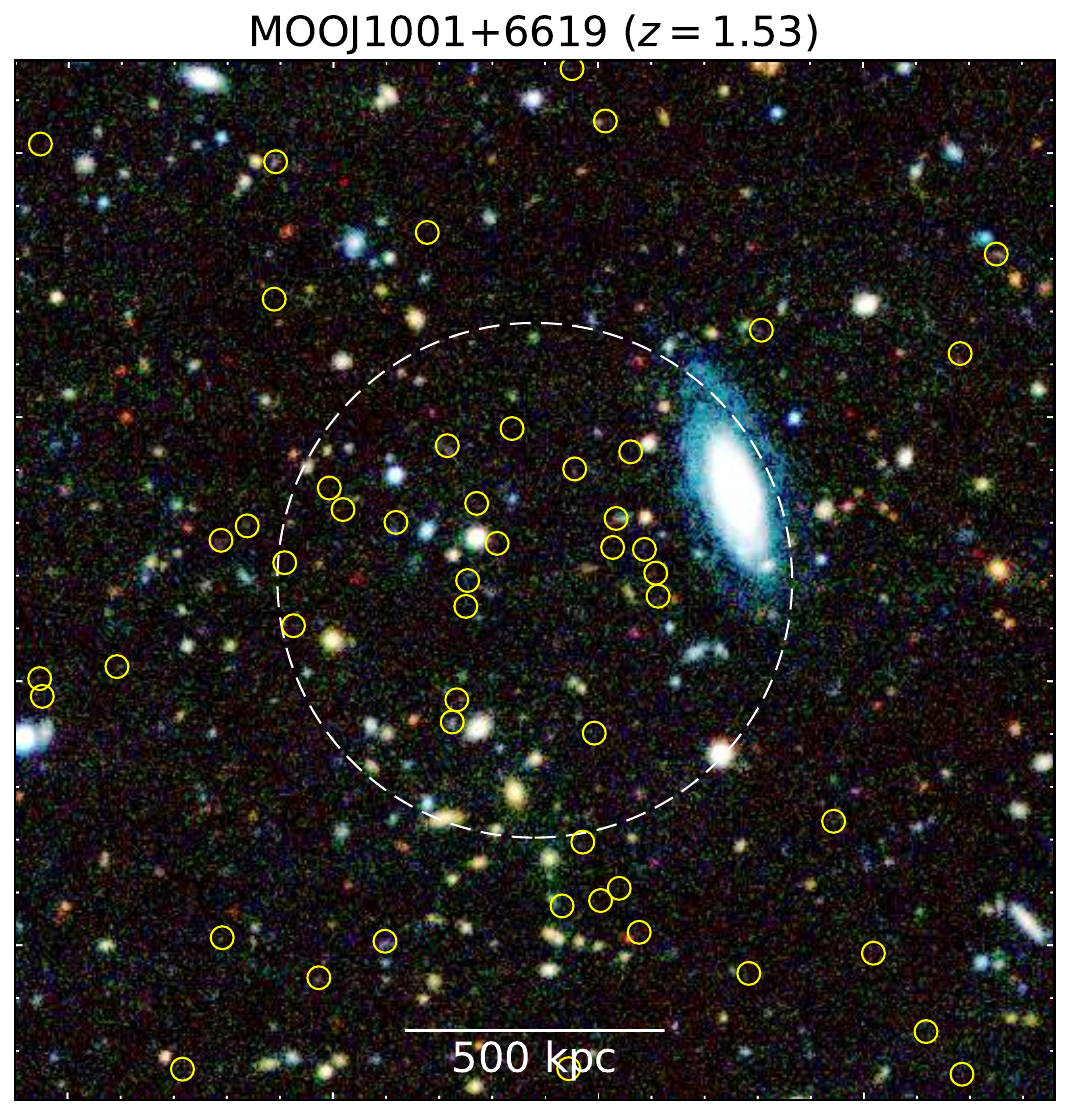}
\includegraphics[width=0.16\textwidth]{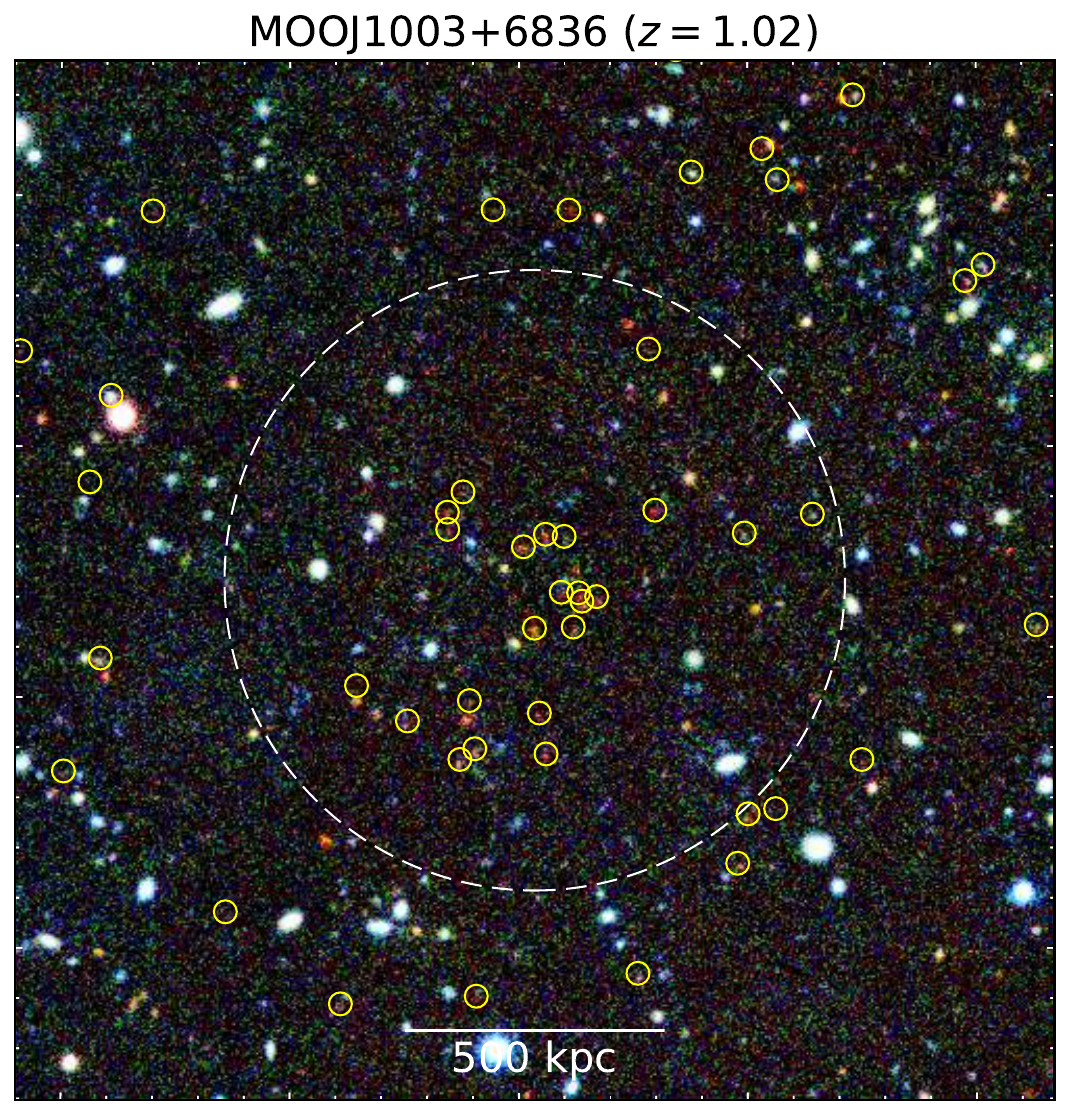}
\includegraphics[width=0.16\textwidth]{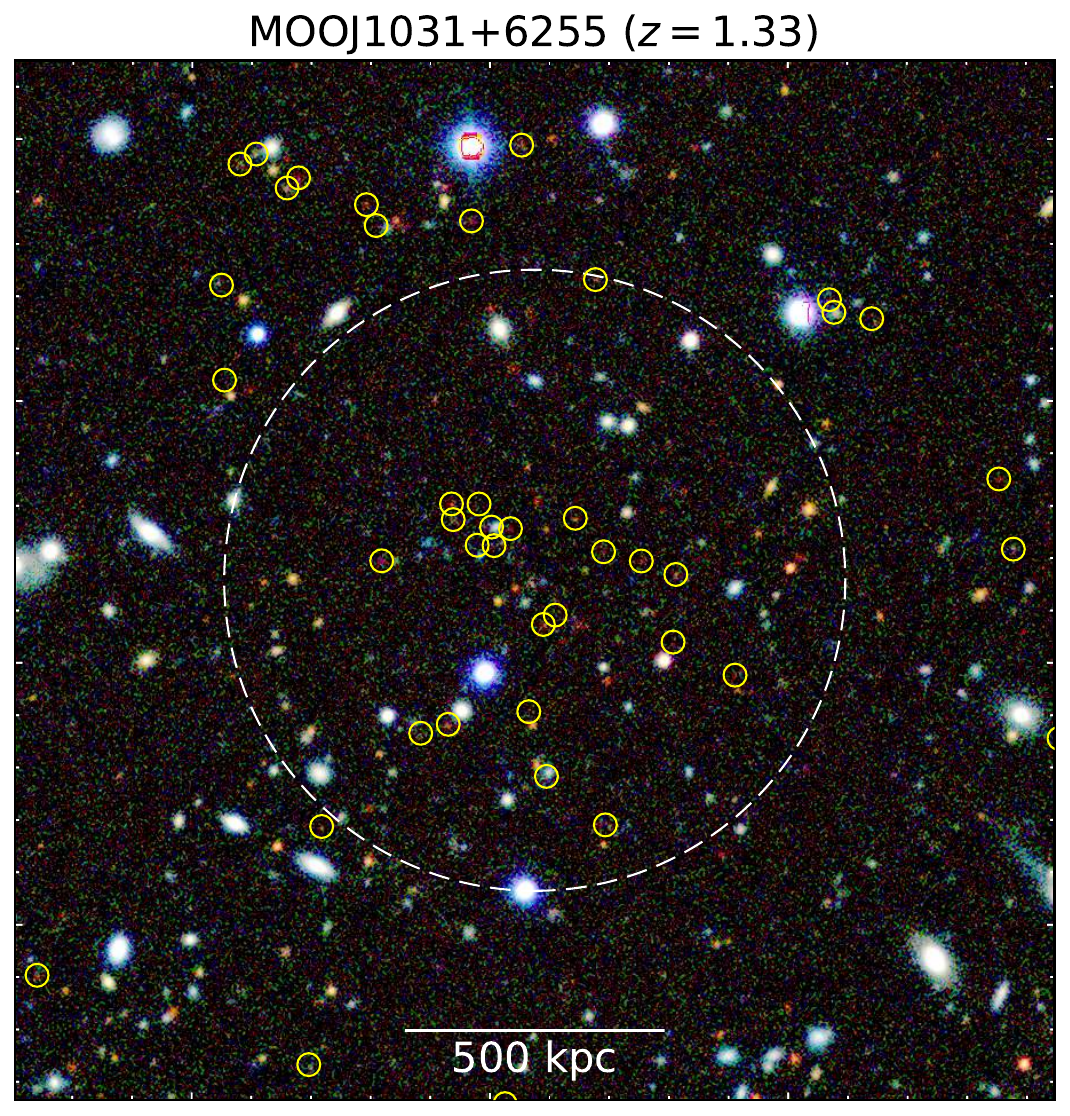}
\includegraphics[width=0.16\textwidth]{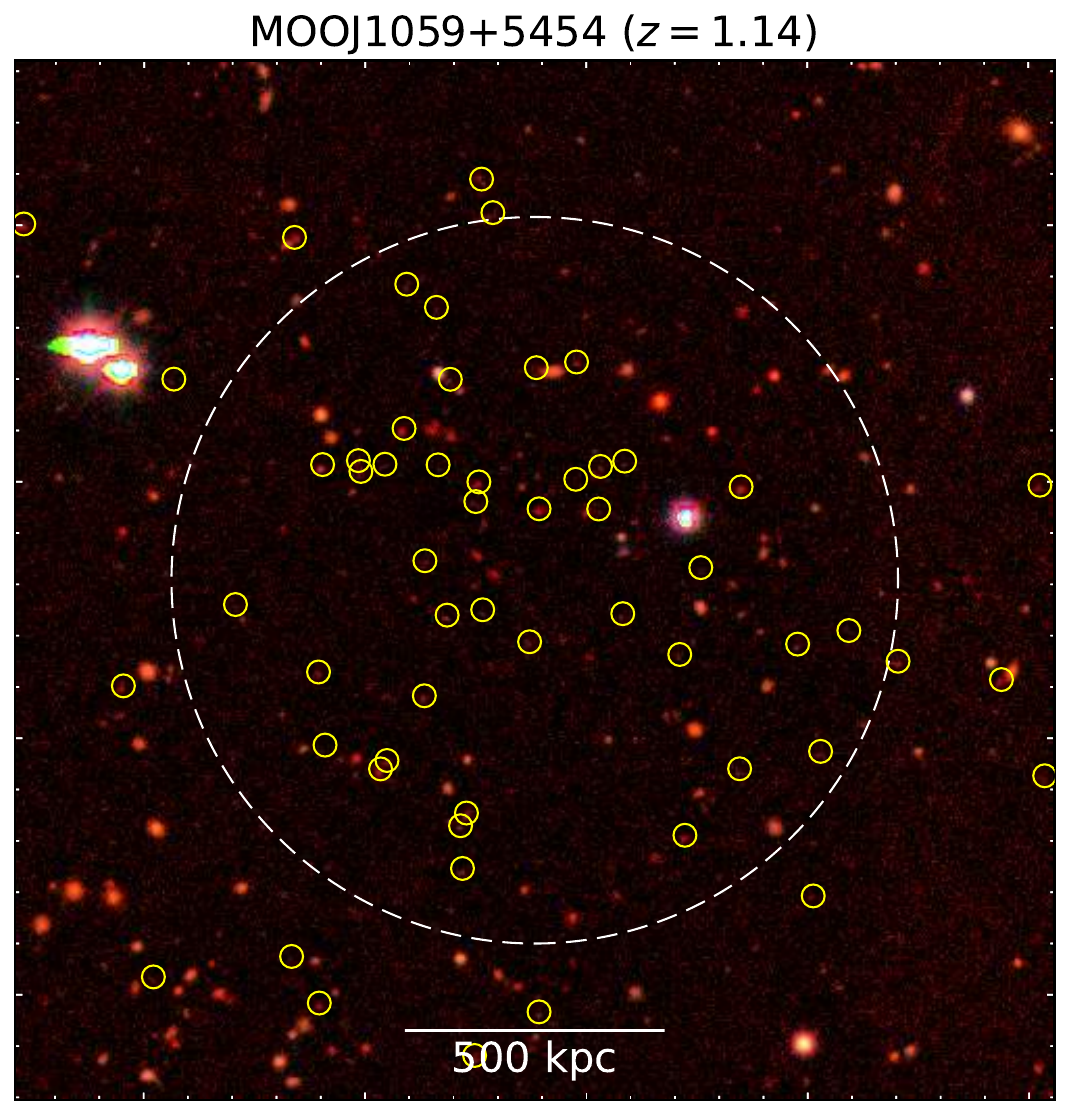}
\includegraphics[width=0.16\textwidth]{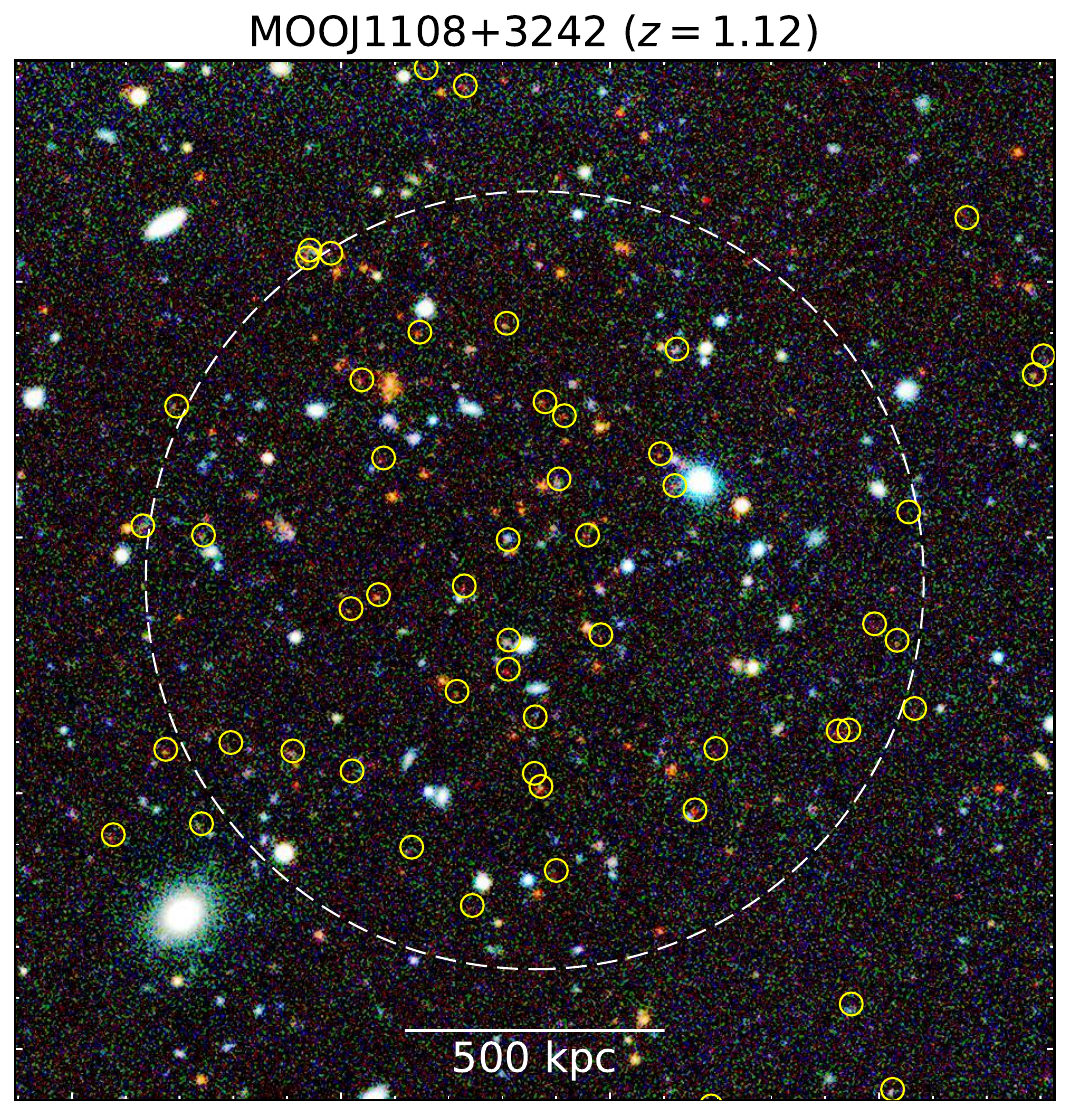}
\includegraphics[width=0.16\textwidth]{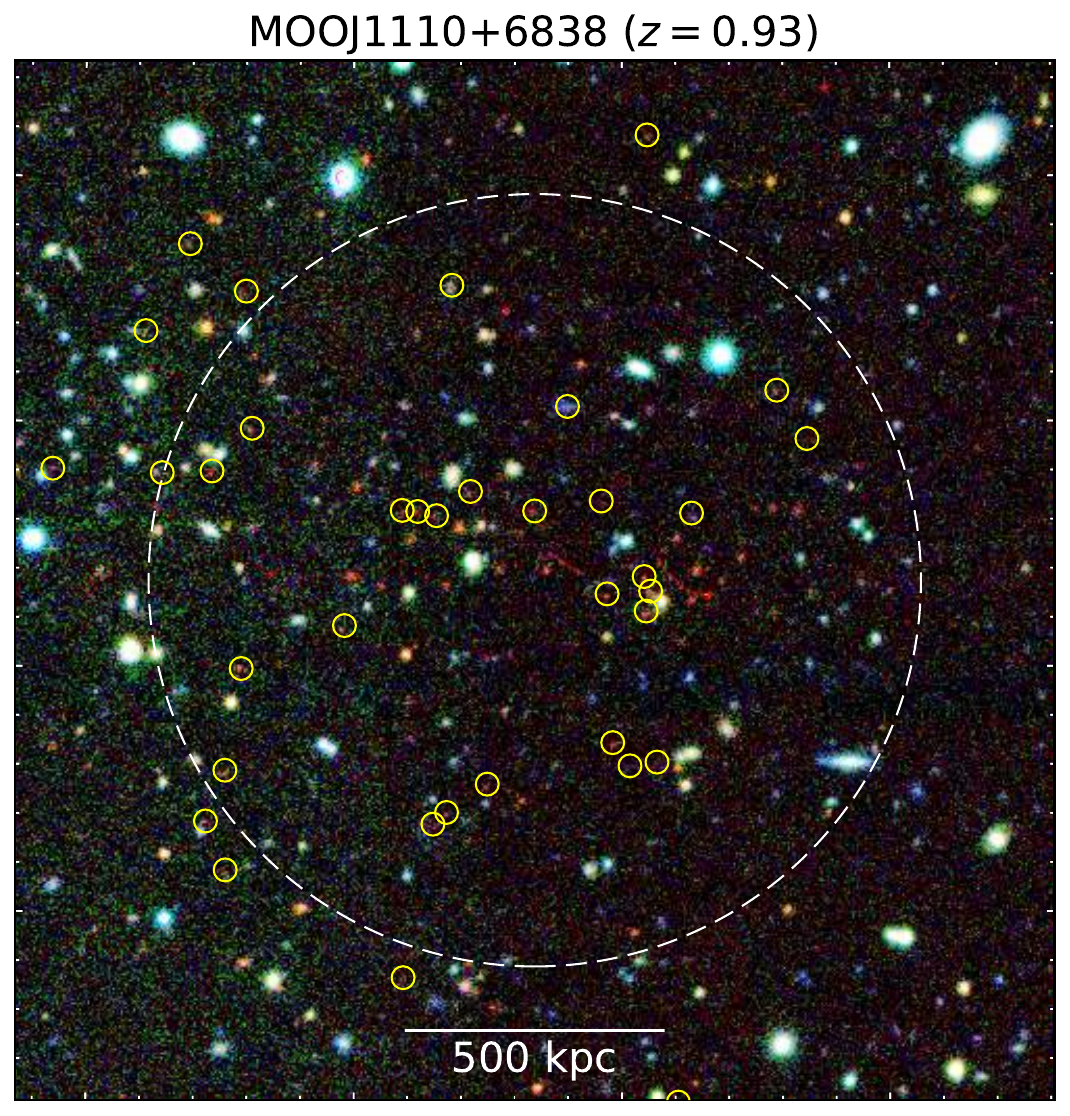}
\includegraphics[width=0.16\textwidth]{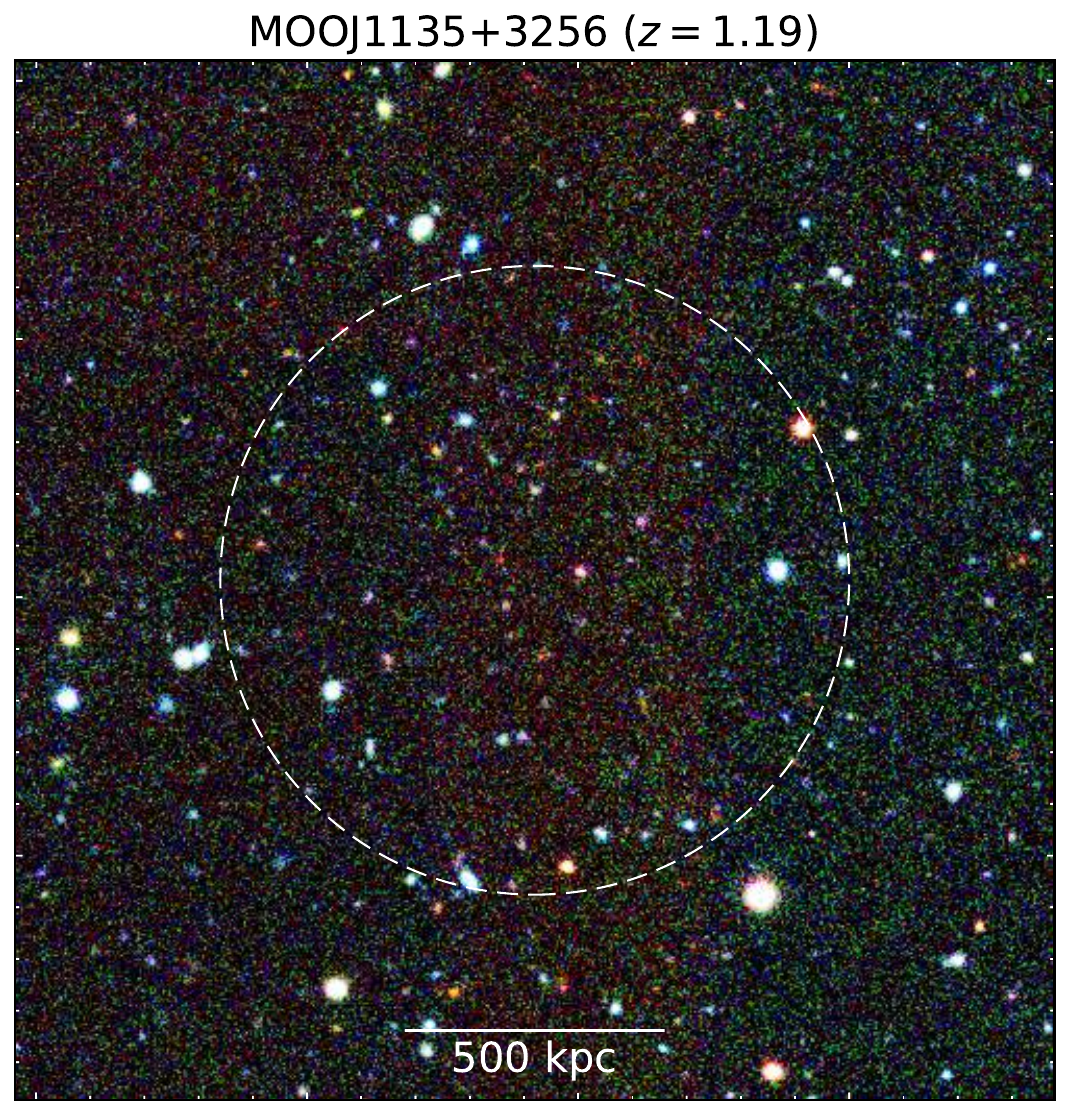}
\includegraphics[width=0.16\textwidth]{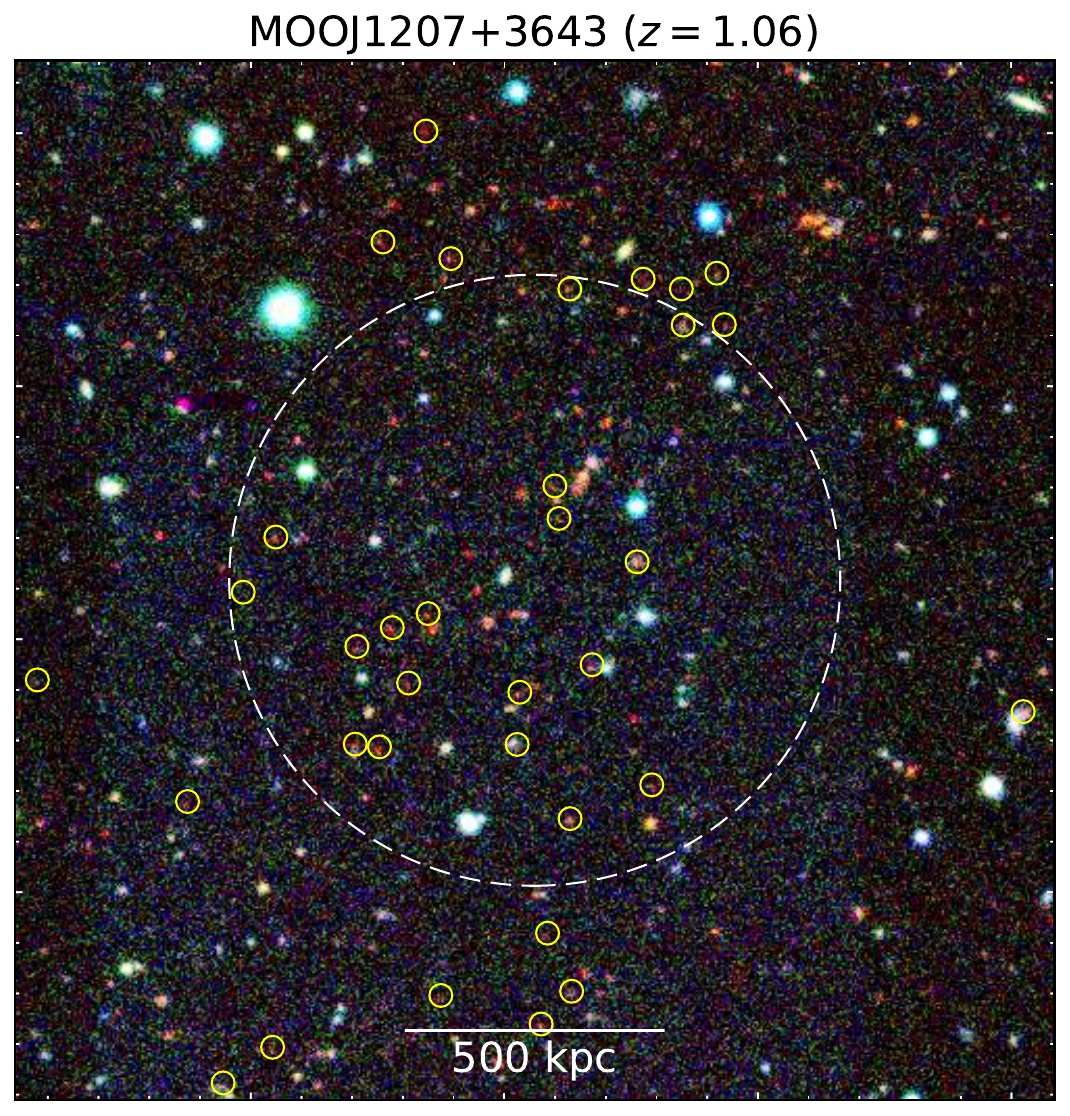}
\includegraphics[width=0.16\textwidth]{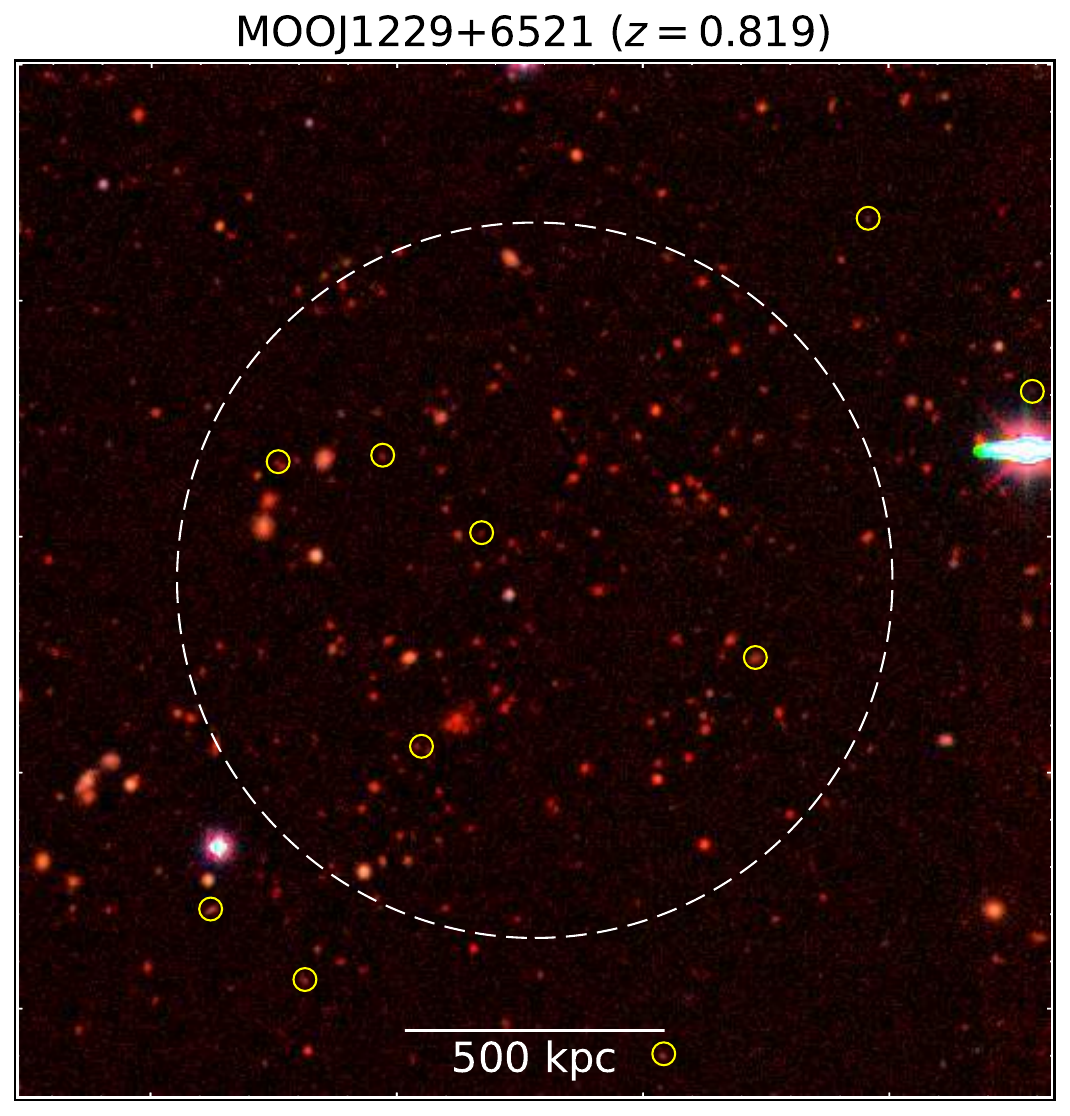}
\includegraphics[width=0.16\textwidth]{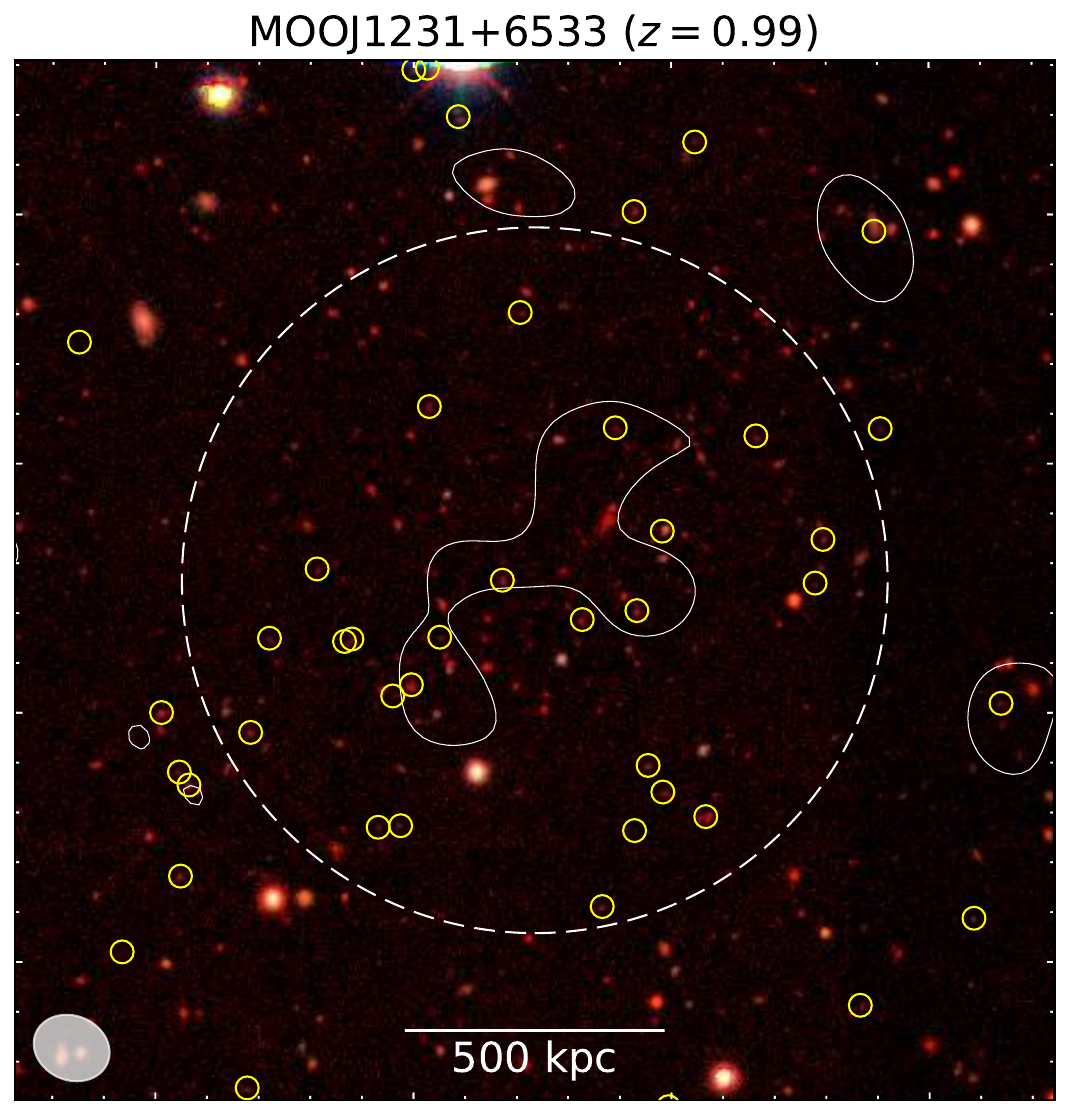}
\includegraphics[width=0.16\textwidth]{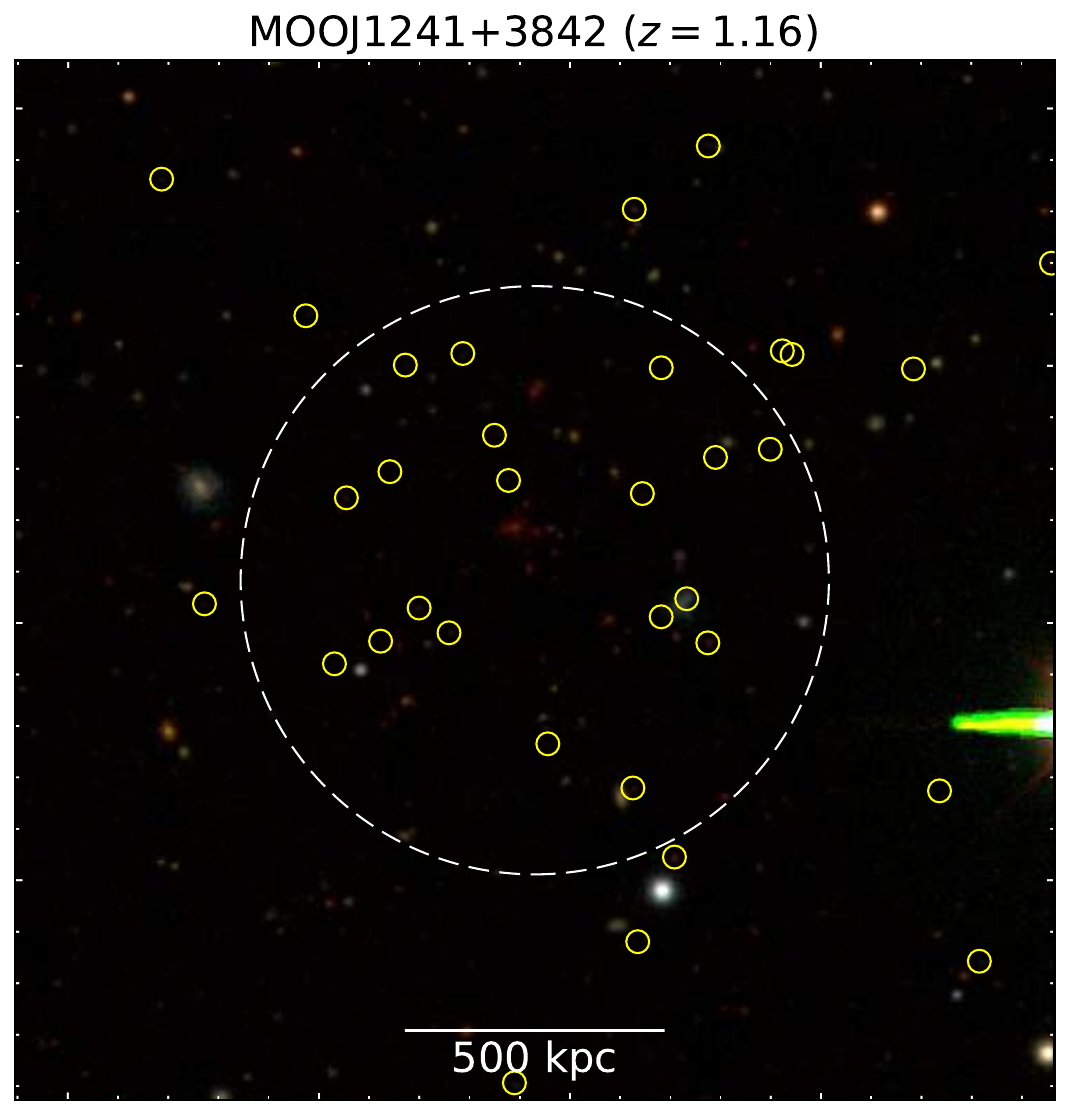}
\includegraphics[width=0.16\textwidth]{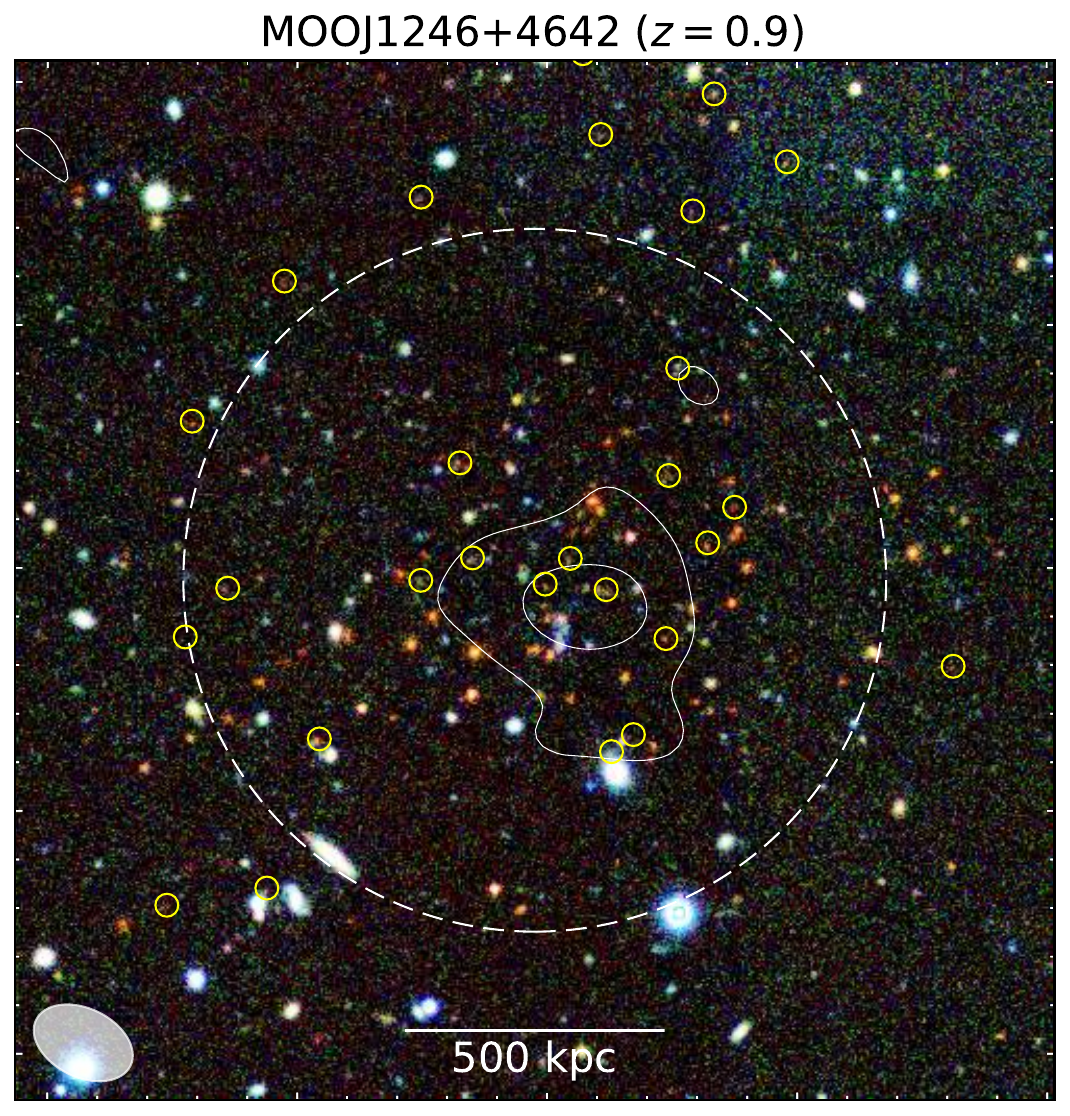}
\includegraphics[width=0.16\textwidth]{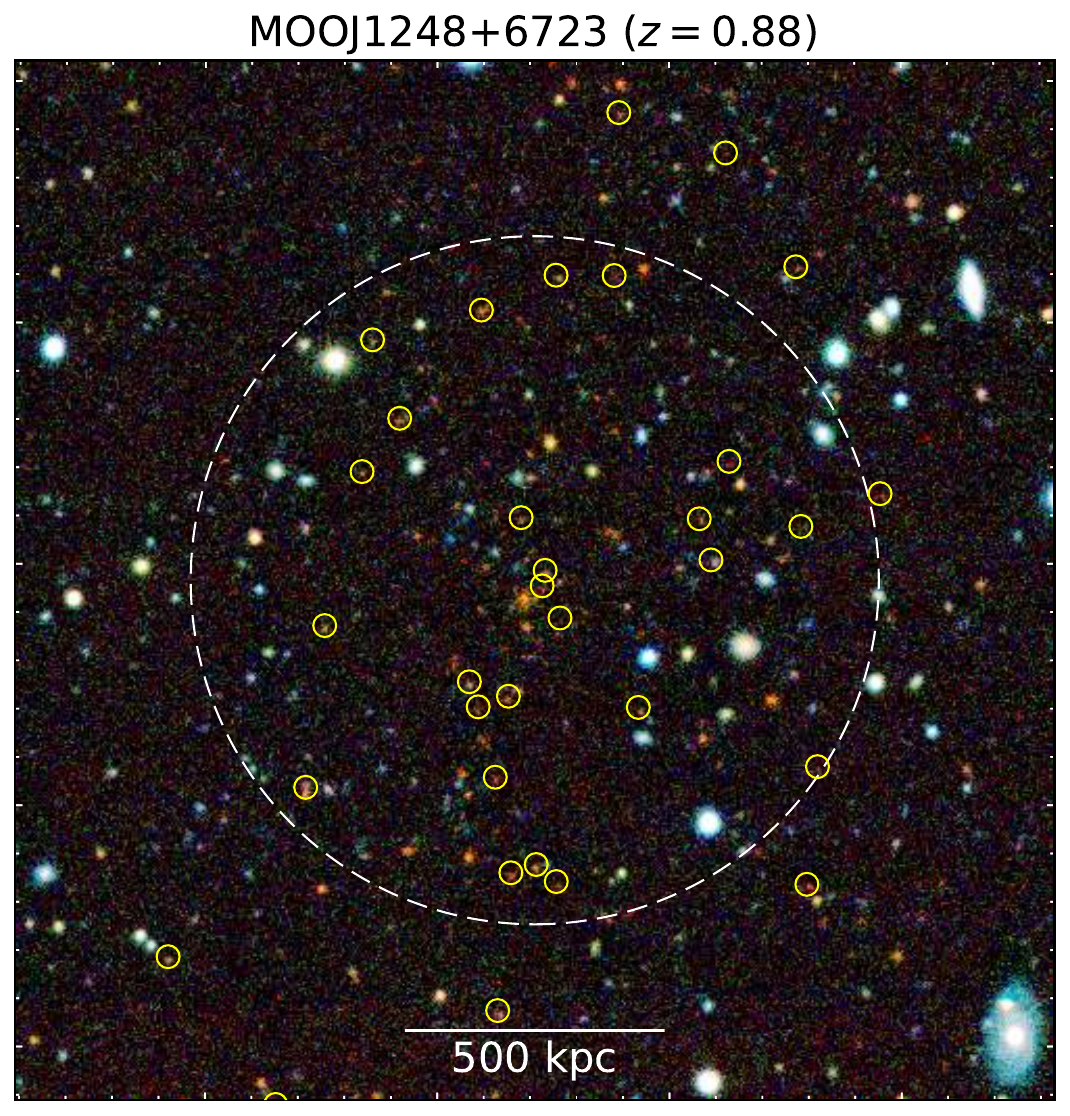}
\includegraphics[width=0.16\textwidth]{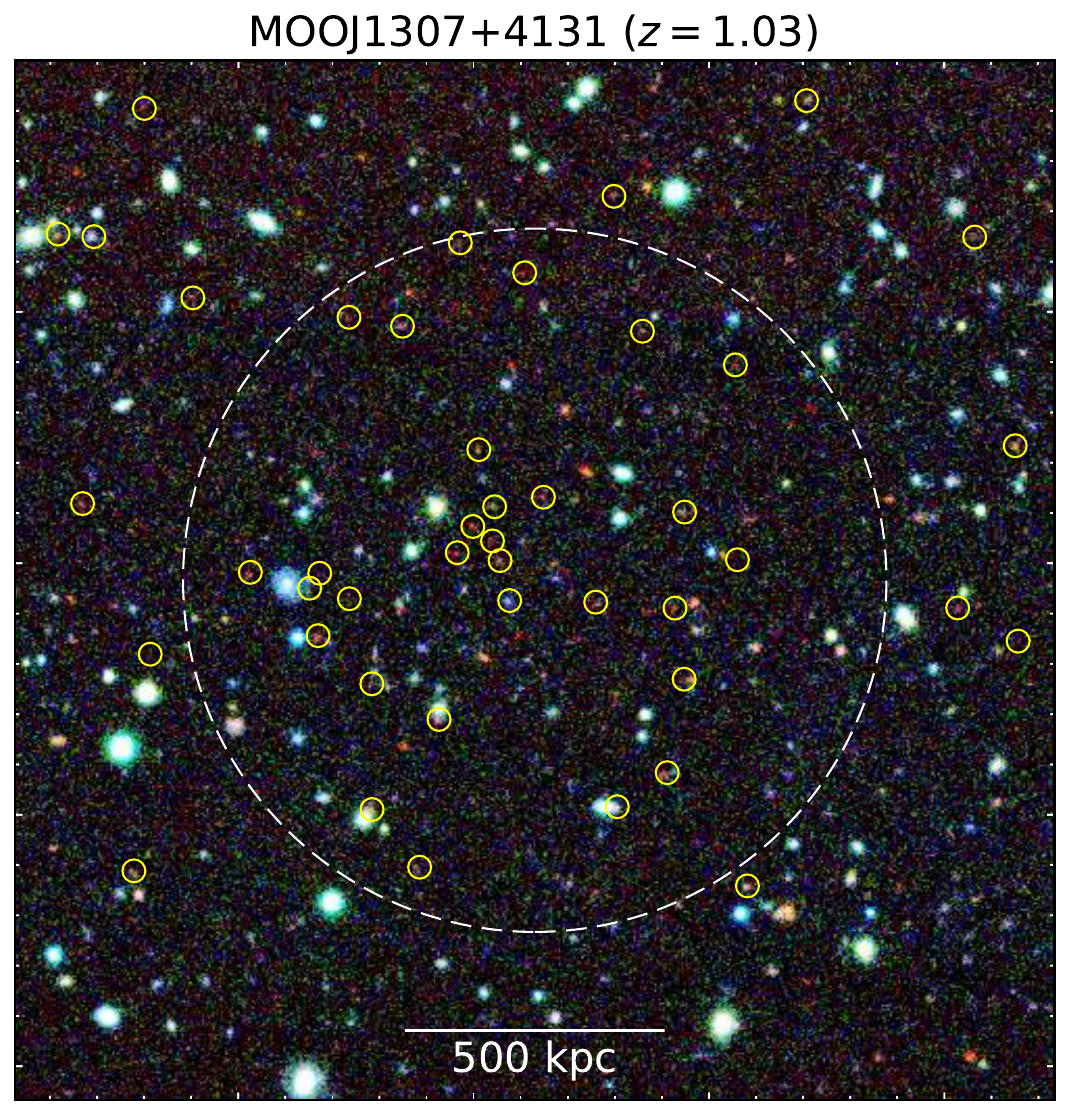}
\includegraphics[width=0.16\textwidth]{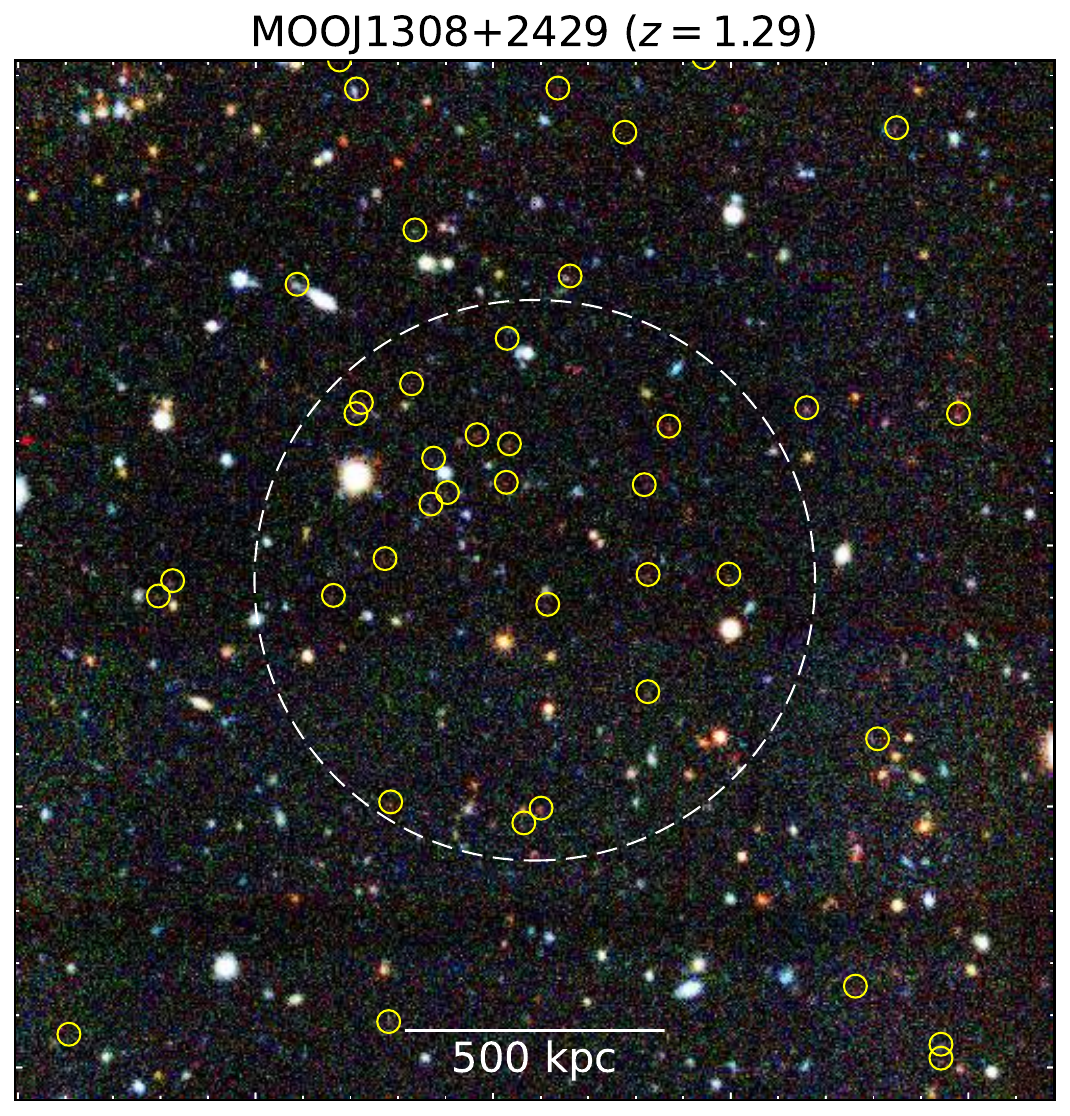}
\includegraphics[width=0.16\textwidth]{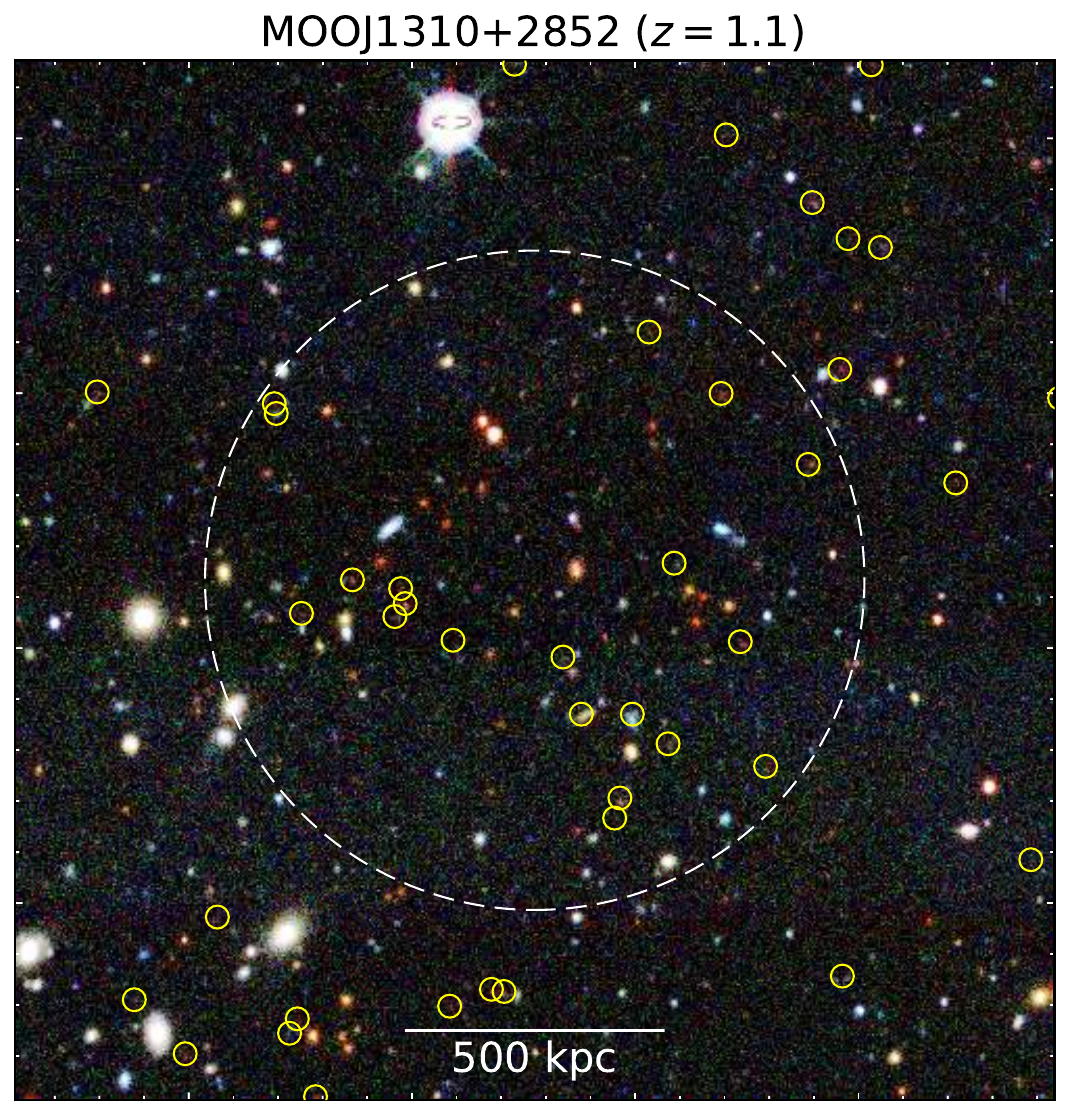}
\includegraphics[width=0.16\textwidth]{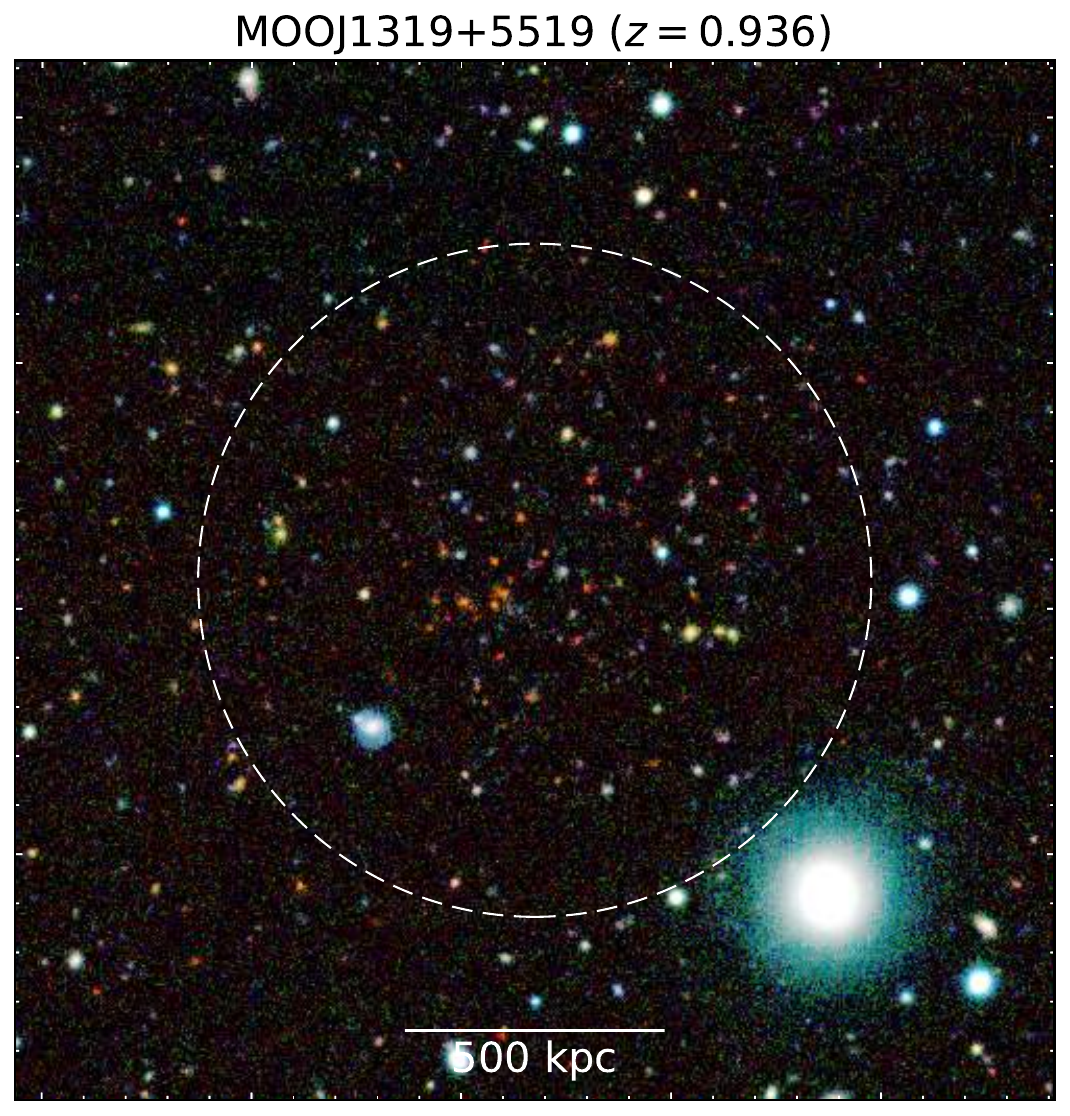}
\includegraphics[width=0.16\textwidth]{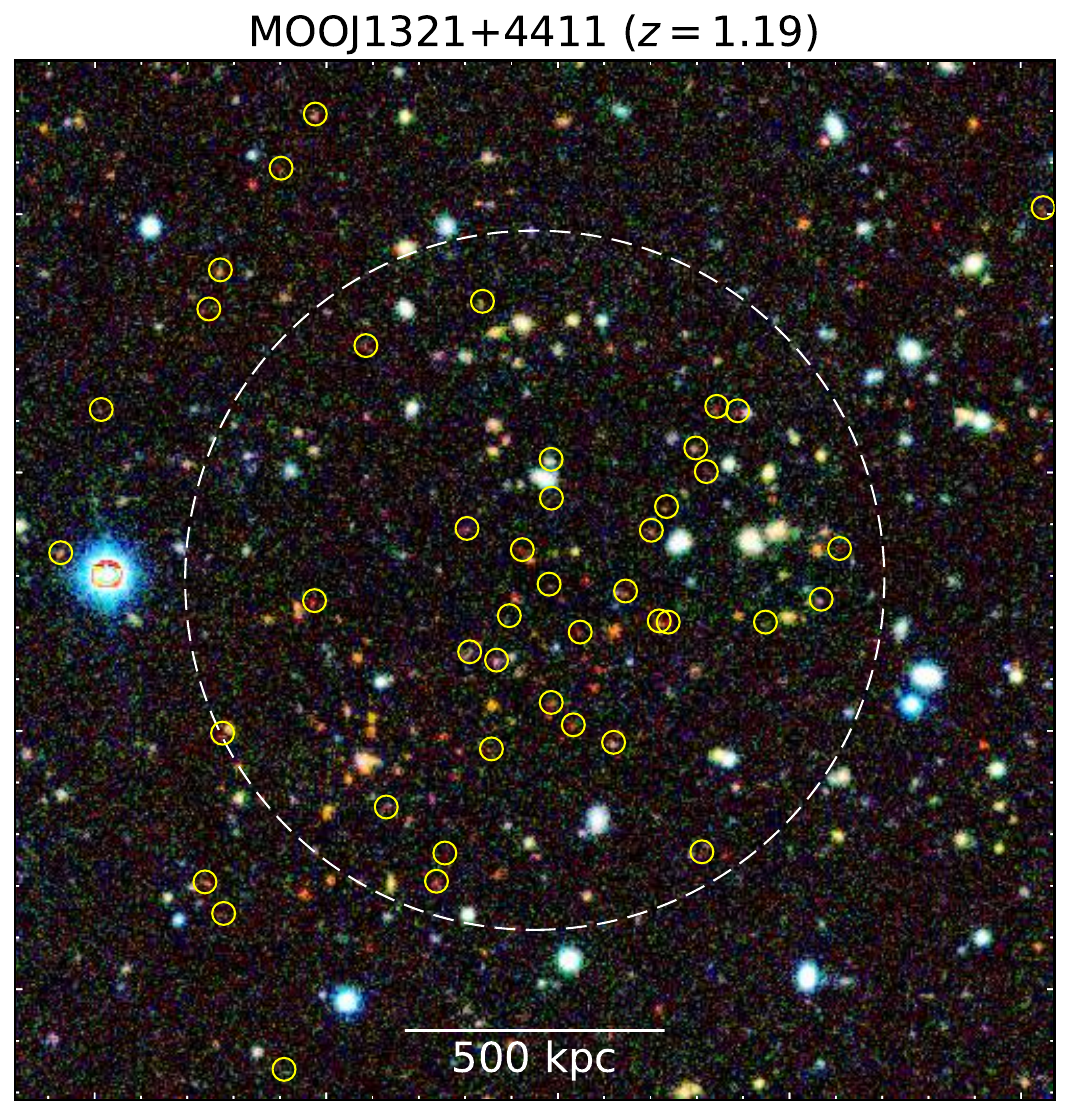}
\includegraphics[width=0.16\textwidth]{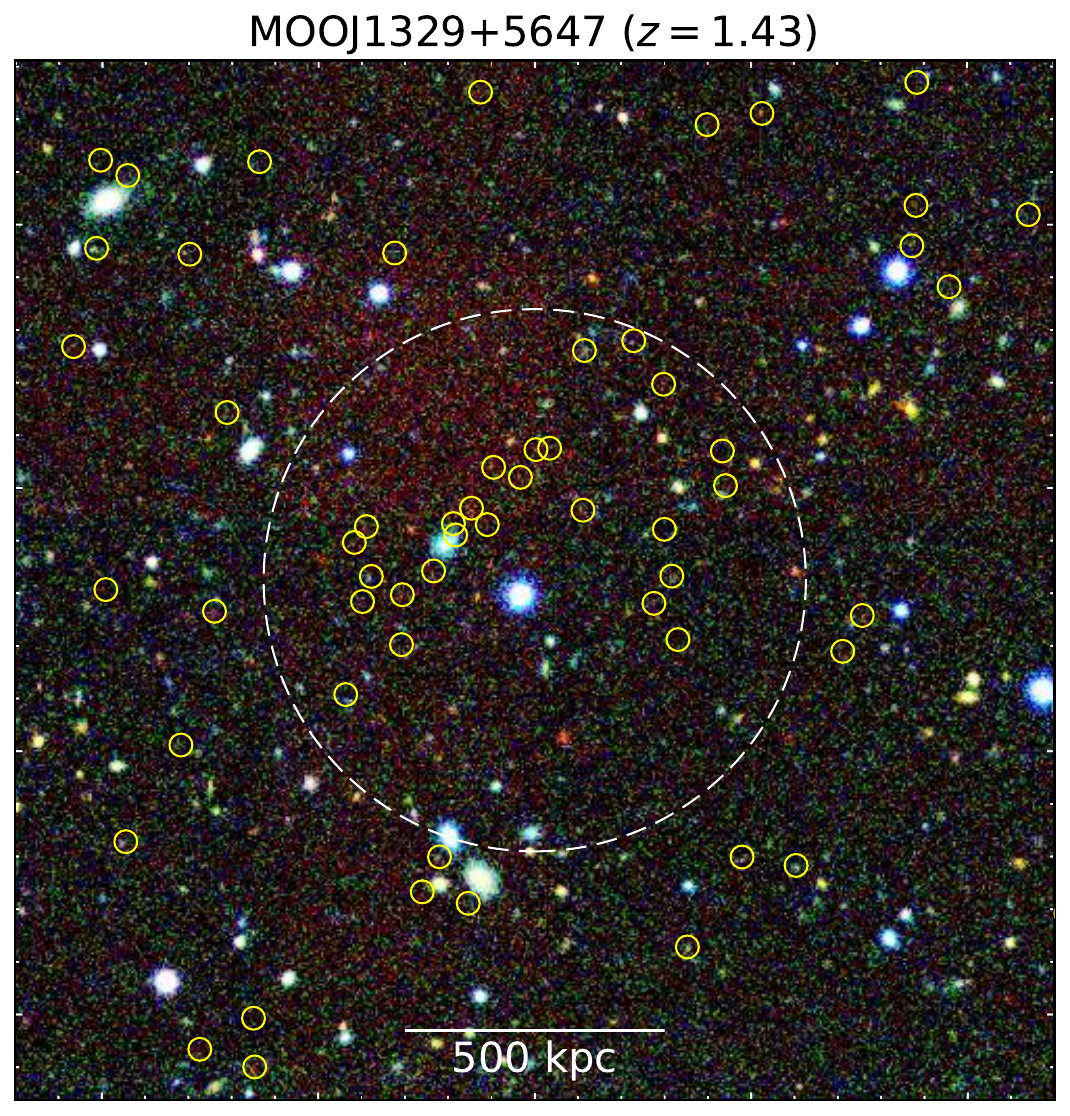}
\includegraphics[width=0.16\textwidth]{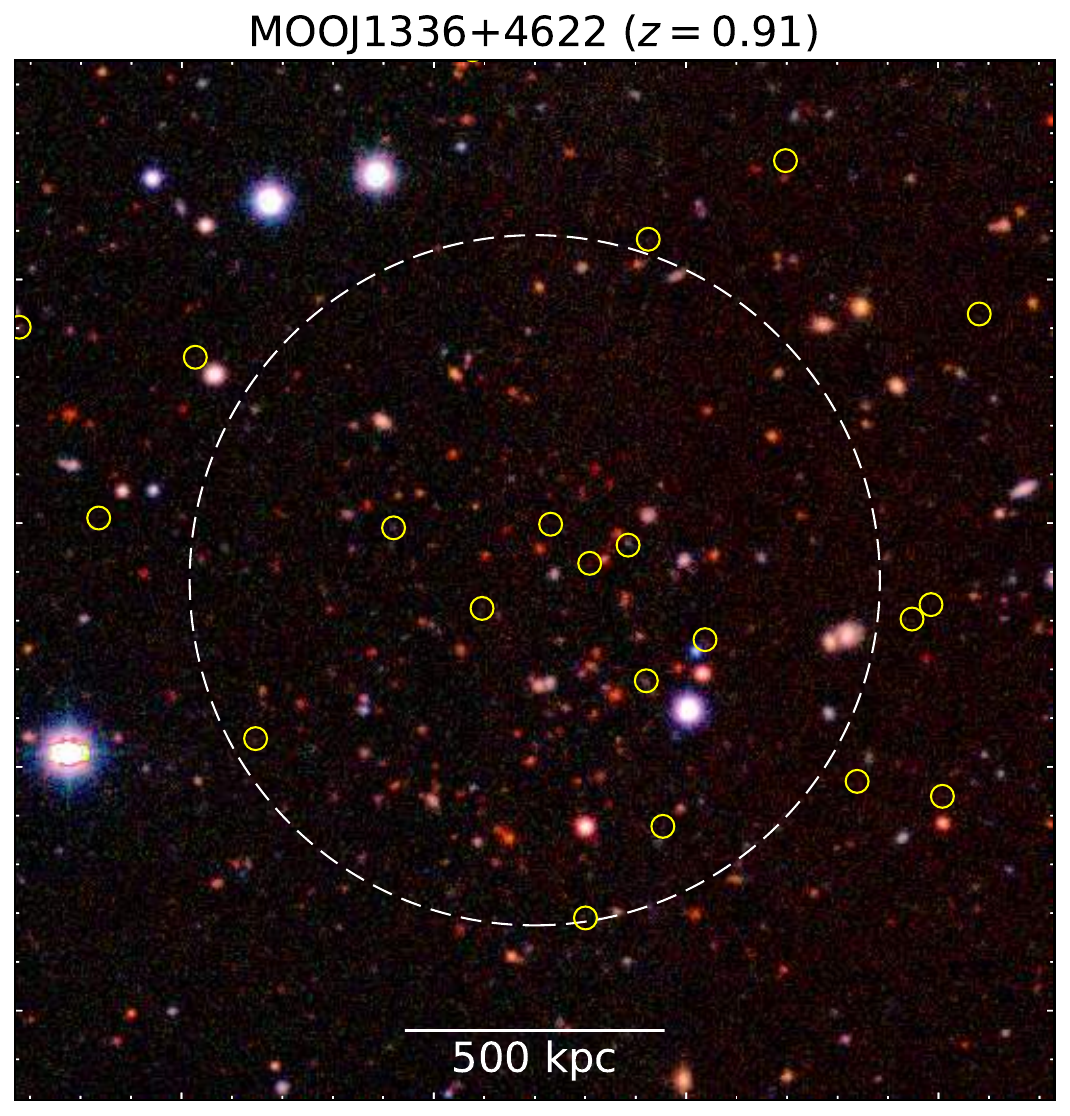}
\includegraphics[width=0.16\textwidth]{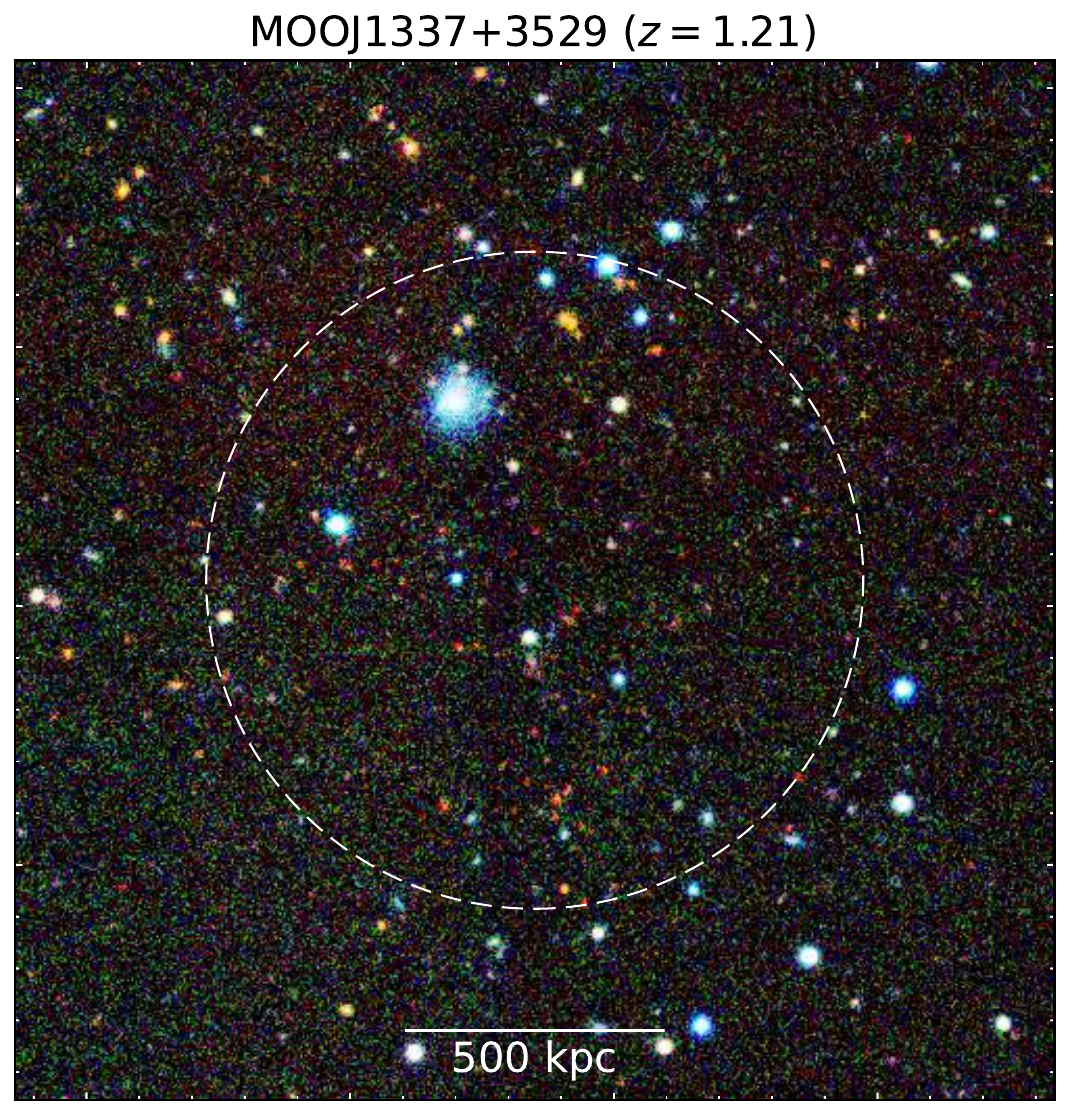}
\includegraphics[width=0.16\textwidth]{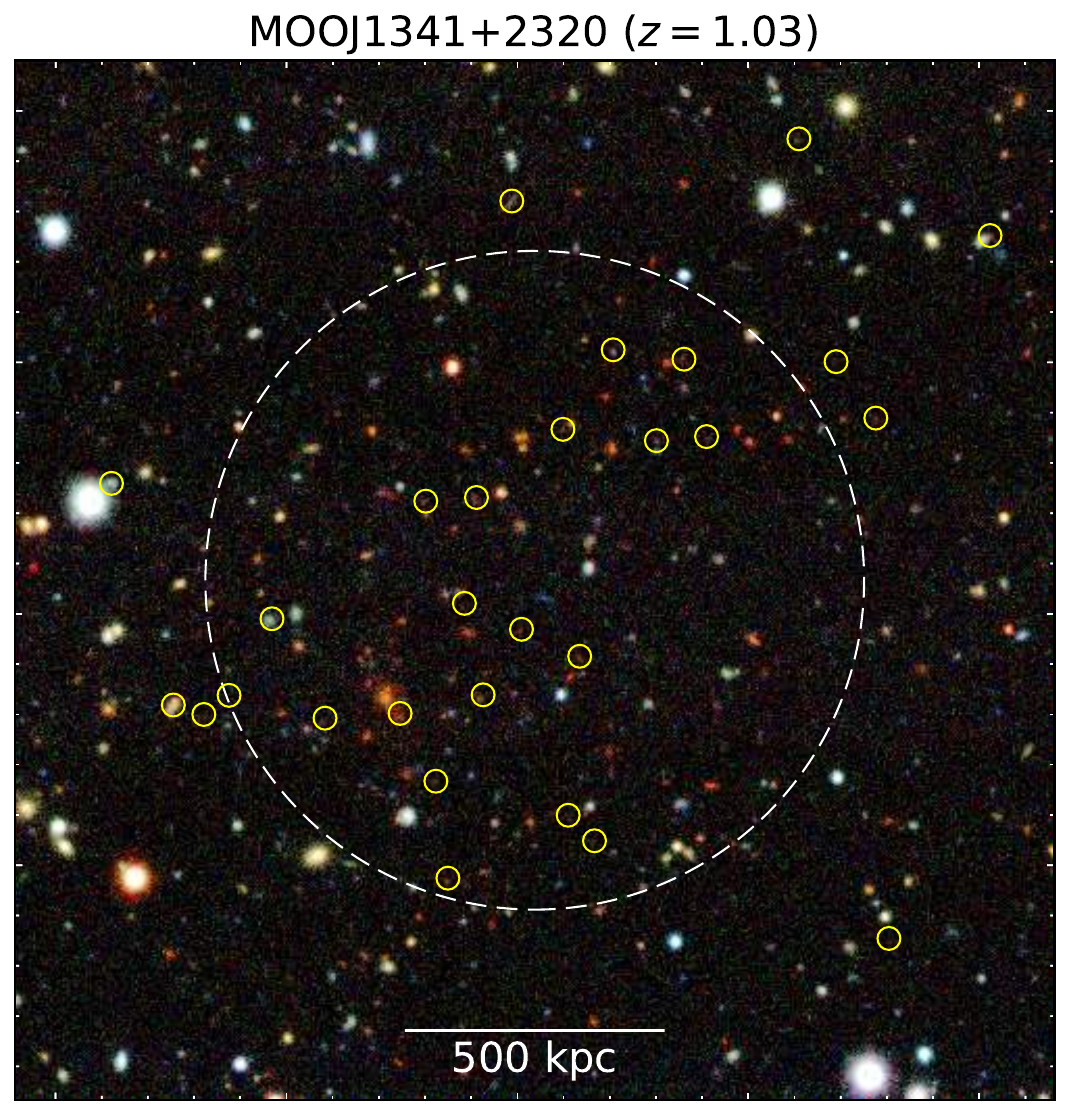}
\includegraphics[width=0.16\textwidth]{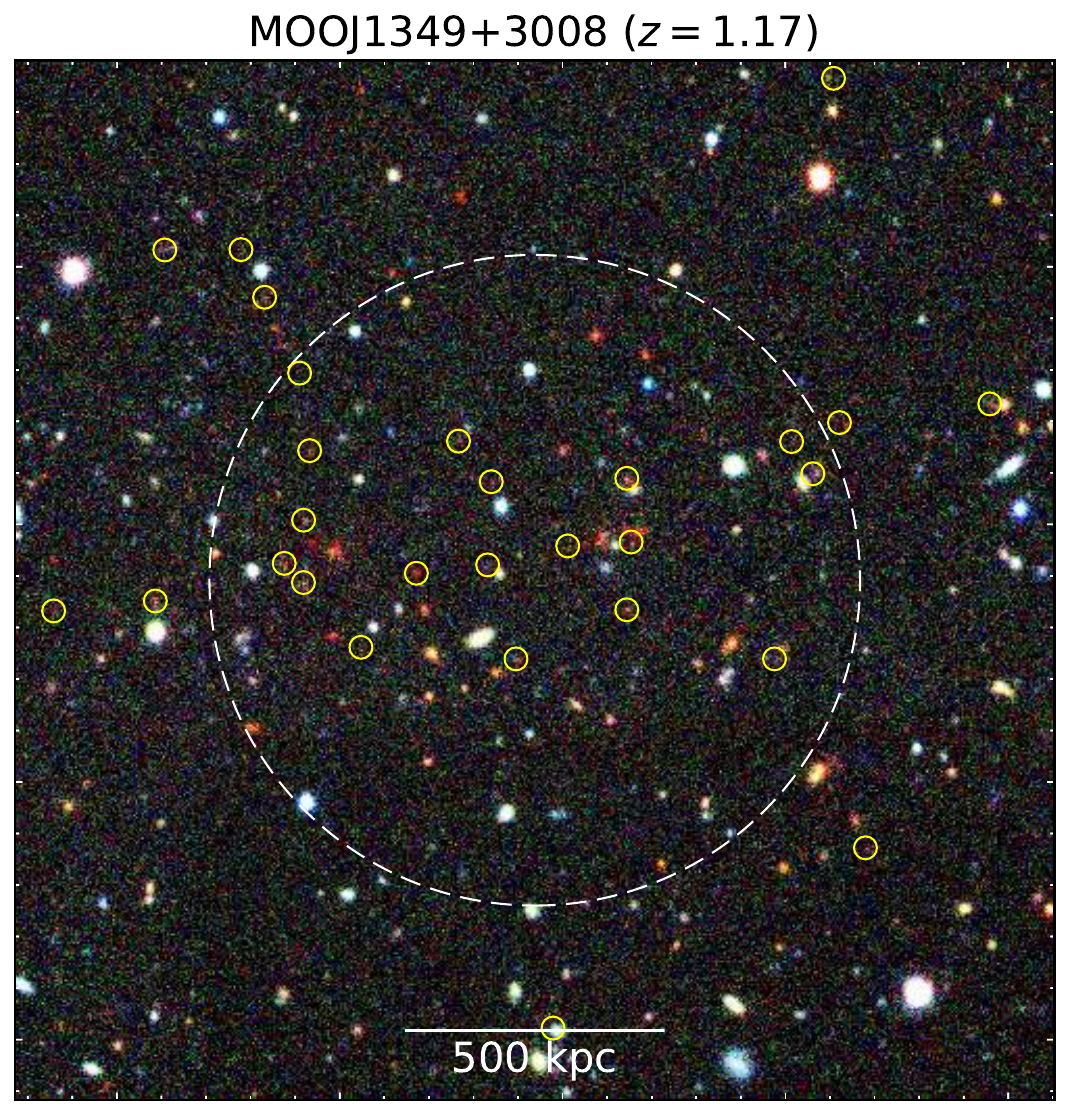}
\includegraphics[width=0.16\textwidth]{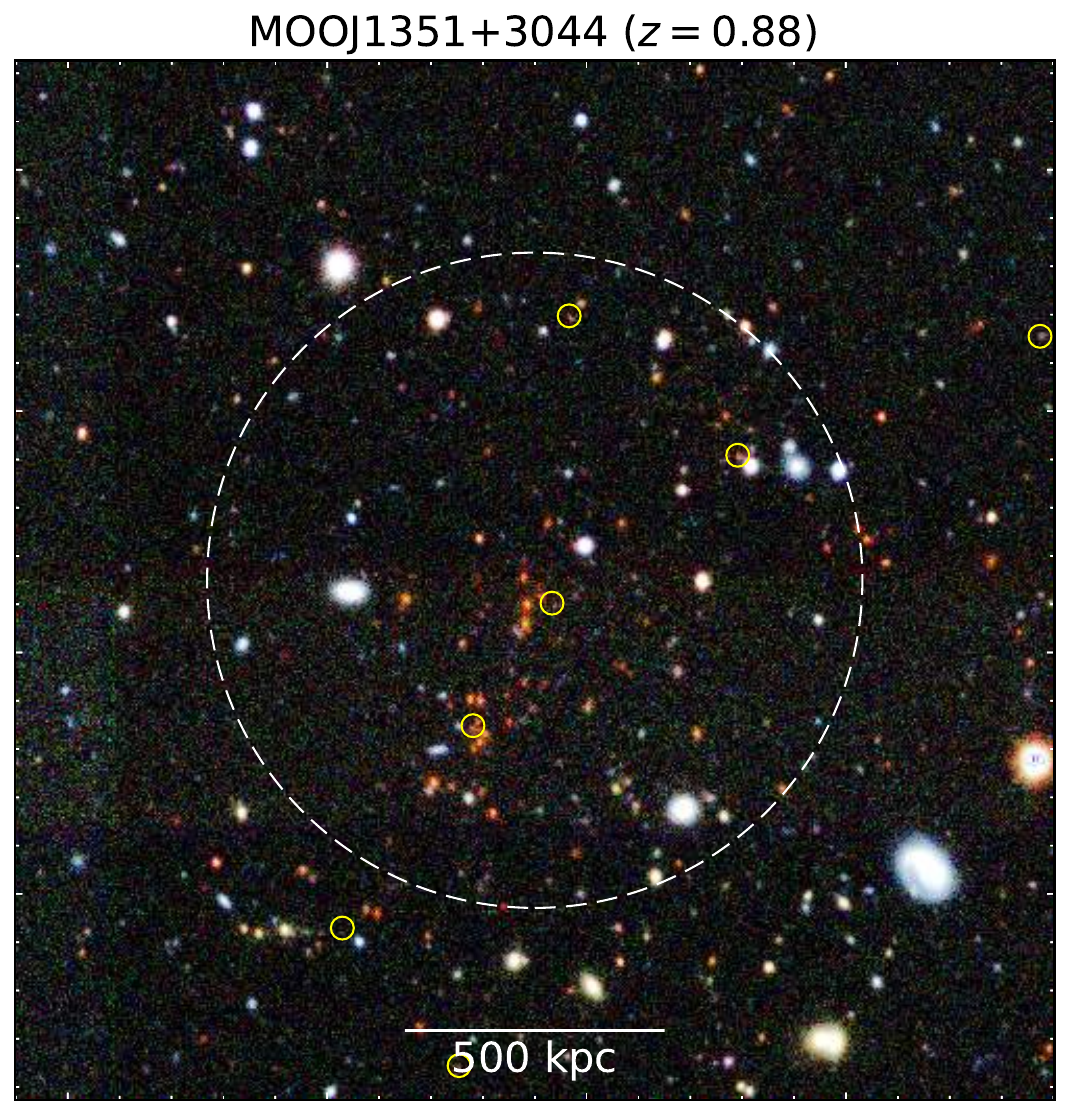}
\caption{Combined {\it grz} images from the Desi Legacy Survey DR10. When available, we mark with a yellow symbol the cluster's galaxies and the source subtracted {\tt taper=100kpc} (see fourth column in Fig. \ref{fig:madcows_lofar_images}).}\label{fig:optical}
\end{figure*}

\begin{figure*}
\centering
\ContinuedFloat
\includegraphics[width=0.16\textwidth]{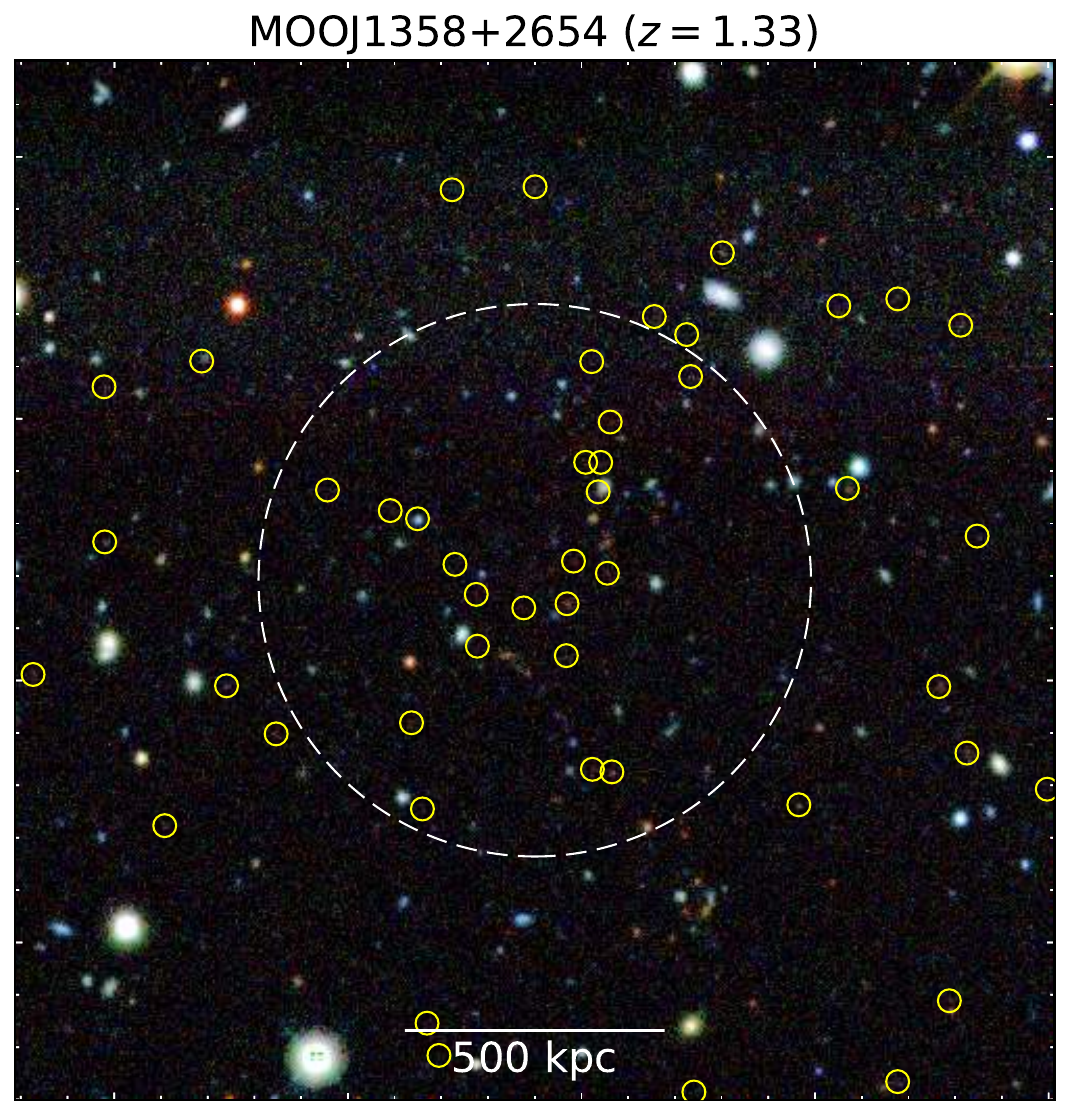}
\includegraphics[width=0.16\textwidth]{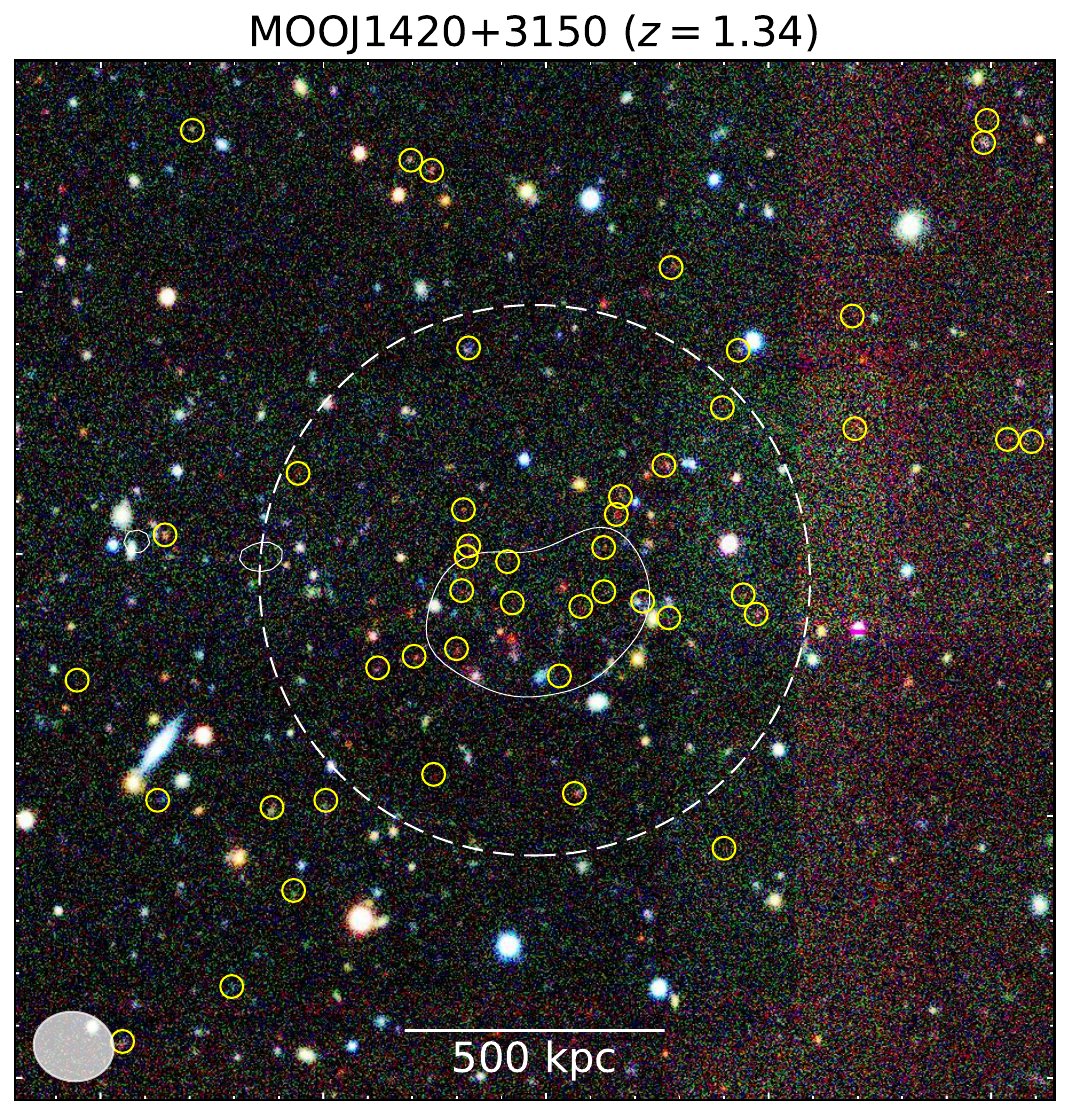}
\includegraphics[width=0.16\textwidth]{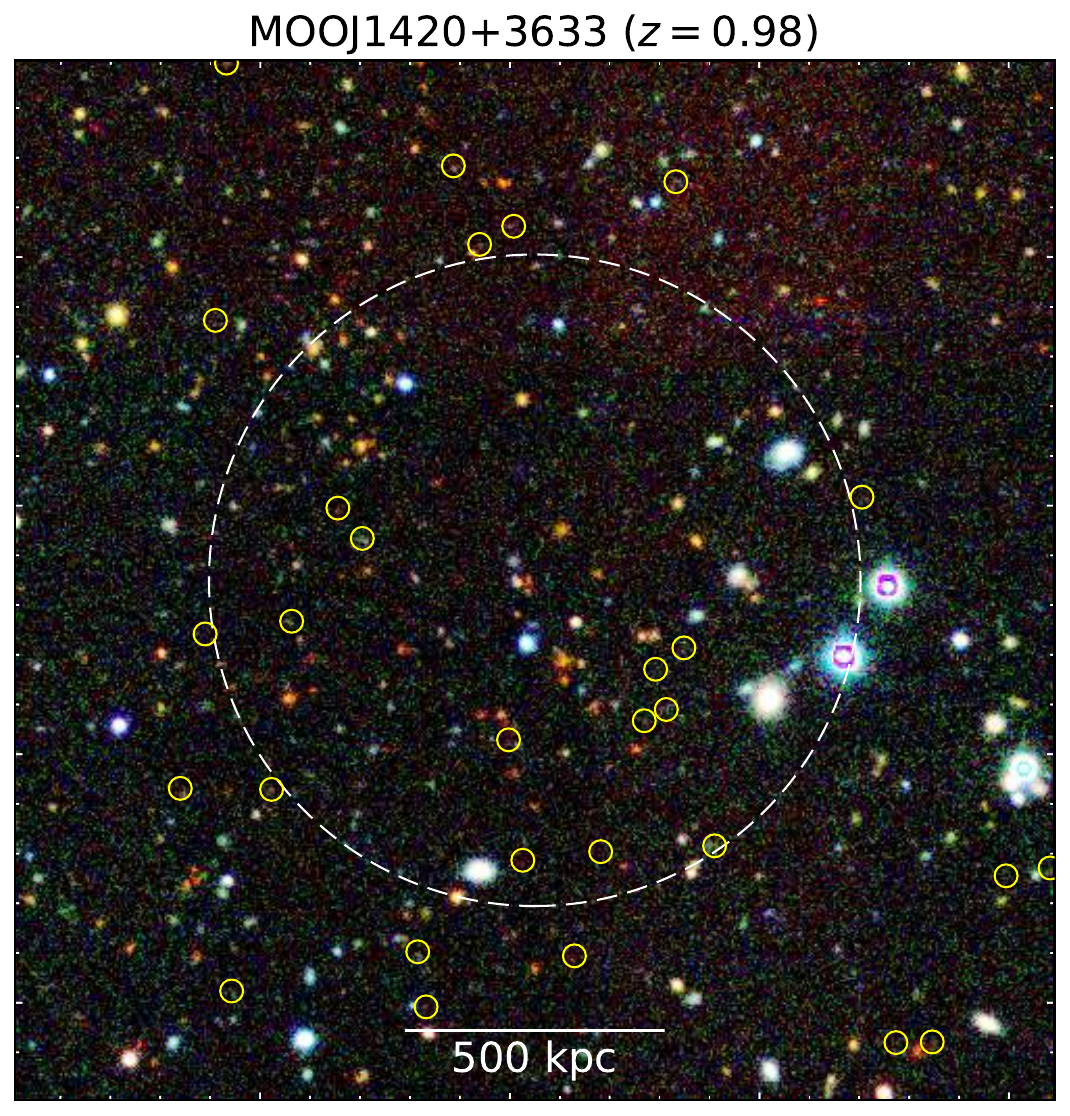}
\includegraphics[width=0.16\textwidth]{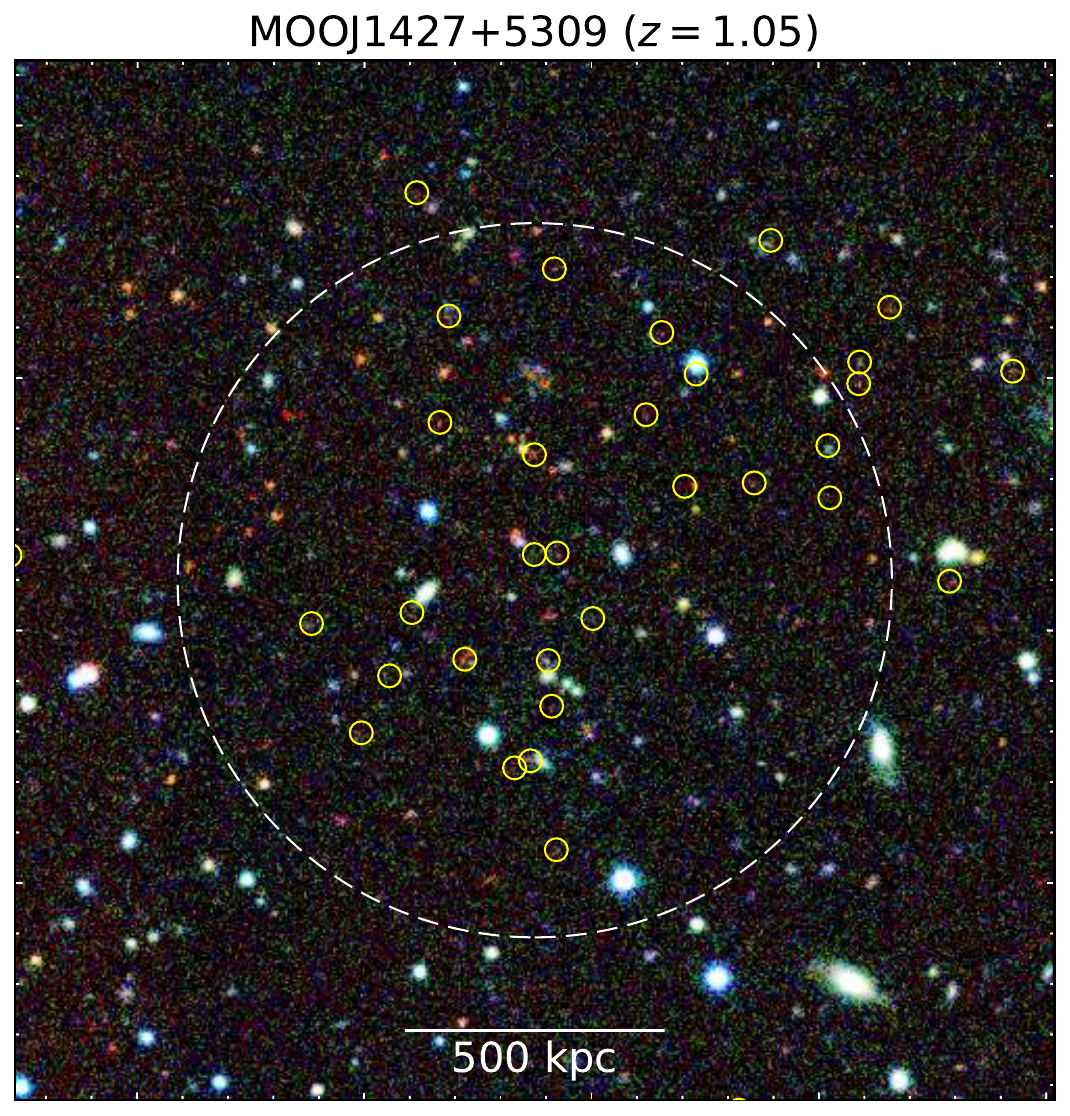}
\includegraphics[width=0.16\textwidth]{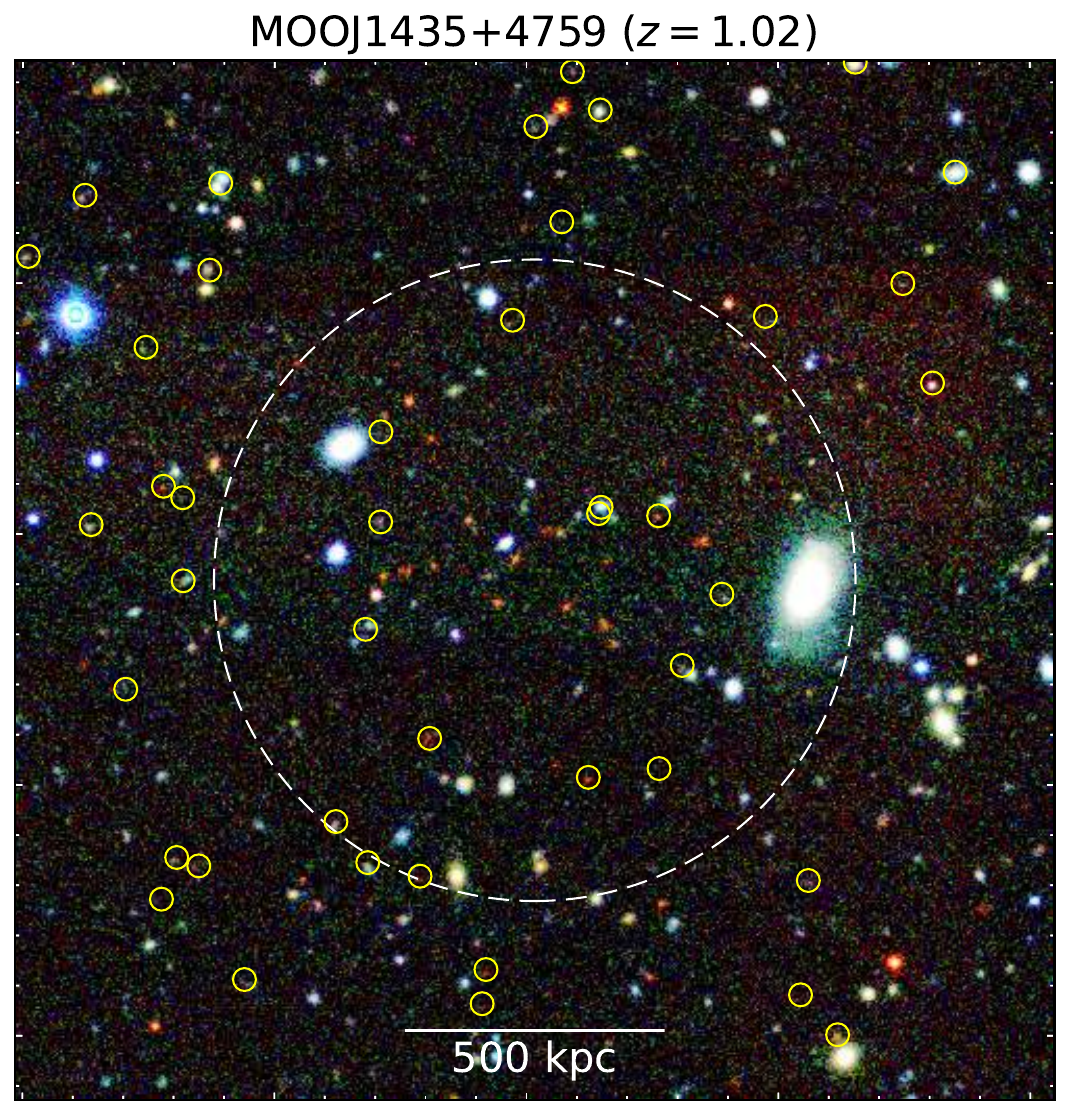}
\includegraphics[width=0.16\textwidth]{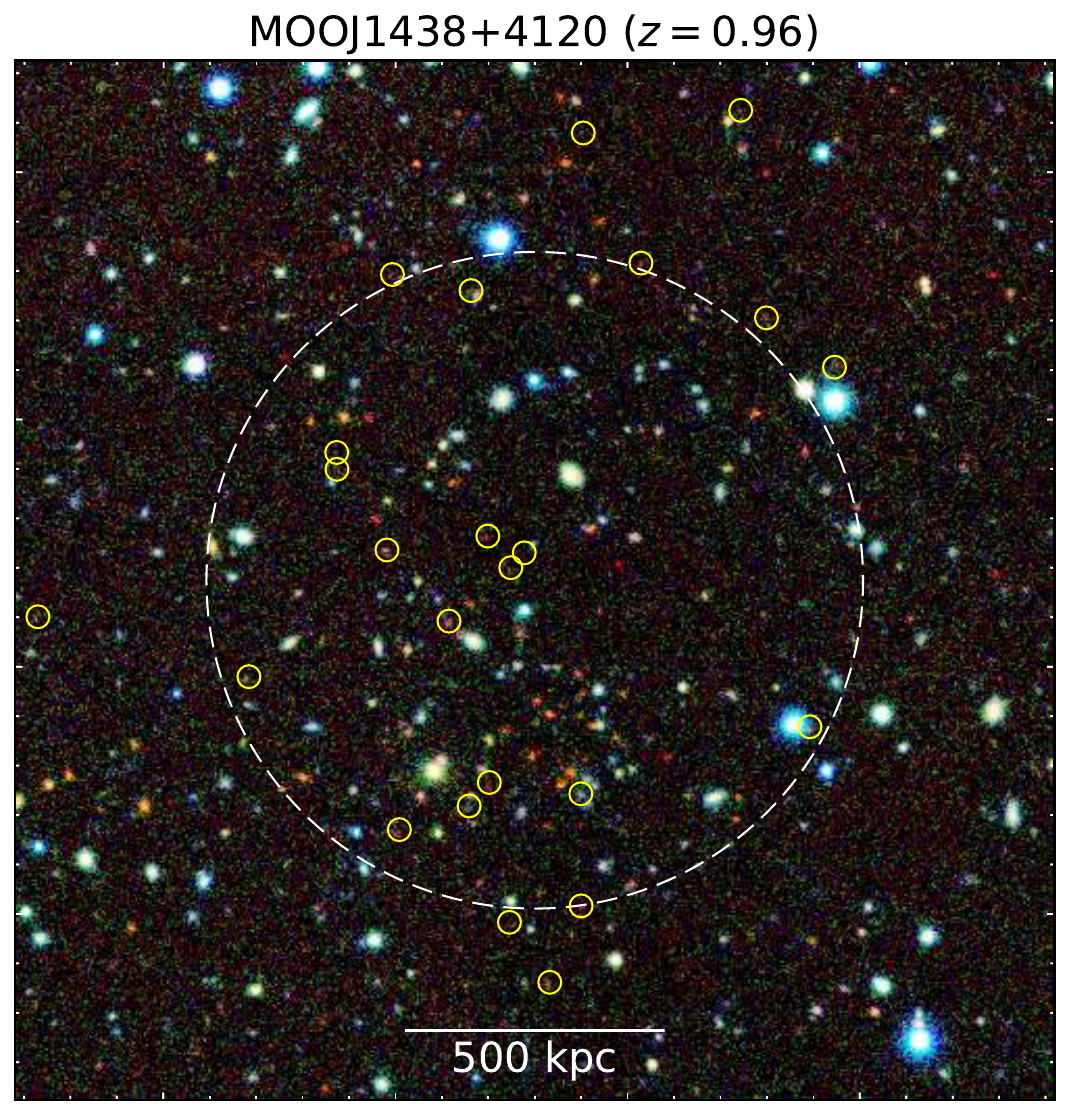}
\includegraphics[width=0.16\textwidth]{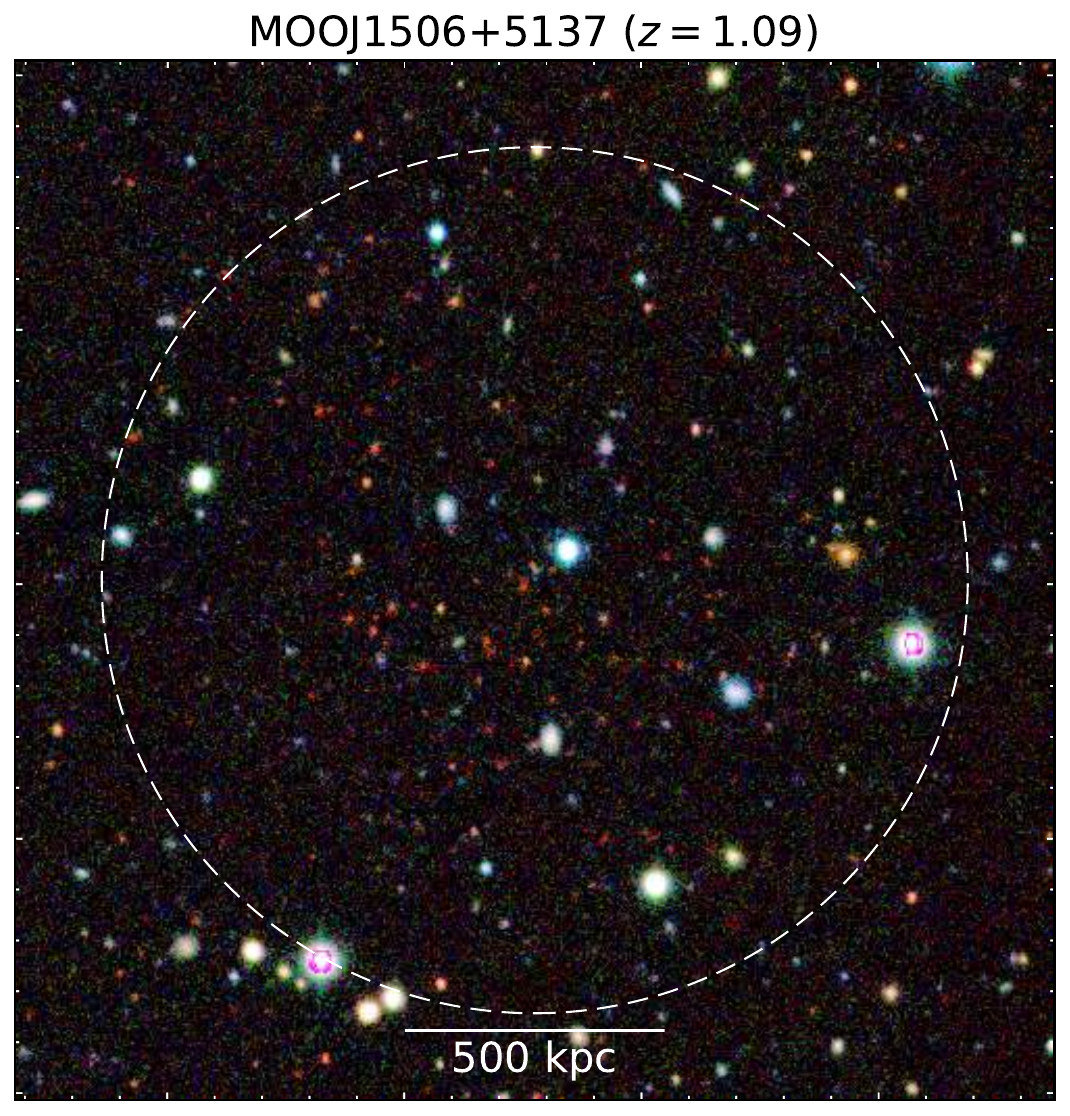}
\includegraphics[width=0.16\textwidth]{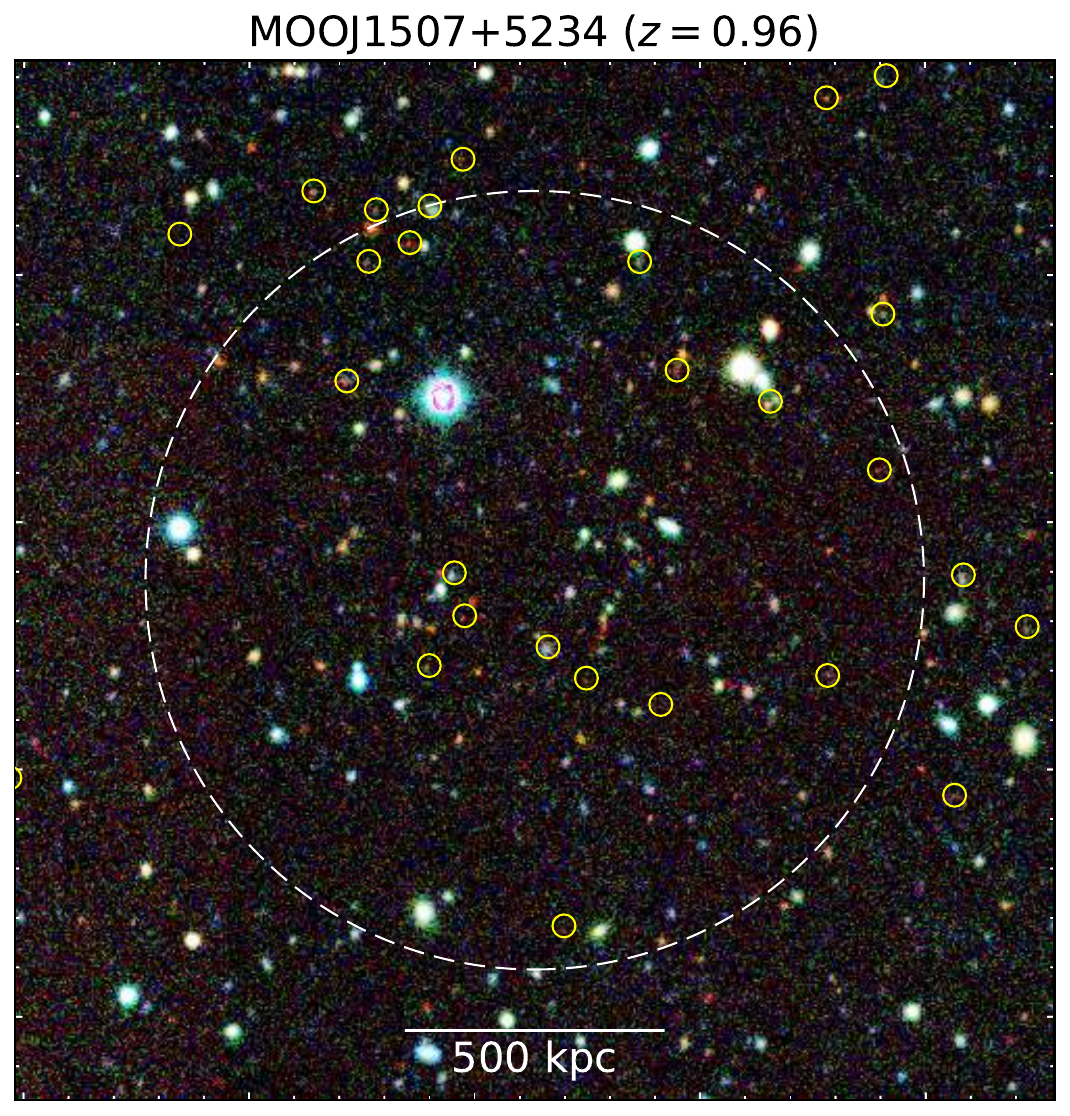}
\includegraphics[width=0.16\textwidth]{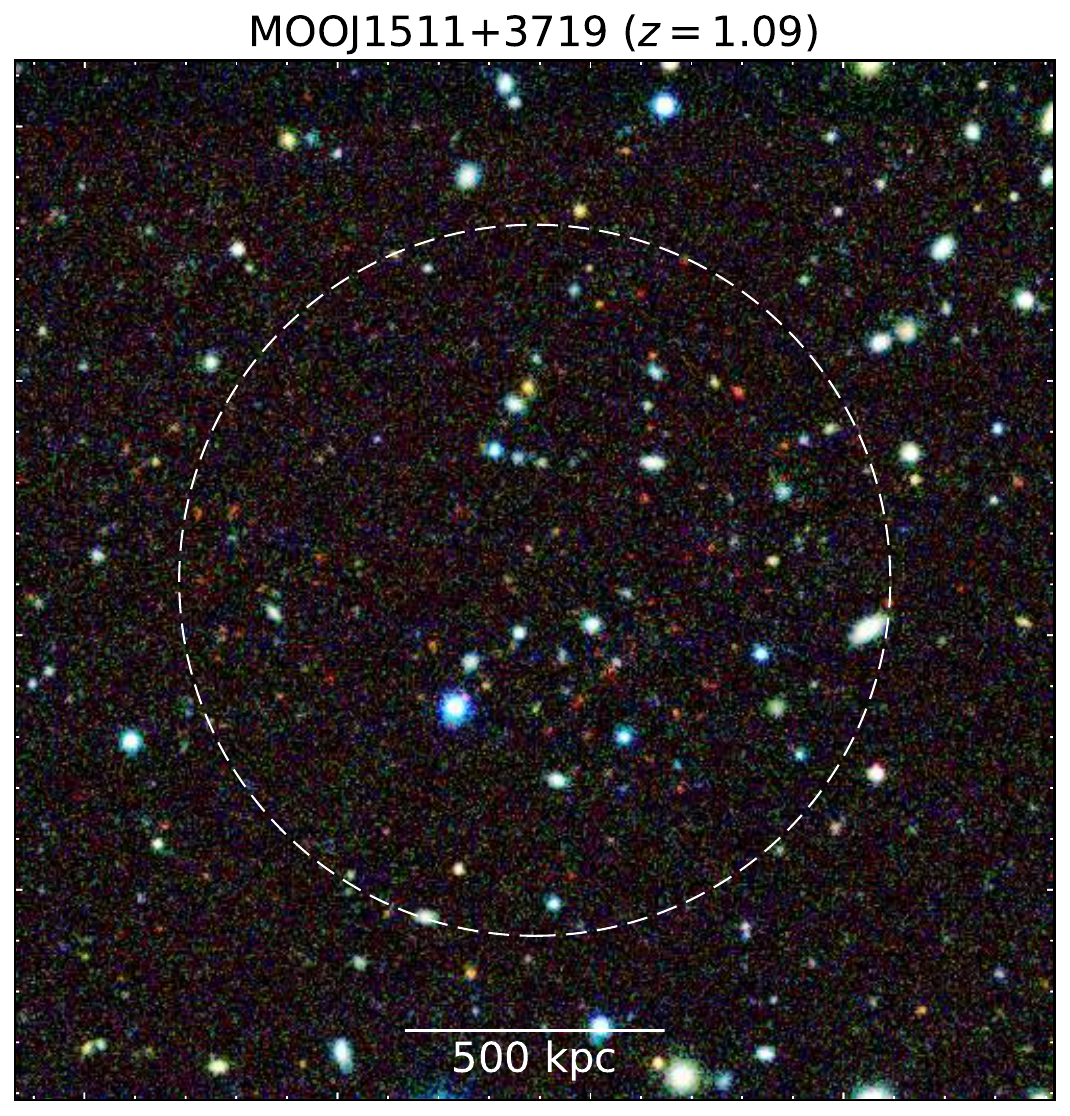}
\includegraphics[width=0.16\textwidth]{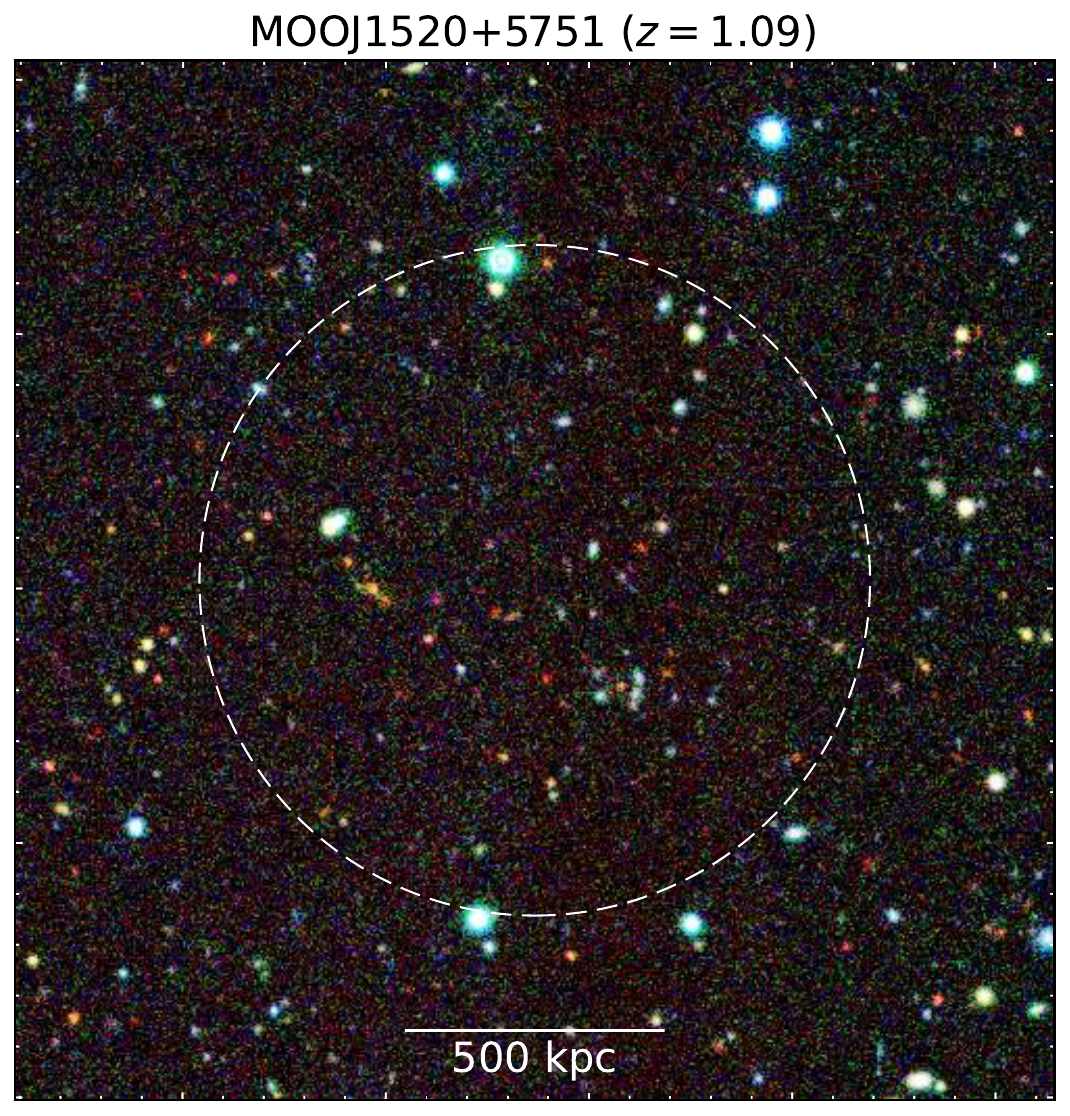}
\includegraphics[width=0.16\textwidth]{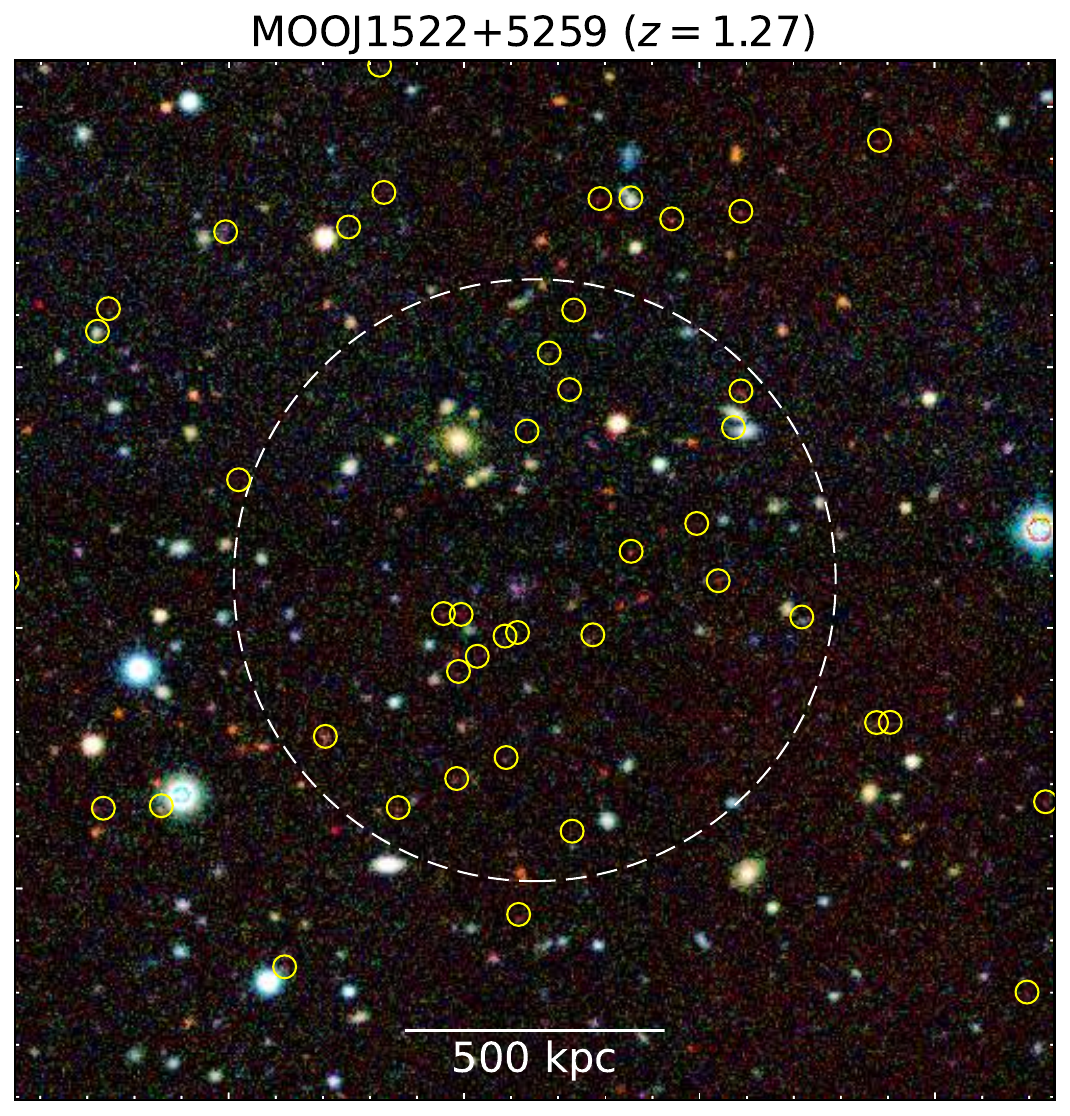}
\includegraphics[width=0.16\textwidth]{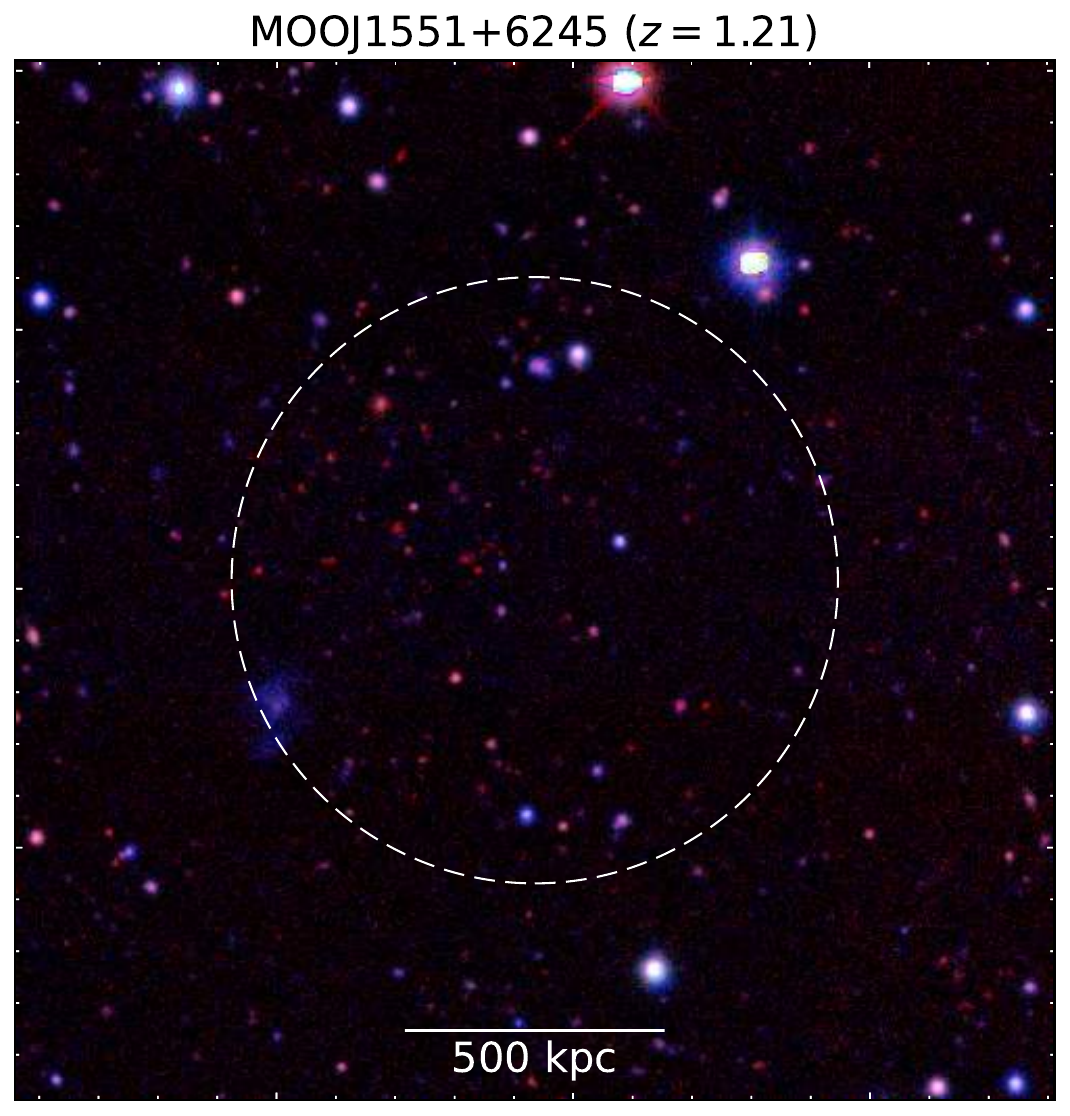}
\includegraphics[width=0.16\textwidth]{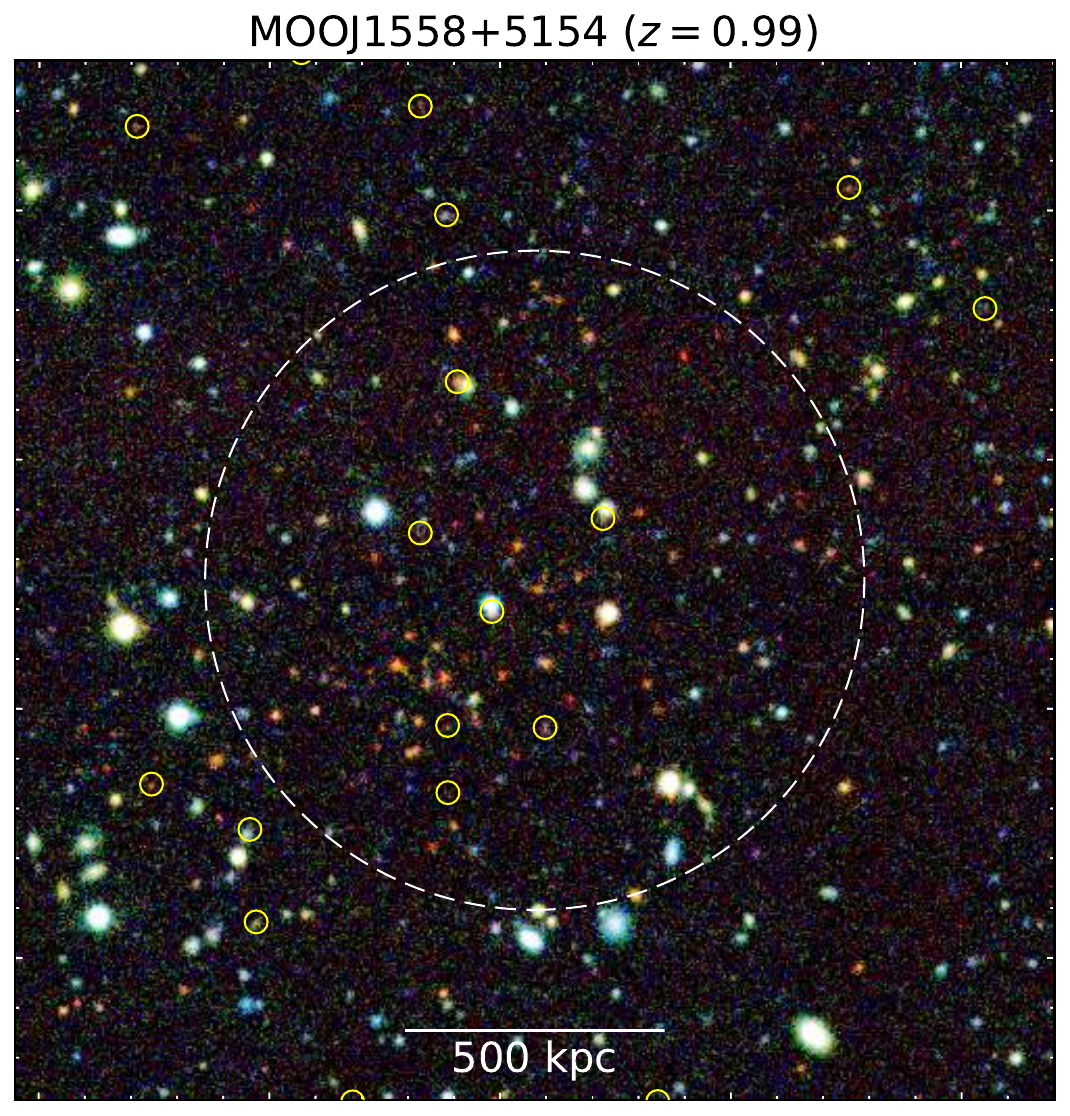}
\includegraphics[width=0.16\textwidth]{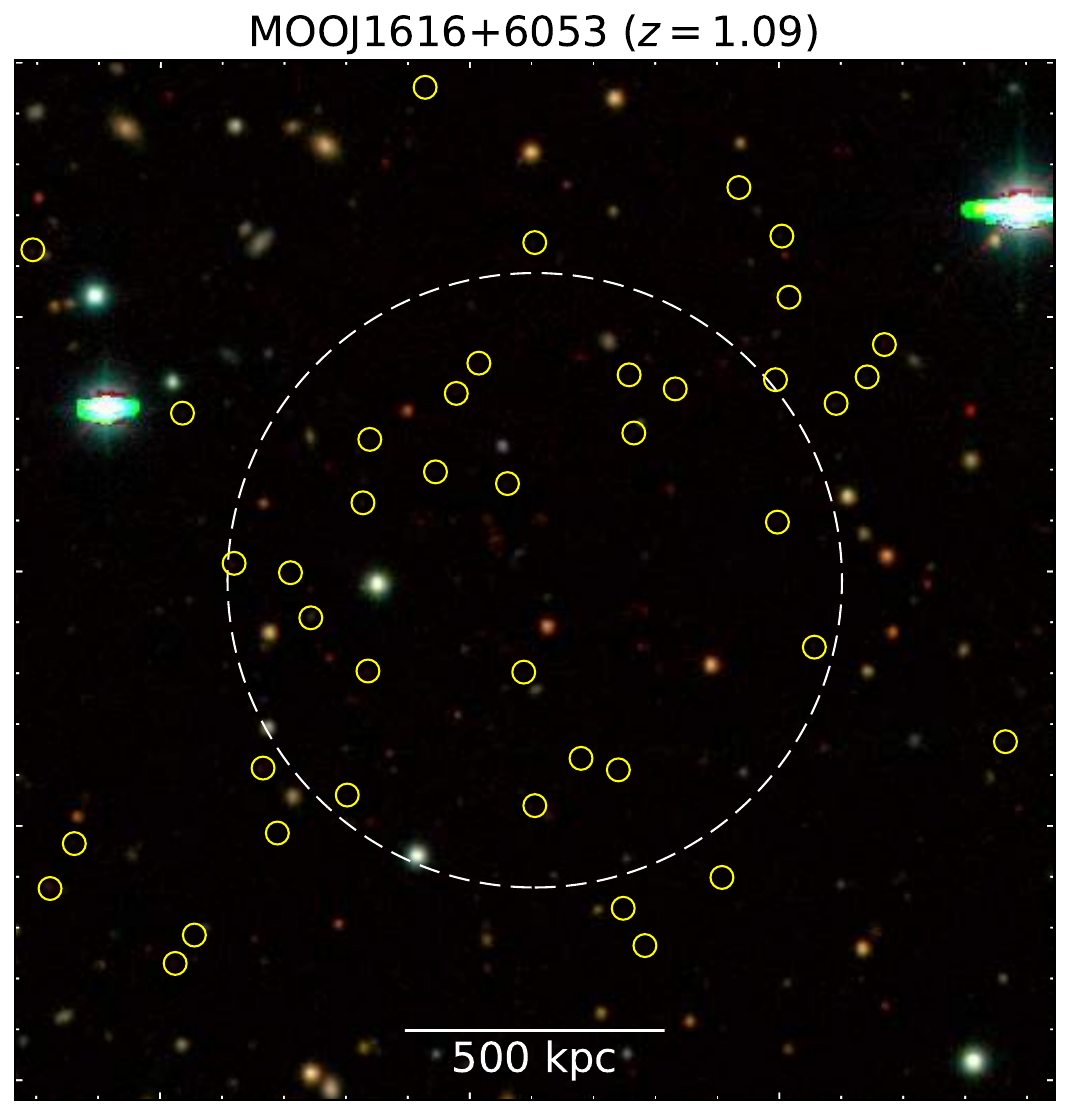}
\includegraphics[width=0.16\textwidth]{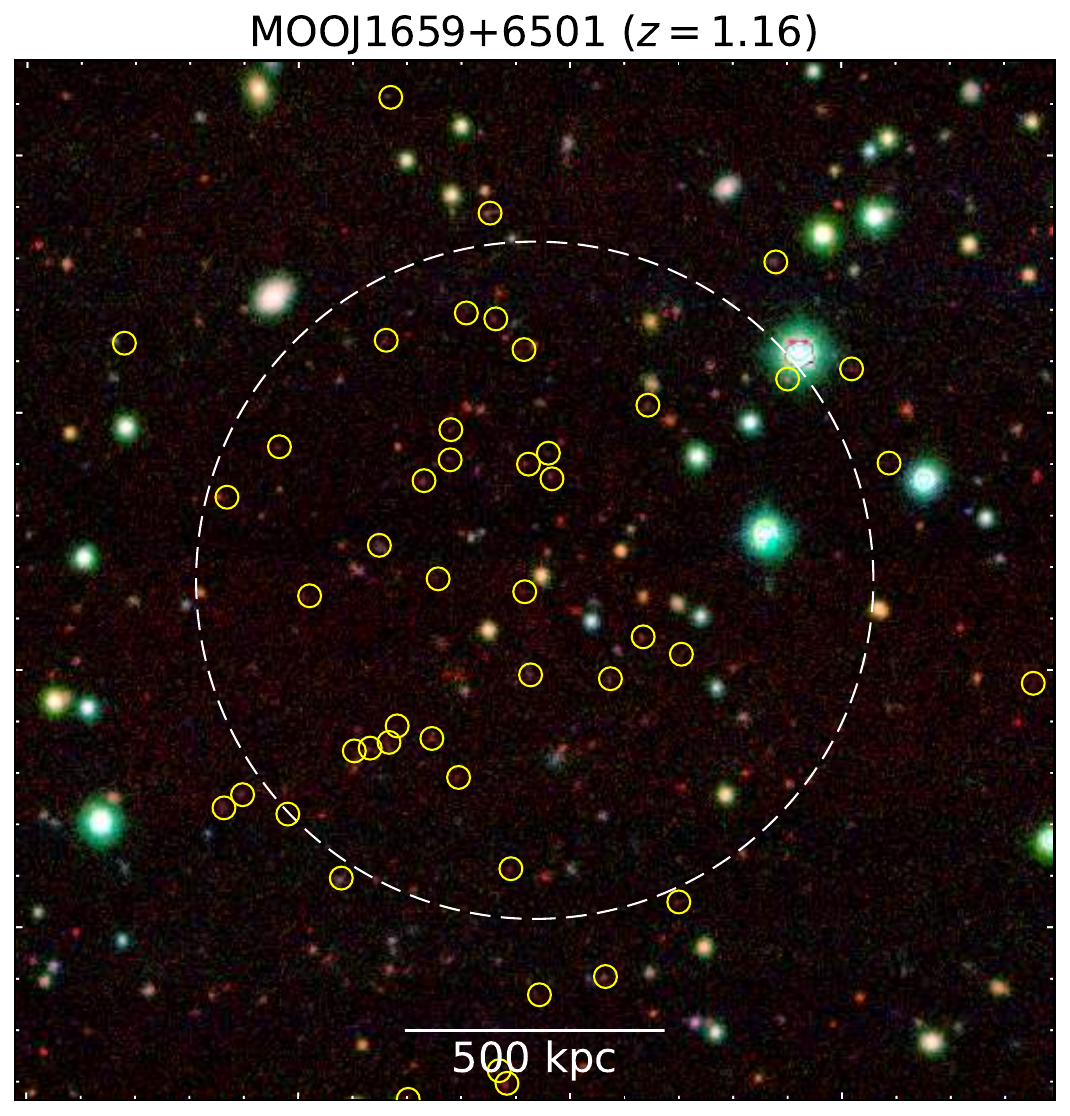}
\includegraphics[width=0.16\textwidth]{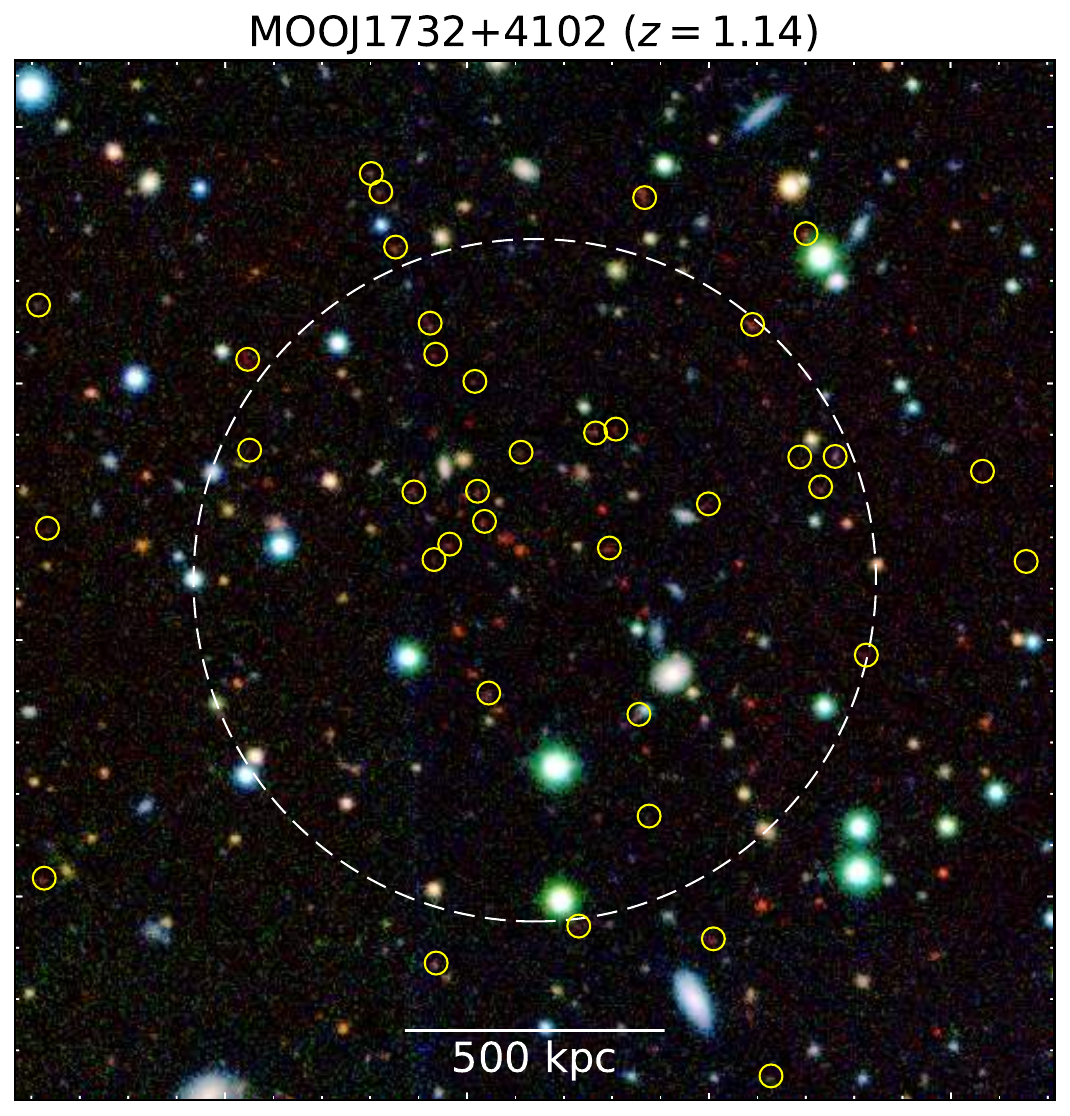}
\includegraphics[width=0.16\textwidth]{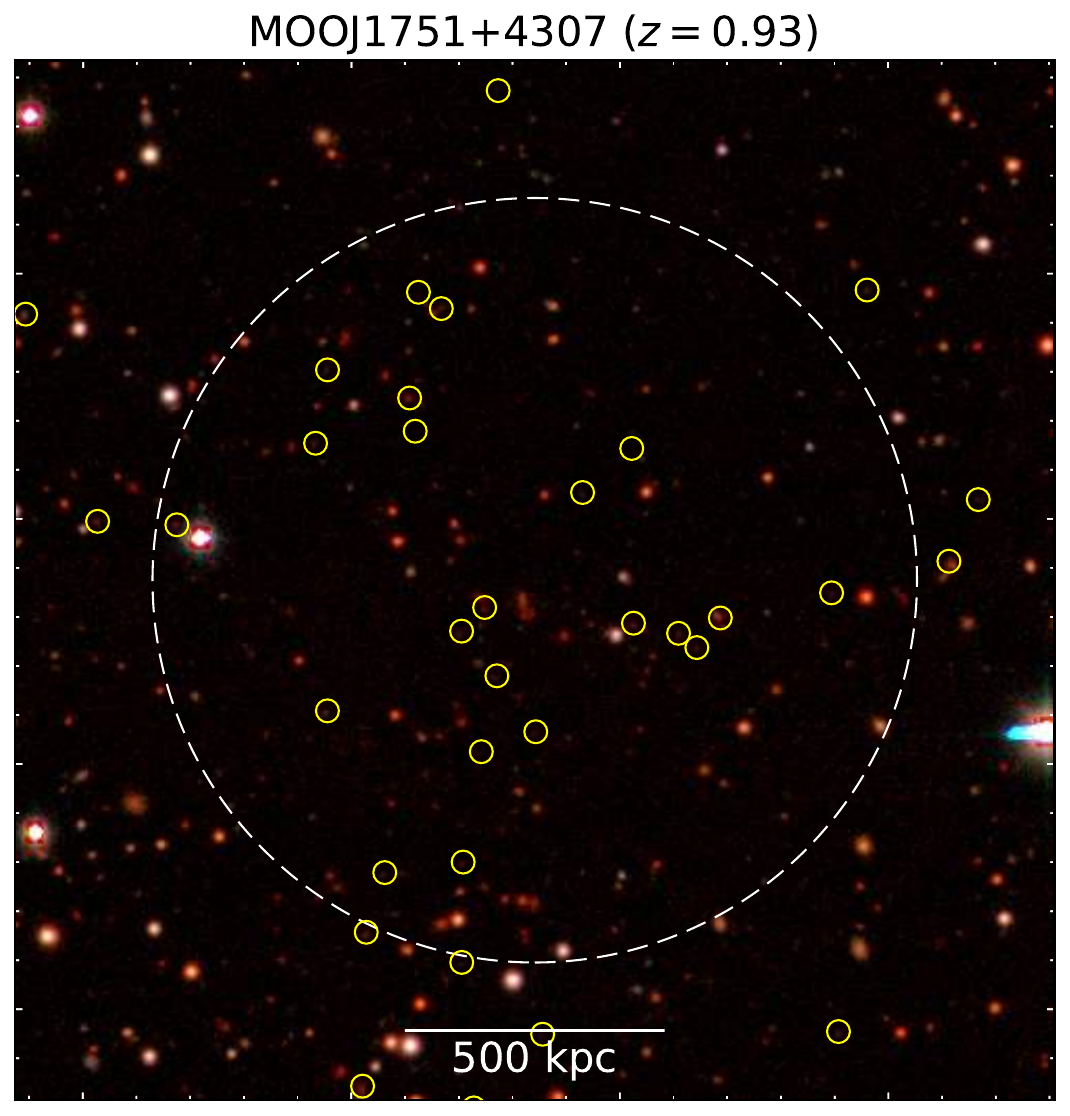}
\includegraphics[width=0.16\textwidth]{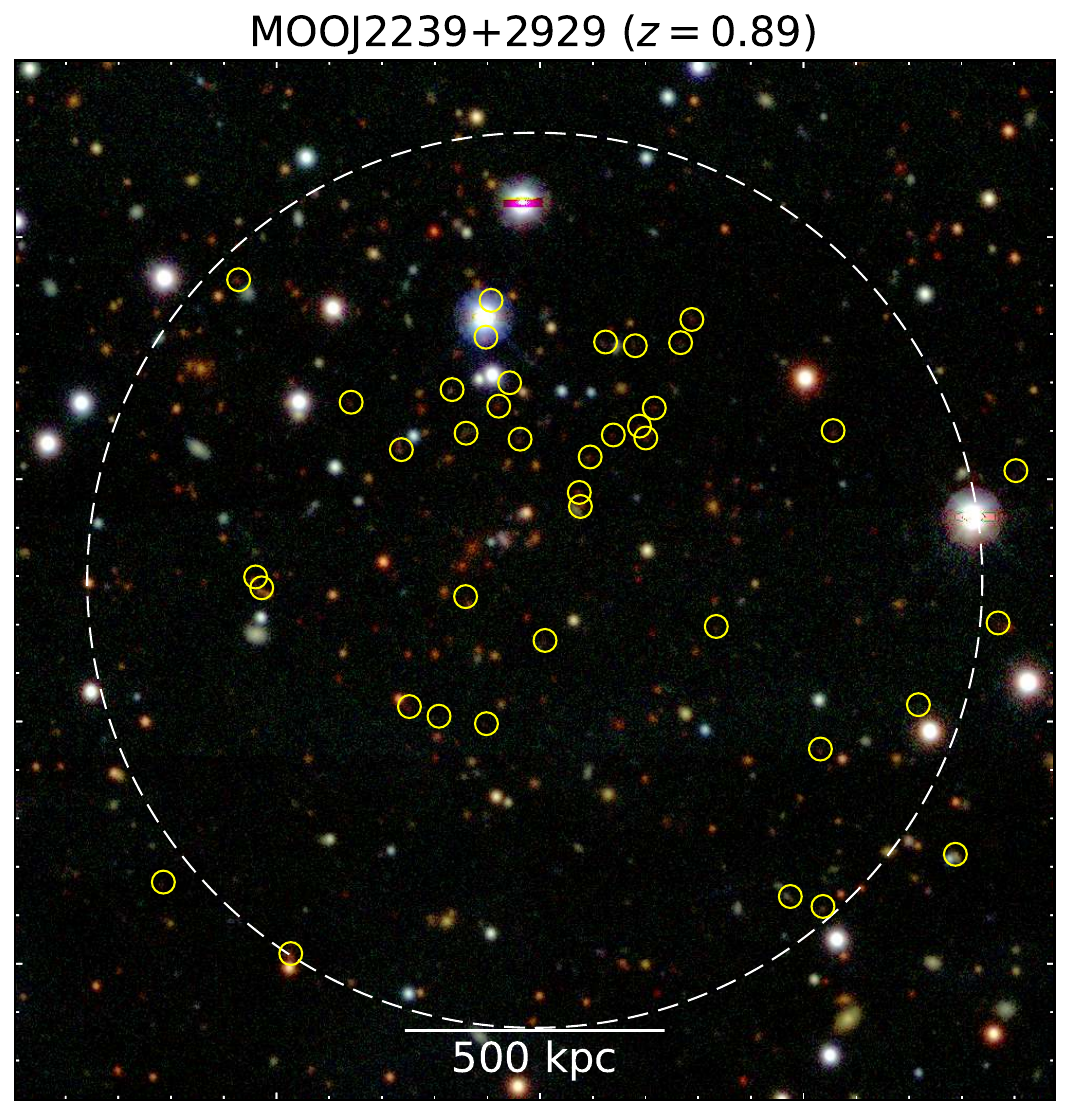}
\includegraphics[width=0.16\textwidth]{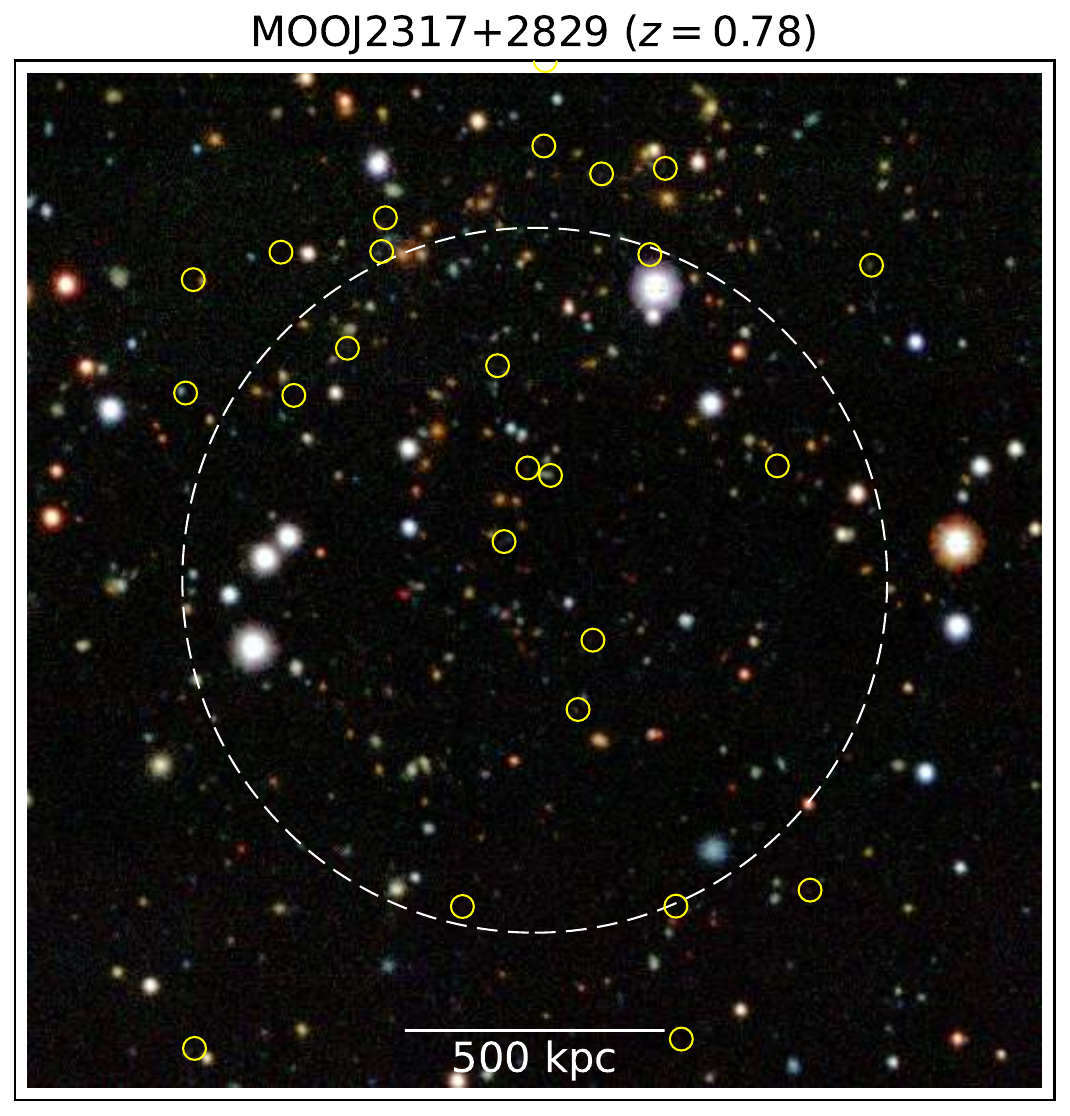}
\caption{Continued.}
\end{figure*}

\clearpage
\section{Upper limits}
In this section, we present $P_\nu-M_{500}$ diagrams at 150 MHz (left) and at 1.4 GHz (right) including also the upper limits on the non-detection.

\begin{figure*}[h!]
\centering
\includegraphics[width=\textwidth]{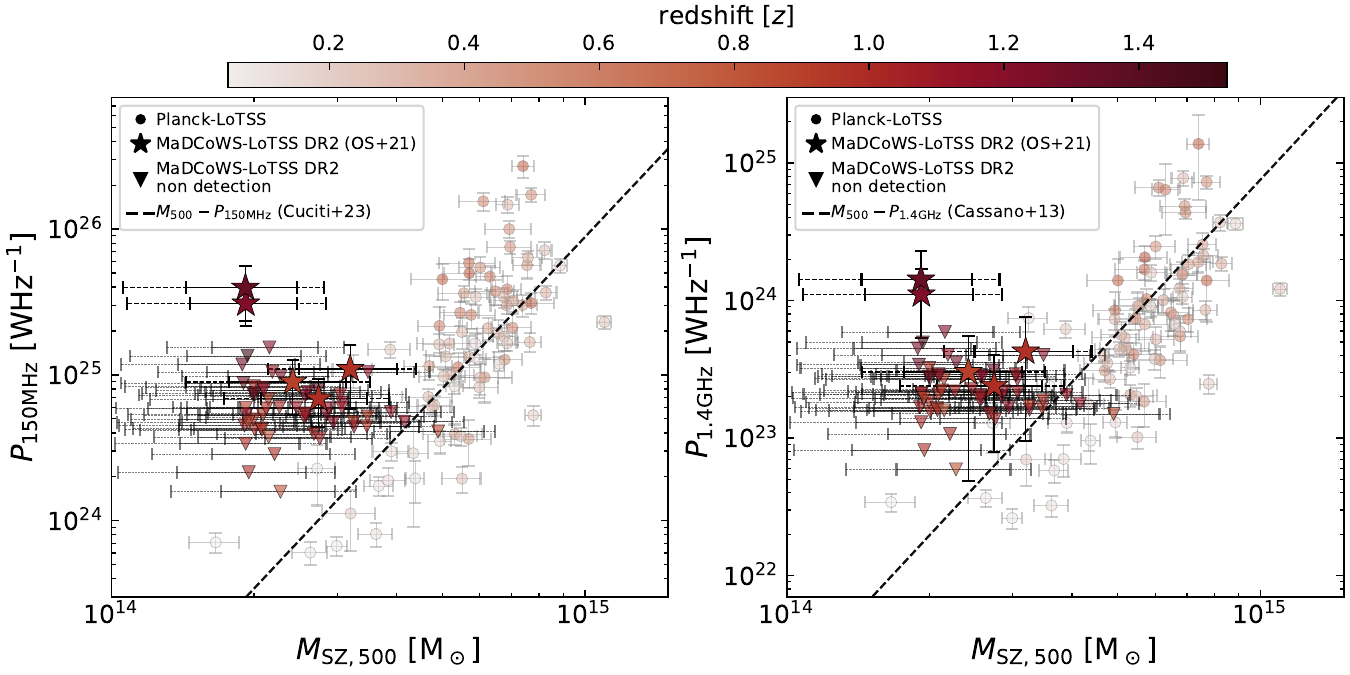}
\caption{As for Fig. \ref{fig:powermass}, but including also the upper limits (low-vertix triangles).}
\label{fig:powermassUL}
\end{figure*}

\section{Limits from the LOFAR observations}\label{apx:powerlimit}
In this section, we present the comparison of the minimum radio power detectable by a standard LoTSS (i.e. 8 hours per pointing) and a deep LOFAR HBA \citep[i.e. 100 hours][]{tasse+21} observations. 

\begin{figure*}[h!]
\centering
\includegraphics[width=0.48\textwidth]{LOFAR150MHzpower_limits_sigma200.pdf}
\includegraphics[width=0.48\textwidth]{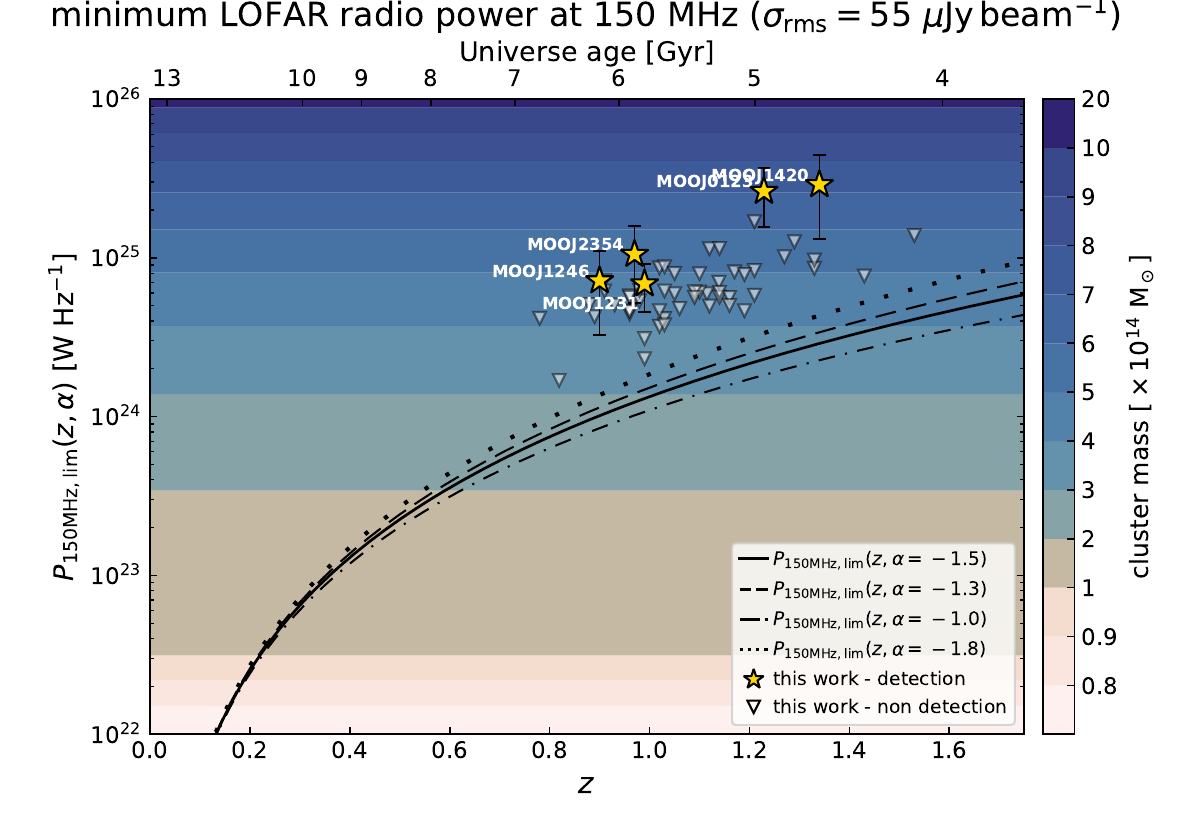}
\caption{Comparison of the observable minimum radio power from ``standard'' (e.g. LoTSS, left) and deep (right) LOFAR observations. The deep observations are taken to reproduce the LoTSS deep fields observations \citep[$>100$ hours][]{tasse+21}
}
\label{fig:deepobspower}
\end{figure*}

\end{appendix}

\end{document}